\documentclass[traditabstract]{aa} 
\usepackage{graphicx} \usepackage{epstopdf}
\usepackage{txfonts}

\usepackage{natbib,twoopt}
\bibpunct{(}{)}{;}{a}{}{,} 
\makeatletter
\newcommandtwoopt{\citeads}[3][][]{\href{http://adsabs.harvard.edu/abs/#3}%
{\def\hyper@linkstart##1##2{}%
\let\hyper@linkend\@empty\citealp[#1][#2]{#3}}}
\newcommandtwoopt{\citepads}[3][][]{\href{http://adsabs.harvard.edu/abs/#3}%
{\def\hyper@linkstart##1##2{}%
\let\hyper@linkend\@empty\citep[#1][#2]{#3}}}
\newcommandtwoopt{\citetads}[3][][]{\href{http://adsabs.harvard.edu/abs/#3}%
{\def\hyper@linkstart##1##2{}%
\let\hyper@linkend\@empty\citet[#1][#2]{#3}}}
\newcommandtwoopt{\citeyearads}[3][][]%
{\href{http://adsabs.harvard.edu/abs/#3}
{\def\hyper@linkstart##1##2{}%
\let\hyper@linkend\@empty\citeyear[#1][#2]{#3}}}
\makeatother
\usepackage{lscape}
\newcommand{\lsim}{\ \raise -2.truept\hbox{\rlap{\hbox{$\sim$}}\raise
    5.truept\hbox{$<$}\ }} \newcommand{\gsim}{\ \raise
  -2.truept\hbox{\rlap{\hbox{$\sim$}}\raise 5.truept\hbox{$>$}\ }}
  
\newcommand{\gi}{$(g{-}i)$\ }
\newcommand{\ui}{$(u{-}i)$\ }
\newcommand{\feh}{$[Fe/H]$\ }
\begin{document}

   \title{VEGAS-SSS II: Comparing the globular cluster systems in
     NGC\,3115 and NGC\,1399 using VEGAS and FDS survey data}

   \subtitle{The quest for a common genetic heritage of globular
     cluster systems}
 
   \author{Michele Cantiello\inst{1}\and
Raffaele D'Abrusco\inst{2}\and
Marilena Spavone\inst{3}\and
Maurizio Paolillo\inst{4}\and
Massimo Capaccioli\inst{4}\and
Luca Limatola\inst{3}\and
Aniello Grado\inst{3}\and
Enrica Iodice\inst{3}\and
Gabriella Raimondo\inst{1}\and
Nicola Napolitano\inst{3}\and
John P. Blakeslee\inst{5}\and
Enzo Brocato\inst{6}\and
Duncan A. Forbes\inst{7}\and
Michael Hilker\inst{8}\and
Steffen Mieske\inst{9}\and
Reynier Peletier\inst{10}\and
Glenn van de Ven\inst{11}\and
Pietro Schipani\inst{3}
}

\institute{INAF Osservatorio Astr. di Teramo, via Maggini,I-64100,
  Teramo, Italy \email{cantiello@oa-teramo.inaf.it}
  \and Smithsonian Astrophysical Observatory, 60 Garden Street, 02138 Cambridge (MA);
  \and INAF Osservatorio Astr. di Capodimonte Napoli, Salita Moiariello, 80131, Napoli, Italy
  \and Dip. di Fisica, Universit\'a di Napoli Federico II, C.U. di Monte  Sant'Angelo, Via Cintia, 80126 Naples, Italy
  \and NRC Herzberg Astronomy \& Astrophysics, Victoria, BC V9E 2E7, Canada;
  \and INAF Osservatorio Astronomico di Roma, Via di Frascati, 33, I-00040 Monteporzio Catone, Italy
  \and Centre for Astrophysics \& Supercomputing, Swinburne University, Hawthorn, VIC 3122, Australia
  \and European Southern Observatory, Karl-Schwarzschild-Str. 2, D-85748, Garching bei M\"unchen, Germany
  \and European Southern Observatory, Alonso de Cordova 3107, Vitacura, Santiago, Chile
  \and Kapteyn Astronomical Institute, University of Groningen, P.O. Box 72, 9700 AV Groningen, The Netherlands
  \and Max Planck Institute for Astronomy, K\"onigstuhl 17, 69117 Heidelberg, Germany}
   \date{}

\authorrunning{Cantiello et al.} \titlerunning{VEGAS-SSS: Genetics of
  globular cluster systems}

\abstract{We analyze the globular cluster (GC) systems in two very
  different galaxies, NGC\,3115 and NGC\,1399. With the papers of this
  series, we aim at highlighting common and different properties in
  the GC systems in galaxies covering a wide range of parameter
  space. We compare the GCs in NGC\,3115 and NGC\,1399 as derived from
  the analysis of one square degree $u-$, $g-$, and $i-$band images
  taken with the VST telescope as part of the VST early-type galaxy
  survey (VEGAS) and Fornax deep survey (FDS). We selected GC
  candidates using as reference the morpho-photometric and color
  properties of confirmed GCs. The surface density maps of GCs in
  NGC\,3115 reveal a morphology similar to the light profile of field
  stars; the same is true when blue and red GCs are taken
  separately. The GC maps for NGC\,1399 are richer in structure and
  confirm the existence of an intra-cluster GC component. We confirm
  the presence of a spatial offset in the NGC\,1399 GC centroid and
  find that the centroid of the GCs for NGC\,3115 coincides well with
  the galaxy center. Both GC systems show unambiguous color bimodality
  in \gi and $(u{-}i)$; the color-color relations of the two GC
  systems are slightly different with NGC\,3115 appearing more linear
  than NGC\,1399. The azimuthal average of the radial density profiles
  in both galaxies reveals a larger spatial extent for the total GCs
  population with respect to the galaxy surface brightness
  profile. For both galaxies, the red GCs have radial density profiles
  compatible with the galaxy light profile, while the radial profiles
  for blue GCs are shallower. As for the specific frequency of GCs,
  $S_N$, we find it is a factor of two higher in NGC\,1399 than for
  NGC\,3115; this is mainly the result of extra blue GCs. By
  inspecting the radial behavior of the specific frequency, $S_N(<r)$,
  for the total, blue, and red GCs, we find notable similarities
  between the trends for red GCs in the two targets. In spite of
  extremely different host environments, the red GCs in both cases
  appear closely linked to the light distribution of field stars. Blue
  GCs extend to larger galactocentric scales than red GCs, marking a
  significant difference between the two galaxies: the blue/red GCs
  and field stellar components of NGC\,3115 appear well thermalized
  with each other and the blue GCs in NGC\,1399 appear to fade into an
  unrelaxed intra-cluster GC population.}
\maketitle


\section{Introduction}

Understanding the formation and evolution of galaxies is fundamental to shed
light on the assembly of the baryonic matter in the Universe during
the cosmic growth of large-scale structures. Although galaxies can be
observed out to extremely large distances, the study of the properties
of local galaxies remains fundamental. The redshift $z\sim0$ is the current
end point of galaxy evolution and an accurate comprehension of all
the properties of local structures is essential for characterizing in
detail the processes that lead to the present organization of matter
\citep{mo10}.

Globular clusters (GCs) are an important tool for understanding the formation
and evolution of galaxies
\citep{harris01,brodie06,peng08,georgiev10,harris13,durrell14}.

Extragalactic, unresolved GCs are possibly the simplest class of
astrophysical objects beyond stars. To a first approximation, GCs host
a simple, single age, and single metallicity stellar
population. However, the last decade has seen a growing amount of
studies proving that the classical paradigm GCs $\equiv$ simple
stellar populations is not valid for a large number of Milky Way (MW)
GCs and for some of the clusters in the Magellanic Clouds
\citep{gratton04,piotto08,carretta09}. Nevertheless, there is no doubt
that GCs host a stellar population that is much simpler than galaxies
in terms of metallicity and age distributions because GCs have a
simpler star formation history.

Two other properties make GCs very useful: old age and high
luminosity.  In the few galaxies beyond the MW for which spectroscopic
or multiband photometric studies of GCs have been carried out, the
results almost uniformly pointed out a population with mean ages
comparable to the GC system of the MW, older than $\sim$10 Gyr
\citep[e.g.,][]{cohen98,cohen03,strader05,chiessantos11}. This makes
GC systems the fossil tracer of the formation of a galaxy and its
environment.

Moreover, extragalactic GCs appear as bright clumps of light on the
otherwise smooth light profile of the galaxy and, under typical
observing conditions from the ground, they appear as point-like
sources. The compactness and high contrast with respect to the
background light from the galaxy and sky make GCs observable out to large
distances. Photometric studies have been carried out for a GC
system at $z\sim0.2$ \citep[$d\sim800$ Mpc, HST/ACS data;][]{alamo13}
and, more recently, at $z\sim0.3$ \citep[$d\sim1250$ Mpc, HST/ACS, and
  WFC3 data;][]{janssens17}. The spectroscopic observations are much
more limited and feasible only for the nearest, brightest GCs even with
the largest 8-10 m class telescopes \citep{brodie14}.

The systematic study of GC systems (GCSs hereafter) in galaxies has
highlighted a wealth of properties that are used to trace the physical
characteristics of the GCS and its host galaxy; these characteristics include the luminosity
function of GCs (GCLF), spatial distribution, projected
surface density, radial color profiles, specific frequency,
 kinematical properties, and color-magnitude relations. All
such properties are effective tracers of the past formation and
evolution history of the galaxy, its physical distance, possible
merging events, mass distribution, etc. \citep{harris01,brodie06}.
\\
\vskip 0.3cm
\noindent
In this paper we present and discuss the results from the analysis of
the properties of the GCSs hosted by \object{NGC\,3115} and
\object{NGC\,1399}. The two galaxies analyzed are very different from
each other and reside in dissimilar environments.  We benefit from
such extreme diversity to search for similarities and differences with
the aim of providing useful constraints to separate intrinsic
properties of GCSs, from the extrinsic properties in a process that
will continue with the future studies of the VST elliptical galaxy
survey (VEGAS) series on small stellar systems (VEGAS-SSS).

Table \ref{tab_props} summarizes the basic properties and
observational details of the galaxies.

NGC\,3115 is one of the closest lenticular galaxies, at $\sim 10$ Mpc,
located far south of its closest group of galaxies, the \object{Leo
  I Group}. The galaxy is very isolated; within a $2\times2~deg^2$
area only one extragalactic source brighter than $m_V\sim15$ mag and
with $cz\leq 1500$ km/s (i.e., $M_V\sim -15$ mag if at the distance of
NGC\,3115) can be found\footnote{Data from the NASA Extragalactic
  Database}. The closest source is the companion galaxy
NGC\,3115-DW01. The total number of extragalactic sources increases
virtually to two objects within an area of $10\times10~deg^2$, but the
second angularly closest and bright object is \object{Sextans A}, a
dwarf spheroidal galaxy in the \object{Local Group}, at $\sim1.4$
Mpc. Hence, on the 100 $deg^2$ area, only one bright galaxy is
spatially close to NGC\,3115.

The GCS of NGC\,3115 is perhaps the best case, beyond our
Galaxy, to reveal a clear-cut bimodal color distribution that has been
unambiguously demonstrated as due to a bimodal metallicity
distribution.  In this case both spectroscopic and (optical and
near-IR) photometric studies consistently indicate the presence of a
 GC system with bimodal metallicity distribution with peaks at \feh
$\sim-1.3$ and $\sim 0.0$ \citep{brodie12,cantiello14}.

The other target, NGC\,1399, is classified as an E1 galaxy located near
the dynamical center of the Fornax cluster and is the second
brightest early-type galaxy of the cluster; the brightest is
\object{NGC\,1316,} which is offset by $\sim3.5$ degrees southwest of
the main body of the cluster \citep{ferguson89,iodice16,iodice17},
although at nearly the same distance \citep{cantiello13}. The
$2\times2~deg^2$ region around NGC\,1399 harbors 19 objects brighter
than $m_V\sim16.5$ mag and $cz\leq3000$ km/s, which is equivalent to $M_V\sim
-15$ mag if the source is at the adopted distance for NGC\,1399; the
number increases to 43 objects within a $10\times10~deg^2$ area.  Thus,
NGC\,1399 resides in a much denser galaxy environment than NGC\,3115.

The GCS of NGC\,1399 is well studied in the literature with a great
variety of studies from ground-based photometric and spectroscopic
data
\citep{geisler90,kp97b,forbes01,richtler04,bergond07,firth07,schuberth08,hilker15}
and space-based X-ray, optical, and near-IR data \citep[][in addition
  to the others cited elsewhere in this
  work]{forbes98,gebhardt99,grillmair99,larsen01,kundu05,villegas10,mieske10,liu11,paolillo11,puzia14,jordan15}.

NGC\,1399 was one of the first cases of an early-type galaxy with the
GCS revealing a pronounced bimodal color distribution in optical bands
\citep{ostrov98,dirsch03}. A more recent study of the inner GC system
in the galaxy, based on ACS optical and WFC3 near-IR data, has
obtained observational evidence for the nonlinearity of its
color-metallicity relations \citep{blake12a}. The study also revealed
inconsistent properties of the color distributions between the optical
and near-IR colors, such as different likelihoods of color bimodality
and different red to blue GC ratios. Such inconsistency between
optical and near-IR color distributions was one of the predictions
proposed as evidence in support of the so-called projection scenario
\citep[][which we discuss further in the next
  section]{yoon06,cantiello07d}. Unfortunately, despite many studies
over the years, spectroscopic metallicities for a large sample of GCs
in NGC 1399 are not available.

A further peculiarity of GCs in NGC\,1399 is the observational
evidence supporting the membership of a fraction of the GCs to the
Fornax cluster rather to the galaxy itself
\citep{grillmair94,kp98,bassino03,schuberth10,dabrusco16}. The
presence of an intergalactic GC population allowed the
  resolution of another peculiarity of the GCS in the galaxy. The
first studies on the GC specific frequency\footnote{A parameter
  relating the galaxy and GCS properties, quantified as the number
  of GCs per unit galaxy luminosity $S_N\equiv N_{GC}\times 10 ^{0.4
    (M_V+15)}$, where $N_{GC}$ is the total number of clusters and
  $M_V$ is the total absolute visual magnitude of the galaxy
  \citep{harris81,harris91}.}, obtained unusually high $S_N$ values
\citep{harris87,bridges91,wagner91,geisler90,ostrov98}. By associating
a faction of GCs to the whole cluster and normalizing the specific
frequency to the total light (galaxy plus halo), the estimated $S_N$
goes back to normal values for a system such as NGC\,1399 and its
environment \citep{peng08,georgiev10}.

Because of the isolation of NGC\,3115 and the much denser cluster
environment of NGC\,1399, together with the presence of an
intra-cluster GC component, our aim is to analyze the properties of
the two GCSs and place their observed properties into a homogeneous
and self-consistent picture their to formulate useful constraints upon
the importance of mergers and interactions on the global properties of
the GC system.
\vskip 0.3cm
\noindent
Characterizing the GCS color bimodality in galaxies hosted in various
environments, with different masses and morphologies, is one of the
science cases for VEGAS-SSS. This science case has strong implications
for GCS formation.  Here we provide some background on the color (and
metallicity) distribution properties of GC systems.

Color bimodality (CB hereafter) is a simple and yet dramatically
important feature observed in nearly all galaxies massive enough to
host a reasonable number of GCs ($N_{GC}\gsim50$). The GC bimodality
consists of the presence of two well-defined peaks in color
distribution separated by $\sim$0.2 up to $\sim$1 mag
\citep{larsen01,kundu01,harris06a}, depending mainly on the
  specific filters involved. Color bimodality is a very ubiquitous feature, is
observed in very massive ellipticals near $M_B\sim-22$ mag (e.g., in
M\,87, the massive elliptical in Virgo), and in spheroidal
galaxies some 5 magnitudes fainter \citep{peng06}. This feature is found in
ellipticals and spirals. It is found in galaxies over a range
of environments from clusters and groups to isolated systems
\citep{brodie06,peng06}.

\citet[][]{ashman92} initiated the interpretation that GC color
  bimodality was a consequence of GC metallicity
  bimodality. The idea was supported by the evidence that our
best-known reference galaxy, the Milky Way, has a GC system with a
bimodal \feh distribution, where the metal-poor GCs (peak \feh
$\sim{-}1.6$) have a spatial distribution and kinematic properties
associated with the Galaxy halo, and the metal-rich GCs (peak \feh
$\sim{-}0.55$) are more centrally concentrated and show kinematic
properties similar to bulge stars \citep{cote99}.  Moreover, as
already mentioned, there is growing evidence based on the data
collected from the 1990s and beyond that GCs are all nearly coeval and old
\citep[t$>$10
  Gyr;][]{puzia04,cohen98,cohen03,strader05,chiessantos11}. As a
consequence of the nearly constant old age and the fact that at fixed
metallicity integrated optical colors for ages older than $\sim$8 Gyr
do not vary significantly because of age\footnote{For example, at
  fixed metallicity the change in V--I color for a simple stellar
  population in the age interval 8-14 Gyr is $|\Delta(V{-}I)_{(8-14
    Gyr)}|\sim0.05$ \citep[SPoT simple stellar population
    models;][]{raimondo05,raimondo09}.}, the metallicity is left as
the only parameter that can explain the observed CB.

Thus it was natural to equate CB with metallicity bimodality (FeB
hereafter).  Such equivalence has dramatic implications for the
processes of formation and evolution of GC systems and their host
galaxies.  If the assumption is correct then the FeB of GC systems is
a very common feature and formation models need to account for the
existence of two GC subpopulations with markedly different mean \feh,
which formed either in two different epochs or with two different
mechanisms, or a combination of both.  The great success of FeB as a
valid interpretation of CB comes also from the fact that, as already
mentioned, it was predicted before it was a commonly observed feature
\citep{ashman92}.

Since the 1990s many authors proposed several other models for GCS
formation accounting for the FeB. All models had their own
ingredients, pros, and cons. All of the models can be basically associated with
one of the following three classes: \\

$\bullet$ FeB forms as a consequence of the dissipative merging of
gas-rich galaxies: Blue GCs are pristine clusters hosted by the halos
of the first galaxies. Red GCs form later, along with intense star
formation due to galaxy mergers. In such a scenario, a $\sim2$ Gyr age
gap exists between blue and the GC subpopulations \citep{ashman92}. \\

$\bullet$ FeB from hierarchical growth: Massive seed galaxies host
intrinsically more metal-rich GC systems because their potential well
was able to capture the metals produced by the first generations of
stars more efficiently. The satellite, lower mass galaxies host a more
metal poor GC system. As a consequence of dissipationless merging
between a massive galaxy and its nearby lower mass companions, the
system of red GCs is joined by more metal-poor, blue GCs
\citep{beasley02,cote98,hilker99c}. There is no age gap required
between GCs in this class of model. \\

$\bullet$ FeB generated in situ: Both blue and red GCs are indigenous
to the galaxy, formed in two distinct phases of star formation, and
emerge from gas of differing metallicity. The blue and metal-poor GCs
formed at an early stage in the collapse of the protogalactic
cloud. The red and metal-rich GCs formed out of more enriched gas,
roughly contemporaneously with the galaxy stars. An age gap of 1--2
Gyr is expected between the two GC subpopulations \citep{forbes97}.\\

The proposed scenarios include predictions and interpretations of
observed quantities; successfully explaining only some of the known
properties of GC systems
\citep{beasley02,kravtsov05,bekki08,griffen10,kruijssen14,li14}. One
common ingredient in all scenarios is the existence of two
subpopulations of GCs with different mean metallicities. However, the
color-to-metallicity bimodality equivalence has been challenged by
\citet{yoon06}. The \citeauthor{yoon06} argument against FeB focuses
on the linearity of the color-metallicity relations (CMR). If CMRs are
close to linear, or such to a good approximation, then the shape of
any distribution from one space, for example, in the color domain, is
preserved when transferred into another space, for example, to \feh.

Since 2006 an increasing amount of observational evidence has been
collected showing that the CMRs for the colors mostly used for the
study of extragalactic GCSs ($V{-}I$, $g{-}z$, and $B{-}I$) exhibit a
degree of nonlinearity
\citep{peng06,richtler06,blake12a,yoon11a,yoon13}\footnote{Earlier
  evidence of CMR nonlinearities was presented in \citet{harris02}
  and \citet{cohen03}.}. Nonlinear CMRs might imply that the observed
CBs have a completely different shape when translated into
metallicity  \citep{richtler06,cantiello07d,blake10a,kim13}.  If the
CMR is strongly nonlinear, then a fraction of the observed CBs could
be explained by the so-called {\it bimodality projection effect},
in which the observed color distribution does not actually match with the
properties of the progenitor \feh distribution. Determining the
fraction of real versus projected FeB will likely provide a key
constraint for galaxy and GCs formation models.

 Recently, \citet{harris17}, using HST/ACS data of BCG galaxies,
 showed that color bimodality for these large GC systems breaks down
 and becomes more complex. In their photometric study of GCs, for
 three out of five BCGs in the series of papers, the authors conclude
 that ``...the imposition of a bimodal Gaussian numerical
   model... begins to look increasingly arbitrary''.

The best method to reveal real FeB in GC distributions, and whether
they are ubiquitous, is to measure accurate spectroscopic
metallicities for a large sample of GCs in individual galaxies.
Unfortunately, this requires good S/N and is very expensive in terms
of telescope time. As an example, even though more than 500 GCs in
NGC\,1399 have published radial velocities \citep{schuberth10}, only a
handful have published metallicities. The \citeauthor{schuberth10}
analysis reveals significantly different kinematics for blue and red
subpopulations.  In a study of NGC\,1399 GCs by \citet{kp98} with the
Keck telescope, these authors measured spectroscopic metallicities for
18 GCs. A comparison with optical V--I colors indicated a slightly
nonlinear CMR relation. Their GC metallicity distribution hinted at
two peaks with \feh $\sim$ --1.5 and --0.5. Hence, NGC\,1399 GC system
hints at metallicity bimodality, but a larger sample is required for
more robust conclusions, as the sample of 18 GCs represents $<<1\%$ of
the entire GC population of $N_{GC}\sim6500$
\citep[][]{dirsch03,bassino06}\footnote{A more recent estimate from
  the ACSFCS survey is $N_{GC}\sim10000\pm3000$ \citep{liu17}.}. Even
adding the sample of $\sim50$ bright Ultra Compact Dwarfs (UCDs) and
massive GCs with estimates from \citet{hilker15}, the total number of
objects with known metallicity is $\lsim1\%$ of the total population.

Claims of GC system FeB have been made for several other galaxies
based on metallicity sensitive lines at optical wavelengths,
for example, NGC\,5128 \citep{beasley08}, NGC\,4594 \citep{ab11}, NGC\,4472
\citep{strader07}, and M\,87 \citep{cohen98}. The interpretation of
such claims, however, are still controversial
\citep{blake10a,blake12a}. Using the near-IR calcium triplet lines
(CaT) \citet{usher12} examined the GC metallicity distributions of
several galaxies from the SLUGGS survey. They concluded that six out of
the eight galaxies with sufficient data revealed metallicity
bimodality. However, the color-CaT relations were often
somewhat nonlinear and the bimodality peaks appeared at different
inferred metallicities. The presence of a third, intermediate
metallicity subpopulation (e.g., NGC\,4365) can complicate the
interpretation. The case of NGC\,3115, which is the nearest galaxy in their
sample, revealed the best consistency between its color and CaT-based
metallicity distributions.

Here we use VEGAS images to obtain photometric data of the GC
systems around two very different host galaxies - NGC\,1399 and
NGC\,3115.  We focus on the color distributions, color-color
relations, and spatial properties of their GC systems.
\vskip 0.3cm
\noindent The paper is organized as follows:\ The next section
describes the observations and the procedures for data reduction and
for the selection of GC candidates. The analysis of the surface
distribution of GC candidates, color distribution and color-color
relations, radial density profiles, and local specific frequency are
presented in Section \ref{sec_results}.  Section \ref{sec_discussion}
provides a discussion of the results. In Section \ref{sec_summa} we
summarize our findings. Finally, Appendix \ref{appendix} contains a
list of several interesting objects, mostly in the field of NGC\,3115.

\section{Observations, data reduction, and analysis}
\label{sec_obs}

\subsection{Observations and image processing}

The data used in this study are from two VST GTO surveys: VEGAS (P.Is:
M. Capaccioli, E. Iodice), and the Fornax Deep Survey (FDS; P.Is:
M. Capaccioli and R. Peletier). We analyzed $ugi$-band data of
NGC\,3115, from VEGAS, and of the $\sim1$ sq. degree area centered on
NGC\,1399, a target common to both VEGAS and FDS.

A detailed description of the surveys and procedures adopted for data
acquisition and reduction can be found in \citet{grado12},
\citet{capaccioli15}, and \citet{iodice16}. Therefore, we only provide
a brief overview here.

The data were processed with VST-tube \citep{grado12}, a pipeline
specialized for the data reduction of VST-OmegaCAM executing
pre-reduction (bias subtraction, flat normalization), illumination and
(for the $i$-band) fringe corrections, photometric, and astrometric
calibration. Unlike our previous study on compact stellar systems in
NGC\,3115 \citep{cantiello15}, based on only $g$ and $i$ data, here we
did not apply any selection cuts to the input images adopted for the
final mosaics of the two targets. A summary of the VST observations is
provided in Table \ref{tab_props}. For both galaxies we assumed
Galactic extinction from the \citet{sf11} recalibration of the
\citet{sfd98} infrared-based dust maps.

The VEGAS survey has been obtaining deep optical imaging of galaxies with
heliocentric redshifts $cz\leq 4000~km/s$ and total magnitude
$M_B^{tot}\leq-19.2$ mag to study the signatures of diffuse stellar
components and of compact stellar systems, out to poorly constrained
galactocentric radii. At completion, it is expected the survey will
collect data for $\sim$100 early-type galaxies across a range of
environments and masses. To date, data for about 30 galaxies have
been collected. The status of the survey is regularly updated at the
URL \url{http: //www.na.astro.it/vegas/VEGAS/VEGAS_Targets.html.}

The FDS is a joint effort of the VST FOCUS (Fornax ultra-deep survey,
P.I: R. Peletier) and VEGAS surveys. Similar to VEGAS, the project
plan is to obtain deep optical imaging data. The target is the Fornax
cluster, which is the second closest galaxy cluster after Virgo, and the
scientific objectives of the survey are numerous: the study of the
galaxy luminosity function, derivation of galaxy scaling
relations, determination of the properties of compact stellar
systems (from GCs to UCDs and cEs), an accurate determination of
distances and 3D geometry of the Fornax cluster, and analyses of
diffuse stellar light and galaxy interactions, etc. Some first
results based on FDS data have already been presented in
\citet{iodice16} to study the diffuse stellar halo of the cluster
core to very faint limits, and in \citet{dabrusco16}, where a first
analysis of the cluster-wide globular cluster population is
described. The survey also includes several follow-up programs, such as
the MUSE observations of NGC\,1396 \citep{mentz16} and the
spectroscopic follow-up of $\sim2000$ compact sources
\citep[observations carried out with VIMOS;][in preparation]{pota17}.

Images were calibrated in the SDSS photometric system using several
\citet{landolt92} standard fields with calibrated SDSS photometry.  By
comparison with the literature data available for the $g$ and $i$
bands, we found median differences below $\sim0.05$ mag with
$\leq0.05$ mag $rms$ in available bands - the data for NGC\,3115 are
from the catalogs of \citet{jennings14} and, for NGC\,1399 GCs, the
$g$-band photometry is from \citet{jordan15}. The median offset did
not show any trend with color.

%
\begin{table}
\caption{\label{tab_props} Properties of the targets}
\centering
\begin{tabular}{lcc}
  \hline\hline
     &  NGC\,3115 & NGC\,1399 \\
\hline
Gal. longitude ($l$, deg) &  247.782502  &  236.716356 \\
Gal. latitude  ($b$, deg) &   36.780999  &  -53.635774  \\
$cz$ (km/s)&  663$\pm$4   & 1425 $\pm$     4 \\
$(m{-}M)$      & 29.87$\pm$0.09  & 31.51$\pm$0.03 \\
$M_V^{tot}$(mag) & -20.9    & -23.4 \\
Type         &   S0        &   E1 \\
E(B${-}$V)   &  0.078       &  0.013 \\
$R_{eff}$(arcsec) &   57        &   49 \\
$T_{type}$                &  $-2.9\pm0.6$ & $-4.6\pm0.5$  \\
$\sigma$ (km/s)  & $259\pm3$    & $334\pm5$ \\
$Mg_2$ (mag)     & $0.288\pm0.002$ & $0.335\pm0.002$ \\
\hline
\multicolumn{3}{c}{Observational details}\\
\hline
Exp. t. $u$ (s)&  14800 &   10200    \\
Exp. t. $g$ (s)&  8675 &  6300 \\
Exp. t. $i$ (s)&  6030 &  4960 \\
PSF ($u/g/i$, arcsec)    &1.1/1.0/0.9  & 1.3/1.2/1.1  \\
\hline 
\end{tabular}
\tablefoot{For NGC\,3115 we adopted the distance from \citet{tonry01},
  using the updated calibration zeropoint from \citet{cantiello13};
  the total $V$ magnitudes is from \citet{devaucouleurs91}, obtained
  with $B{-}V\sim1$ from the HyperLeda archive
  \url{http://leda.univ-lyon1.fr/}; the effective radius is from
  \citet{capaccioli87}. For NGC\,1399 we adopted the mean Fornax
  cluster distance from \citet{blake09}; the total magnitude is the
  $g$-band from \citet{iodice16}, transformed to $V$ using
  $g{-}V\sim0.5$ from \citet{cook14}, derived assuming $B{-}V\sim1$
  from HyperLeda; the effective radius is from \citet{iodice16}. The
  morphological type code $T_{type}$, velocity dispersion $\sigma$ and
  $Mg_2$ are from HyperLeda, the remaining properties are taken from
  NED \url{http://ned.ipac.caltech.edu}.}
\end{table}

\subsection{Galaxy modeling and subtraction, and point source photometry}
To study GCs, we needed to minimize the contamination due to light
from bright galaxies in the fields, i.e., NGC\,3115 and
NGC\,3115-DW01 in one case, and NGC\,1399, NGC\,1379, NGC\,1380,
NGC\,1381 NGC\,1382, NGC\,1386, NGC\,1387, NGC\,1389, NGC\,1396, and
NGC\,1404 for Fornax.  To model and subtract the galaxies, we used the
ISOPHOTE/ELLIPSE task in IRAF/STSDAS
\citep{jedrzejewski87}\footnote{IRAF is distributed by the National
  Optical Astronomy Observatory, which is operated by the Association
  of Universities for Research in Astronomy (AURA) under cooperative
  agreement with the National Science Foundation.}.  After modeling
and subtracting the profiles of galaxies, to produce a complete
catalog of all sources in the VST field of view, we independently ran
SExtractor \citep{bertin96} on the galaxy-model-subtracted frame for
each filter. We obtained aperture magnitudes within a diameter
aperture of eight pixels ($\sim1\farcs68$ at OmegaCAM resolution) and
applied aperture correction to infinite radius. The aperture
correction was derived from the analysis of the curve of growth of
bright isolated point-like sources. The photometric catalogs in the
three bands were then matched adopting $0\farcs5$ matching radius and
are available on the project web-page and on the CDS archive.

\subsection{Globular cluster selections}
\label{sec_sel}

%
\newpage
\begin{table*}
\caption{\label{tab_selparam} Photometric and morphometric selection criteria adopted}
\centering
\tiny
\begin{tabular}{lccc}
  \hline\hline
Quantity        &  Band  & NGC\,3115           & NGC\,1399 \\
\hline
$\Delta X_{6-12}$&$g$/$i$ & $\pm 0.15$                              & $\pm 0.15$                          \\   
CLASS\_STAR     &   $g$  &$\geq0.4$                               &$\geq0.4$                            \\            
CLASS\_STAR     &   $i$  &$\geq0.7$                               &$\geq0.2$                            \\          
PSF FWHM        &   $g$  &$\geq 3.5$, $\leq7.5$                    &$\geq 4.5$, $\leq8.5$                 \\        
PSF FWHM        &   $i$  &$\geq 3.5$, $\leq7.5$                    &$\geq 4.5$, $\leq8.5$                  \\                  
Flux Radius     &   $g$  &$\leq 4.5$                              &$\leq 4.5$                            \\                  
Flux Radius     &   $i$  &$\leq 4.0$                              &$\leq 4.5$                            \\                  
Kron Radius     &$g$/$i$ &$\leq 5$                                &$\leq 5$                               \\                 
Petro. Radius   &   $i$  &$\geq 3.5$, $\leq7.5$                    &$\geq 3.5$, $\leq7.5$                   \\                 
Elongation a/b  &$g$/$i$ &$\leq 1.5$                              &$\leq 1.5$                              \\                
\ui             &        &$\geq1.25$, $\leq3.75$                   &$\geq1.25$, $\leq3.75$                   \\                
\gi             &        &$\geq0.4$, $\leq1.4$                     &$\geq0.4$, $\leq1.4$                    \\                 
$\Delta$\ui     &        &$\leq 0.3$                              &$\leq 0.3$                             \\                 
$\Delta$\gi     &        &$\leq0.2$                               &$\leq0.2$                              \\                 
$\Delta$ mag    &$g$/$i$ &$\leq0.2$                               &$\leq0.2$                             \\                  
$m^{bright}$     &   $g$  &19.5                                      &20.2                                      \\               
$m^{bright}$     &   $i$  &18.5                                      &19.2                                    \\                
color-color     &        & $\mid \gi-[0.380\ui-0.005]\mid\leq0.2$  &$\mid \gi-[0.362\ui-0.0205]\mid\leq0.2$   \\
\hline 
\end{tabular}
\tablefoot{Explanation of listed parameters. $\Delta X_{6-12}$: Interval adopted for the
  magnitude concentration index in $g-$ and $i-$band. SExtractor
  output parameters: CLASS\_STAR: Neural-Network-based star/galaxy
  classifier; PSF FWHM: point spread function full width at half
  maximum; Flux Radius: half light radius; Kron Radius: defined from
  the automatic aperture photometry, intended to give a precise
  estimate of total magnitude, inspired to \citet{kron80} first
  moments algorithm; Petro. Radius: the radius at which the surface
  brightness of the isophote is $\eta$ times the average surface
  brightness within the isophote; Elongation: semi-major over
  semi-minor axis ratio \citep[see][and references therein for more
  details]{bertin96}. Other selection parameters.  \ui and \gi: colour
  intervals adopted from the comparison with empirical data and
  stellar population models; $\Delta$\gi and $\Delta$\ui: maximum
  allowed error on colors; $\Delta$ mag: maximum error on magnitude;
  $m^{bright}:$ bright cut magnitude, at $\sim 3\times\sigma$
  brighter than the turn-over magnitude of the GCLF; color-color:
  equation of color-color region used for selecting GC candidates.}
\end{table*}

\begin{figure*} 
   \centering
   \includegraphics[width=7cm]{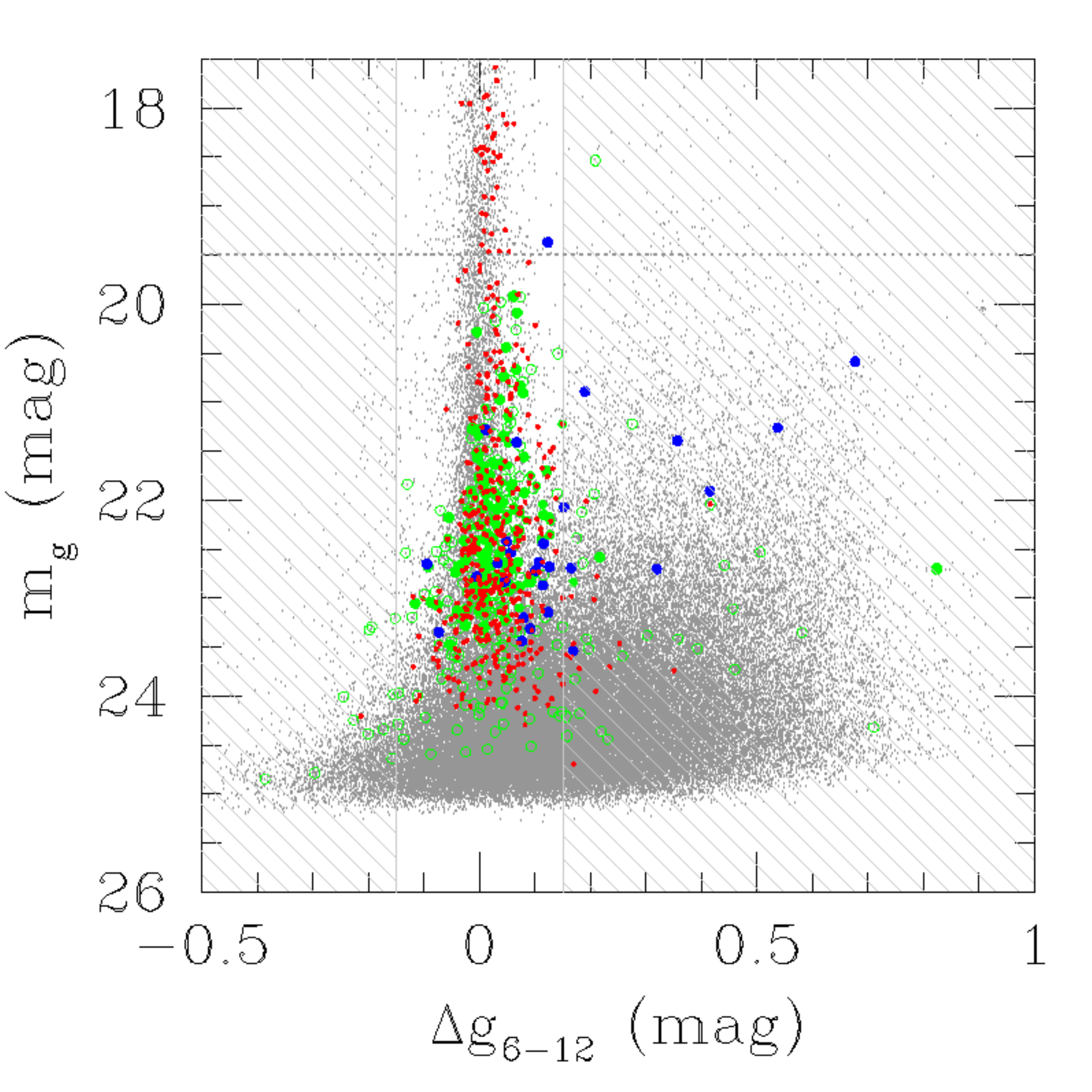}
   \includegraphics[width=7cm]{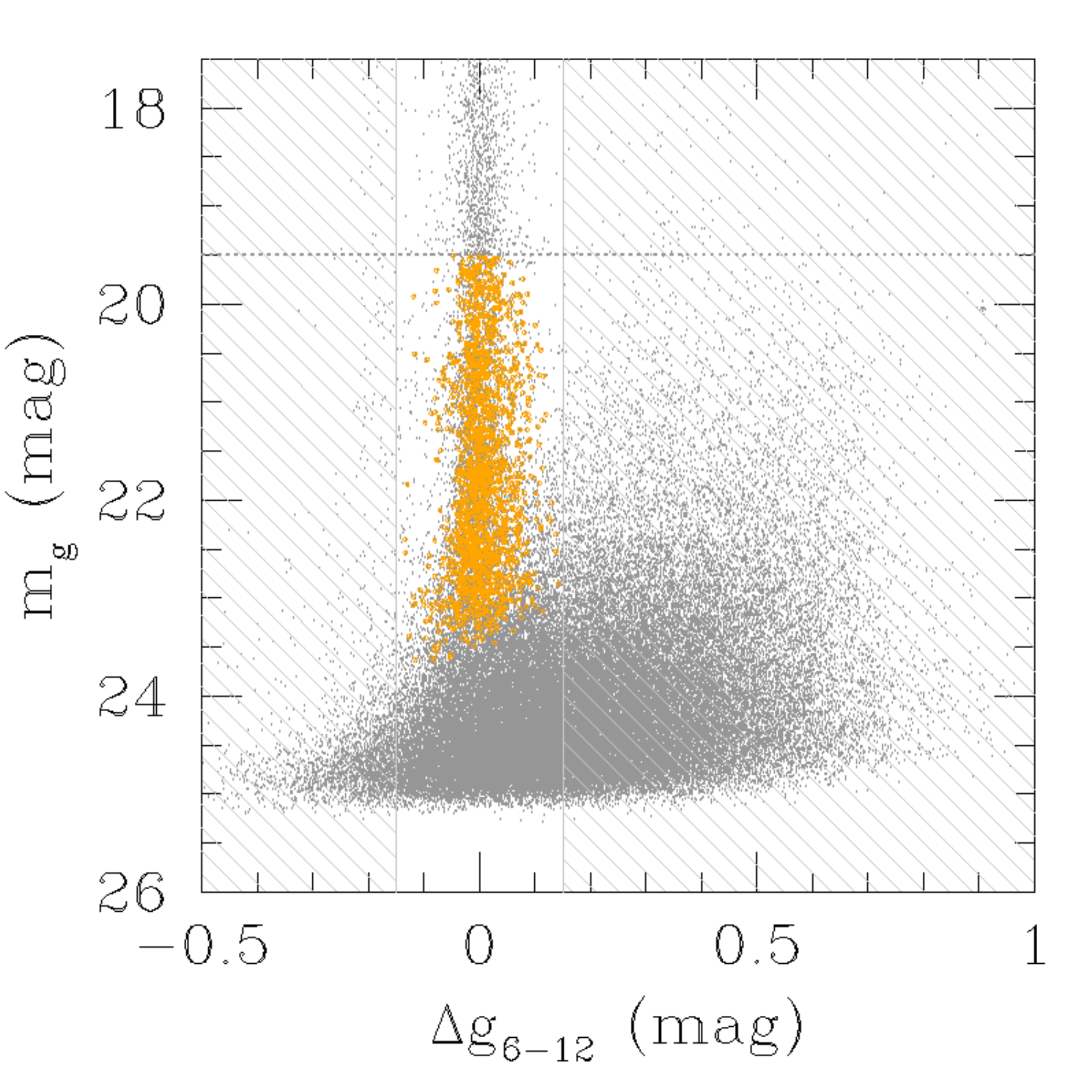}
   \caption{Left panel: $g$ magnitude vs. concentration index
     ($\Delta g_{6-12}\equiv g_{6pix}-g_{12pix}$) for the full $ugi$
     matched sample (in gray) in NGC\,3115. Spectroscopic confirmed
     GCs are shown with green filled circles \citep[from][]{arnold11}and
     photometric confirmed GCs from ACS are shown with empty green
     circles \citep[from][]{jennings14}. Blue circles indicate UCDs from
     \citet{jennings14}. Red dots indicate confirmed stars from both the
     spectroscopic and ACS studies cited. The gray hatched area
     shows the regions of sources rejected based on the concentration
     index selection. The horizontal dotted line indicates the
       bright magnitude cut. Right panel: as left panel, except that
     GC candidates before color-color selection are shown with orange
     empty circles.}
   \label{dmag_g}
   \end{figure*}

\begin{figure*} 
   \centering
   \includegraphics[width=6cm]{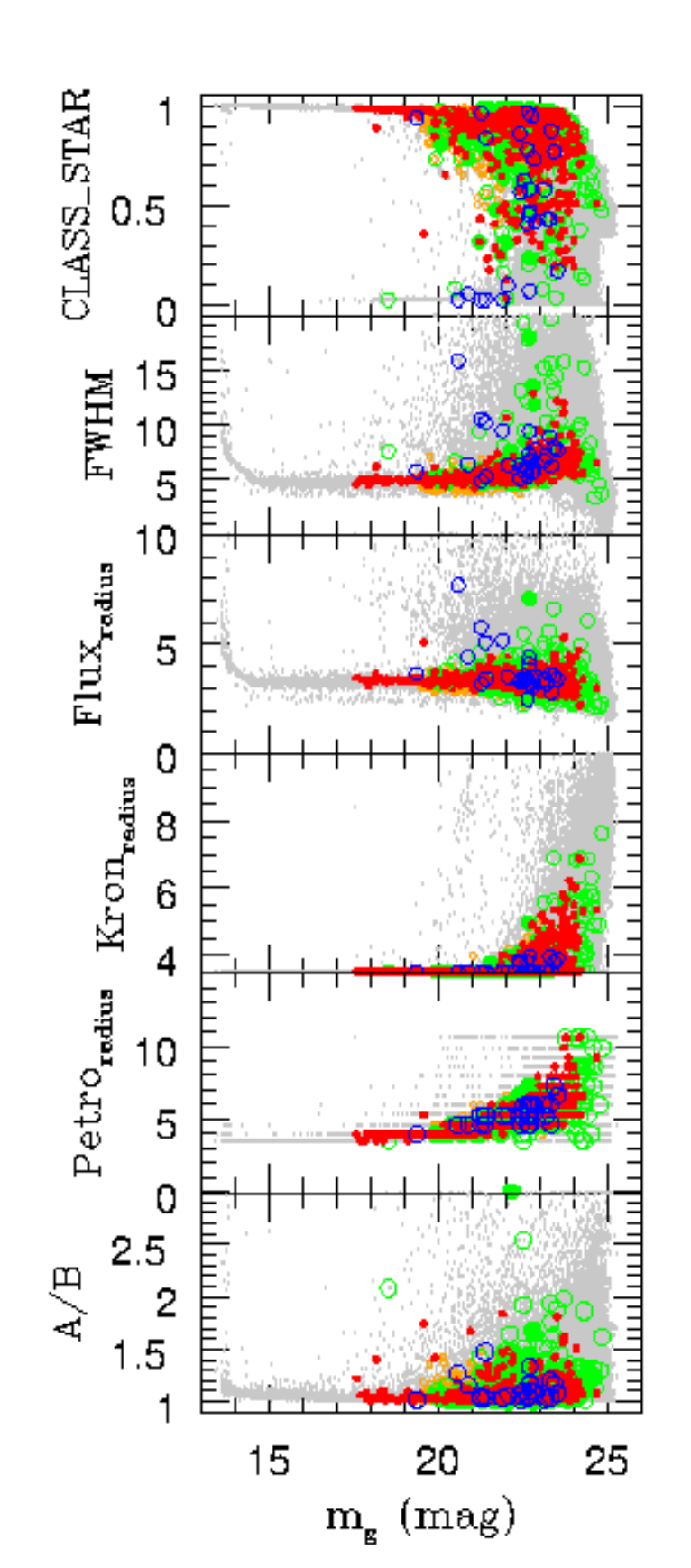}
   \includegraphics[width=6cm]{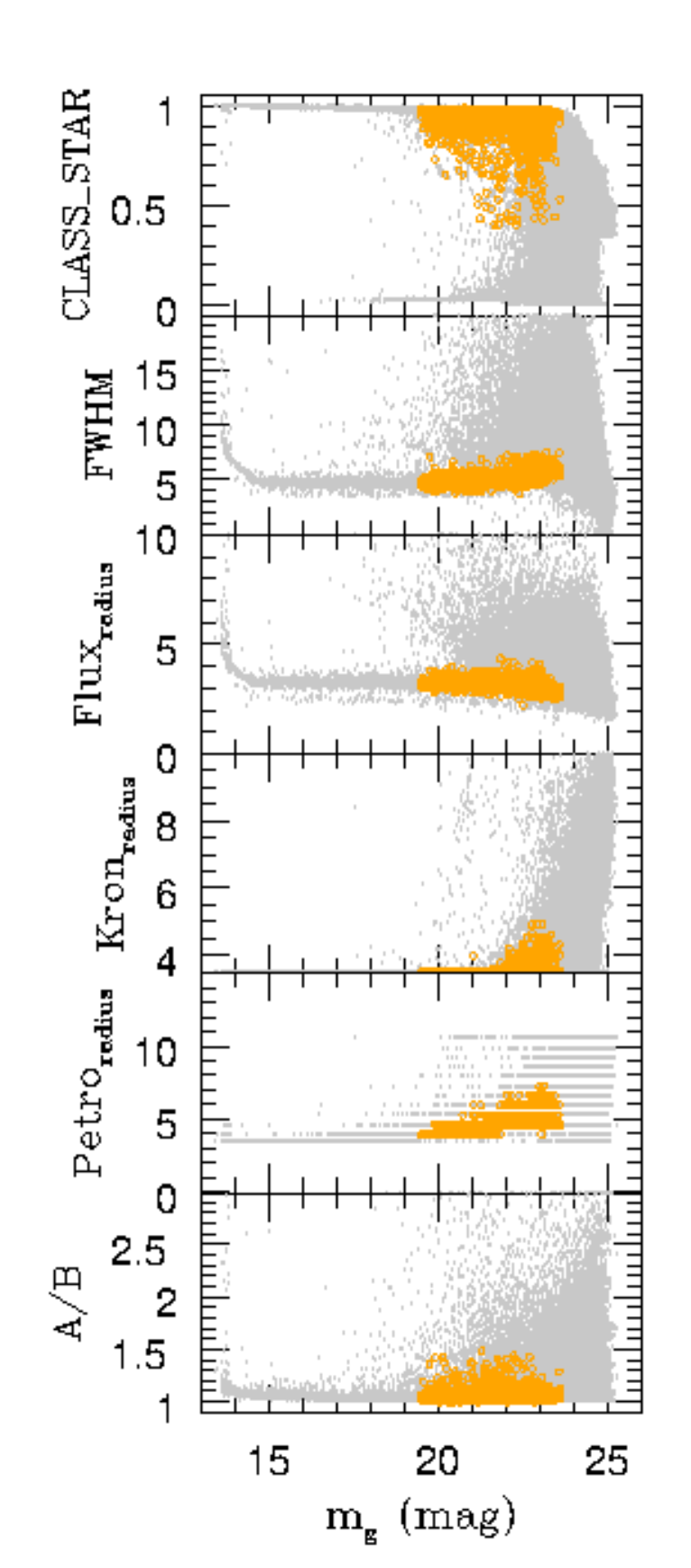}
   \caption{Left panels: SExtractor output parameters in the $g$-band
     for the full sample of detected sources in NGC\,3115 vs.
     corrected aperture magnitude (within 8 pixel diameter, gray
     dots).  Color coding is the same as in Figure \ref{dmag_g}. Right
     panels: as left, but selected GC candidates are shown with orange
     circles.}
   \label{sel_g}
   \end{figure*}

\begin{figure*} 
   \centering
   \includegraphics[width=8cm]{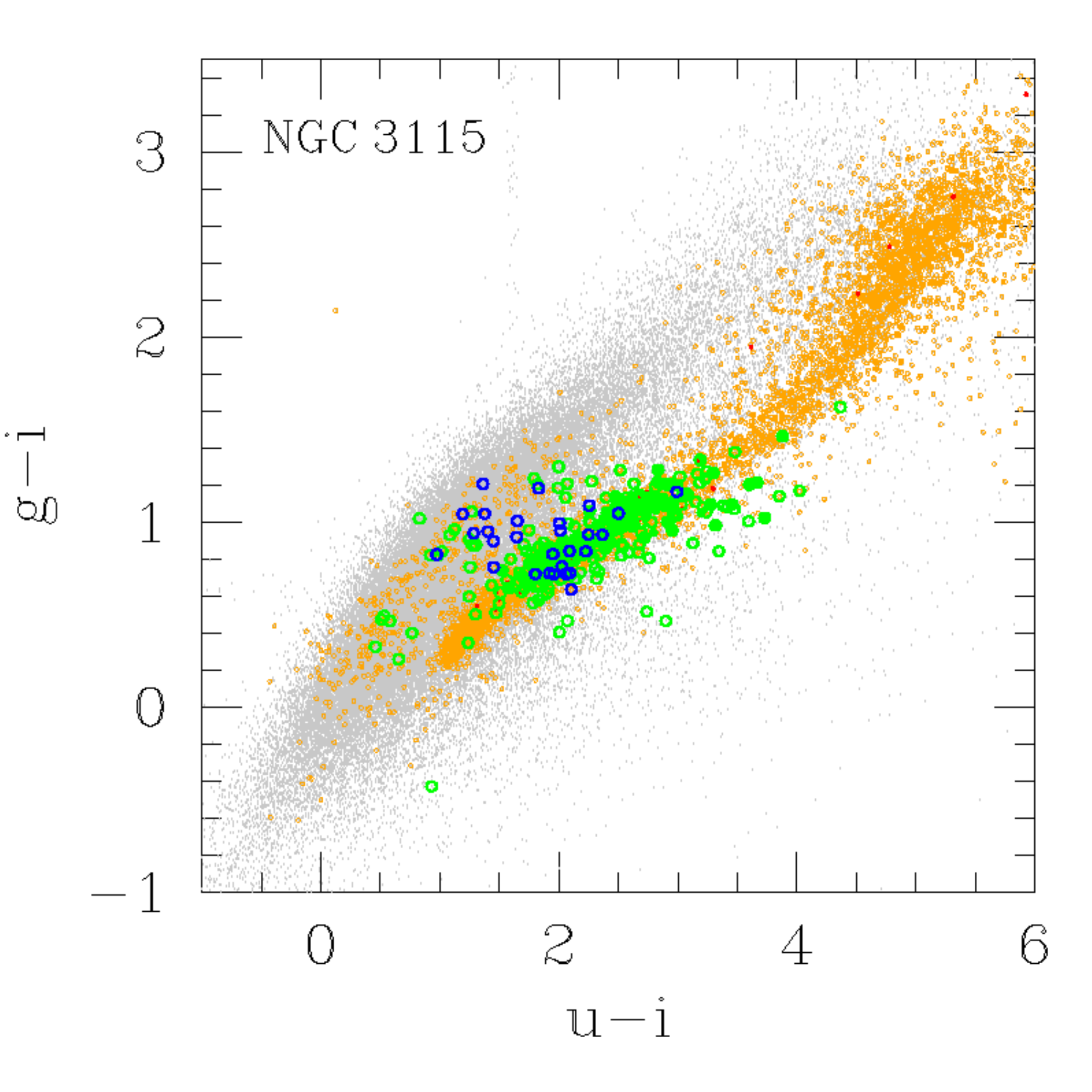}
   \includegraphics[width=8cm]{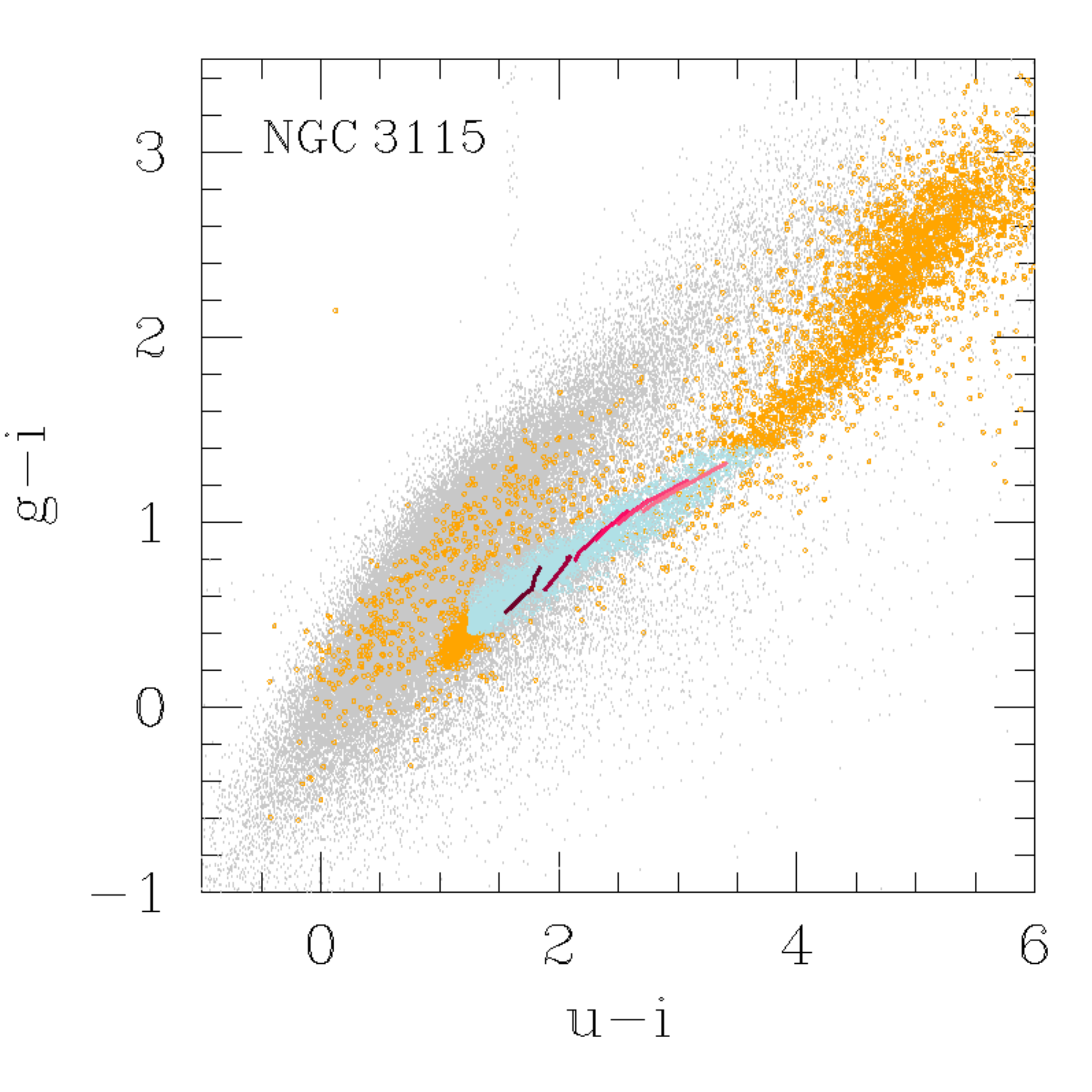}
   \includegraphics[width=8cm]{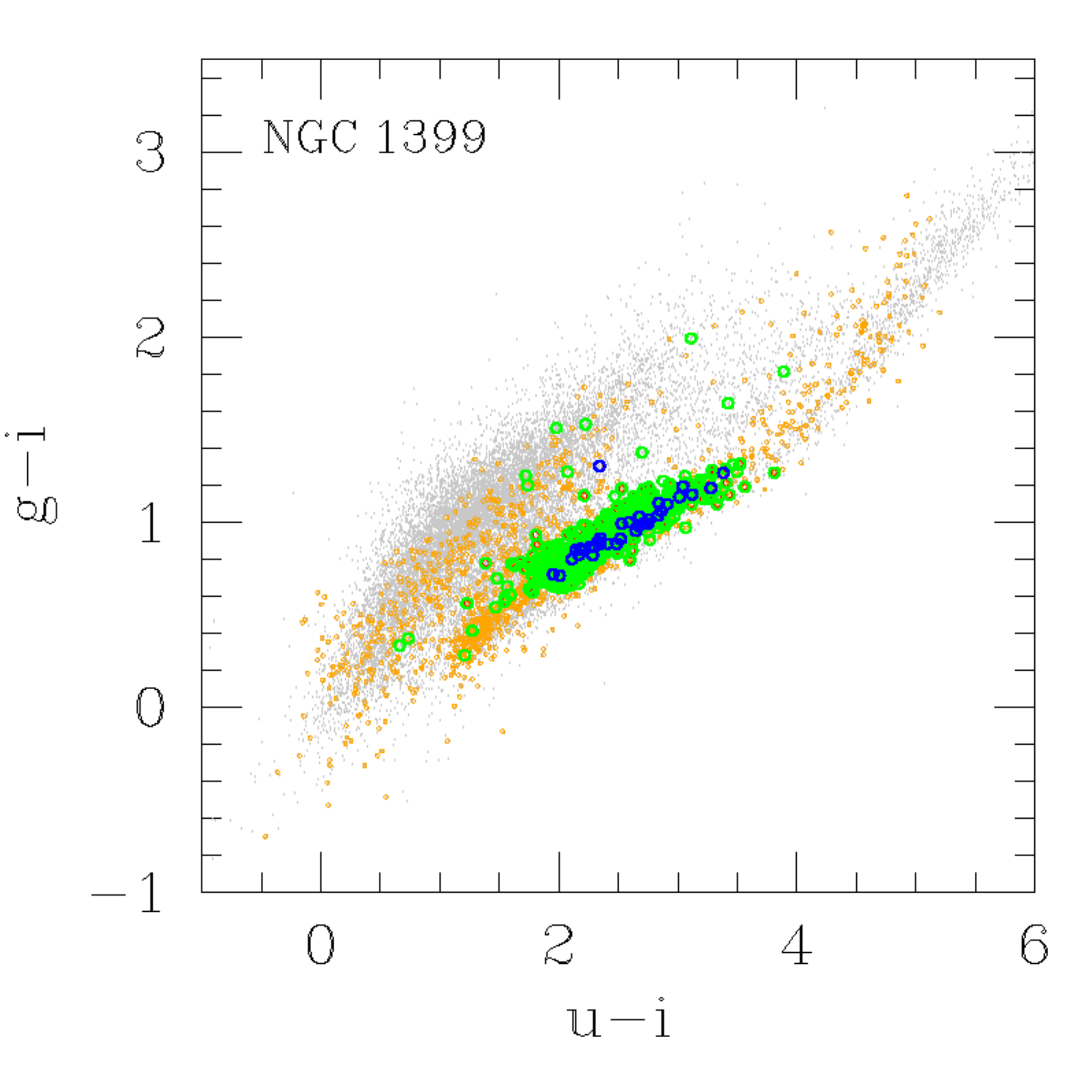}
   \includegraphics[width=8cm]{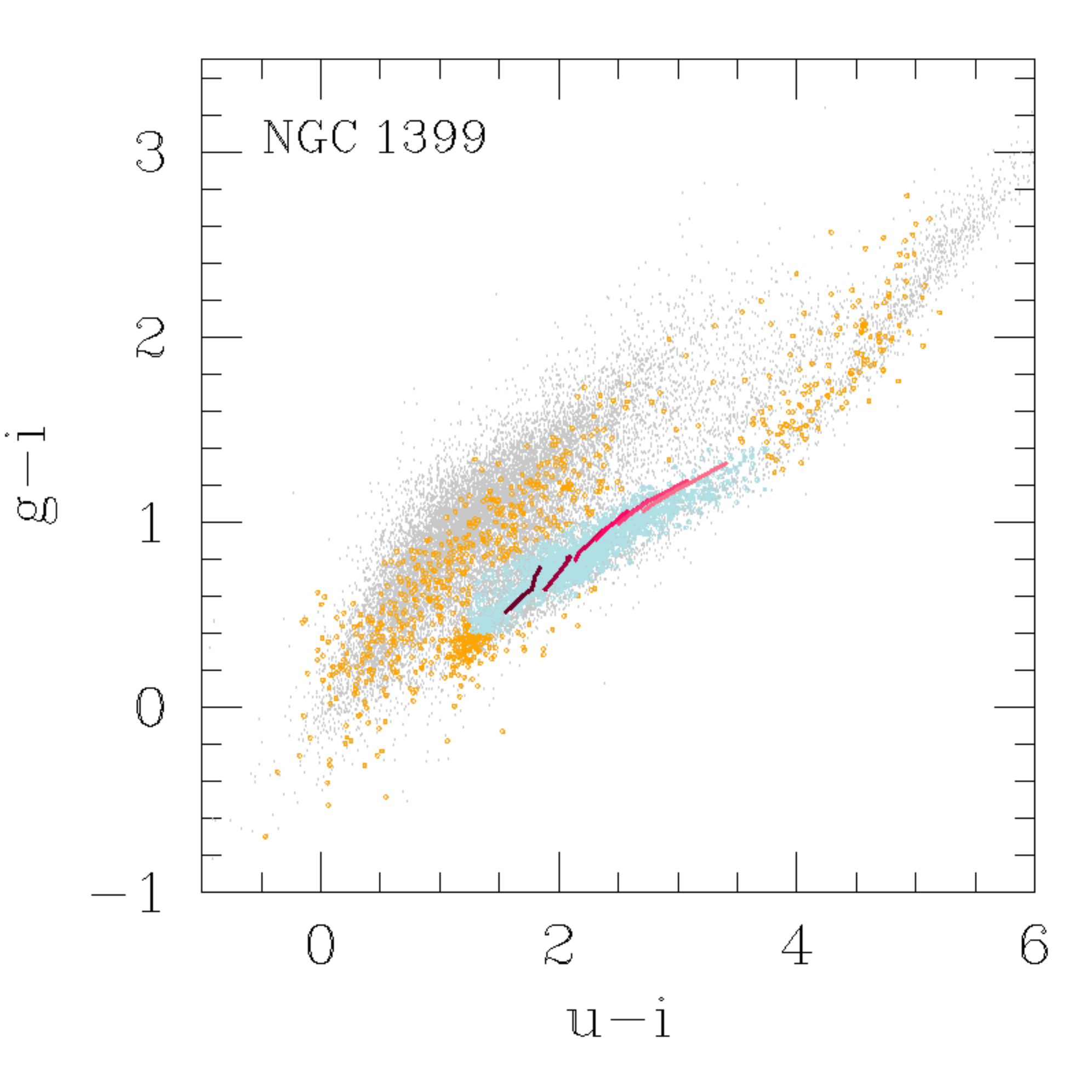}
   \caption{Upper left panel: Color-color diagram for the full catalog
     of matched sources in NGC\,3115 (gray dots) and catalog of GC
     candidates obtained using the photometric and structural
     selection criteria shown in the Figures \ref{dmag_g}-\ref{sel_g}
     (orange circles). Green and blue circles show the GCs and UCDs
     from the literature, as in previous figures.  Even with no color
     selection, the sequence of orange dots concentrates on the
     well-defined sequence of compact sources (stars and GCs). Upper
     right panel: As left, but color selected GC candidates are shown
     with pale-blue circles. SSP SPoT models for ages between 2 and 14
     Gyrs and \feh between $-$2.3 and $+$0.3 are plotted with solid
     lines in various purple shades. Lower left: as in upper left
     panel, but for NGC\,1399. In contrast to the UCDs candidates in
     NGC\,3115, the (spectroscopically confirmed) UCDs in Fornax are
     all well lined up with the color-color sequence of GCs. Lower
     right: as upper right, but for NGC\,1399.}
   \label{ugi}
   \end{figure*}

To select GC candidates we applied photometric, morphometric, and
color selection criteria as listed in Table \ref{tab_selparam},
highlighted in Figures \ref{dmag_g}--\ref{ugi}. In general, the
criteria adopted were based on the parameter space occupied by
confirmed GCs, UCDs, and stars; the latter were rejected as
contaminants from the catalog.  As reference for NGC\,3115 we adopted
the spectroscopic catalog from \citet{arnold11} and the photometric
catalogs of GCs and UCDs from \citet{jennings14}. For NGC\,1399 we
adopted as reference the spectroscopic sample of GCs and stars from
\citet[][]{schuberth10}, and the spectroscopically confirmed UCDs from
\citet[][and references therein]{mieske02,mieske04,mieske08}.  In the
following we explain in more detail the selection criteria adopted.

First, we measured the magnitude concentration index, described in
\citet{peng11}, using the difference in magnitude measured at 6 pixel
aperture diameter and at 12 pixel, $\Delta X_{6-12}\equiv
mag_{X,6pix}-mag_{X,12pix}$, where X is either the $g$ or the $i$ band
aperture corrected magnitude. For point-like sources, after applying
the aperture correction to the magnitudes at both radii, $\Delta
X_{6-12}$ should be statistically consistent with zero.

Figure \ref{dmag_g} shows the $g$-band concentration index for the
full catalog of $ugi$ matched sources in NGC\,3115 (gray
dots). Confirmed sources are shown in the left panel with green for
GCs, where full/empty circles indicate spectroscopic/photometric
confirmed GCs; red for stars; and blue for UCDs. The figure shows that
UCDs tend to have $\Delta g_{6-12}>0$ because of their angular size,
which is not negligible compared to the PSF size even on VST imaging
data\footnote{However, see Figure \ref{ugi} and Appendix
  \ref{app_ucd}, for some comments on the reliability of the list of
  UCD candidates in NGC\,3115.}.

\newpage

Then, we compared the morphometric (FWHM, CLASS\_STAR, flux radius,
Kron radius, petrosian radius, and elongation) and photometric properties
of the full matched $ugi$ catalog with the same properties of
spectroscopically or photometrically confirmed GCs, UCDs, and stars
available from the literature. Such comparison allows us to identify a
membership interval for each class of objects, in particular for GCs,
which are more interesting for this study. We adopted a large set of
parameters so to exclude anomalous or peculiar sources that might be
more efficiently detected with one parameter rather than others, hence
allowing us to minimize stellar contamination.

As further photometric selection criteria we rejected sources
  $\sim3 \sigma_{GCLF}$ brighter than the turn-over magnitude,
  $M^{TOM}$, and with photometric errors on magnitude and colors
  higher than a fixed maximum. The bright limit was obtained adopting
  the $M^{TOM}_{g}$ and $\sigma_{GCLF}-M_z$ relations from
  \citet[][their eq. (9)]{villegas10}, assuming $M_z=-24$ and $-21.5$
  mag, for NGC\,1399 and NGC\,3115, corresponding to $(V{-}z)\sim0.6$
  for both. For NGC\,3115 we extended the bright cutoff farther to
  fainter limits, by 0.5 mag (e.g., from $m_g\sim19.0$ to
  $m_g\sim19.5$), to avoid the expected larger stellar contamination
  from field stars at the expense of gaining only a handful of bright
  GCs. Because of the lower absolute Galactic latitude, NGC\,3115 is
  expected to have higher levels of MW stars contamination with
  respect to NGC\,1399 at any given magnitude. We used the Besancon
  models of the Galaxy \citep{robin03} to obtain an estimate of the
  stellar contamination as a function of magnitude.  The contamination
  for NGC\,3115 is $\sim30\%$ higher than NGC\,1399 at the level of
  $m_g\sim20$ mag and drops to $15\%$ higher at $m_g\sim25$ mag. A
  direct comparison of Figures \ref{colmag3115} and \ref{colmag1399}
  gives visual support to such expectation from models.

In the left panels of Figure \ref{sel_g}, we plot the parameters
analyzed, together with the selection sample, adopting the same color
and symbol coding as in Figure \ref{dmag_g}.

The adopted ranges of $\Delta X_{6-12}$ and of the various
morpho-photometric selection parameters based on the properties of
known GCs are given in Table \ref{tab_selparam}. The differences in
the selection criteria between the two targets are mostly due to
differences in the quality of the imaging data and, also to a lesser
extent, to intrinsic differences between the two galaxies, their
environment, and their position in the sky\footnote{No
  radial cut was imposed to examine the GCs. For NGC\,3115 the choice
  is motivated by the fact that at the galaxy distance the GC system
  could potentially cover the entire VST frame (see
  section \S \ref{radensn}). For NGC\,1399, instead, the choice is
  motivated by the fact that also candidate intergalactic GCs are of
  interest in our analysis (see section \S \ref{sec_results}).}.

The right panels in Figures \ref{dmag_g} and \ref{sel_g} show the
selected GC candidates with orange circles derived from the
morphometric and photometric criteria adopted.

The adoption of the selection described above results in highly
efficient identification of point-like sources. As recognizable in the
color-color diagram shown in Figure \ref{ugi}, the catalog of sources
obtained using the combination of concentration index and
morpho-photometric selections is highly concentrated on the stellar
sequence of the color-color diagram (orange dots in the figure) with
only little contamination from galaxies\footnote{The sequence of
  galaxies is in upper left in the figures, approximately
  centered on $\ui\sim1.0$ and $\gi\sim1.0$ mag.}. The diagram also shows
that the GC sequence overlaps with the sequence of foreground stars
in the MW, ranging from [\ui,\gi]=[$\sim1,0.2$] occupied by
main-sequence turn-off stars, to [$\sim6,3$] for low main-sequence
stars. To narrow down the stellar contamination, we adopted both the
literature GC samples and the simple stellar population models
\citep[from the Teramo-SPoT group;][]{raimondo05,raimondo09,raimondo17}
to identify the region of the \gi--\ui color-color diagram hosting
GCs. The color ranges from both empirical data and stellar population
models appear well defined in the color-color plane. We finally
adopted slightly wider color intervals to be the most inclusive
possible, yet avoiding the  increase in the level of contamination toward
the blue and red maxima of the GC color-color sequence. The \gi--\ui
region adopted for GC selection is also given in Table
\ref{tab_selparam} for both galaxies.

The analysis of the GCS properties could require the
characterization of the completeness functions and the consequent
completeness correction to the data. However, the absolute number of
GCs is not important for our purposes. It is the relative comparison
between the two targets that interest us more. Furthermore,
deriving such a correction goes beyond the scope of the present work,
both because the complexity of deriving such a function, which
would by necessity be color dependent, radial dependent, and also, to
some extent, chip-by-chip dependent and because the results presented in
the forthcoming sections are not be affected by such correction. The
completeness correction would certainly increase the number of GCs,
globally and locally, but it would affect the numbers almost
homogeneously at all radii, except for the innermost galactocentric
radii $\lsim1\arcmin$ \citep[e.g.,][]{kim1399}, which are mostly
populated by red GCs and not of much interest for this work because of the low
GC detection rates with VST data at such galaxy radii characterized
by high surface brightness levels. Hence, the correction leaves
the relative fractions of GCs and the outcome
of our analysis essentially unchanged.

The final sample of selected GC candidates, shown with pale blue dots
in the right panels of Figure \ref{ugi}, contains $N_{GC}=2142$ for
NGC\,3115 and $N_{GC}=1792$ for NGC\,1399.  The full Table with
objects coordinates and photometry is available at the CDS and on the
VEGAS project pages.

\section{Results}
\label{sec_results}

\begin{figure*} 
   \centering
   \includegraphics[width=8cm]{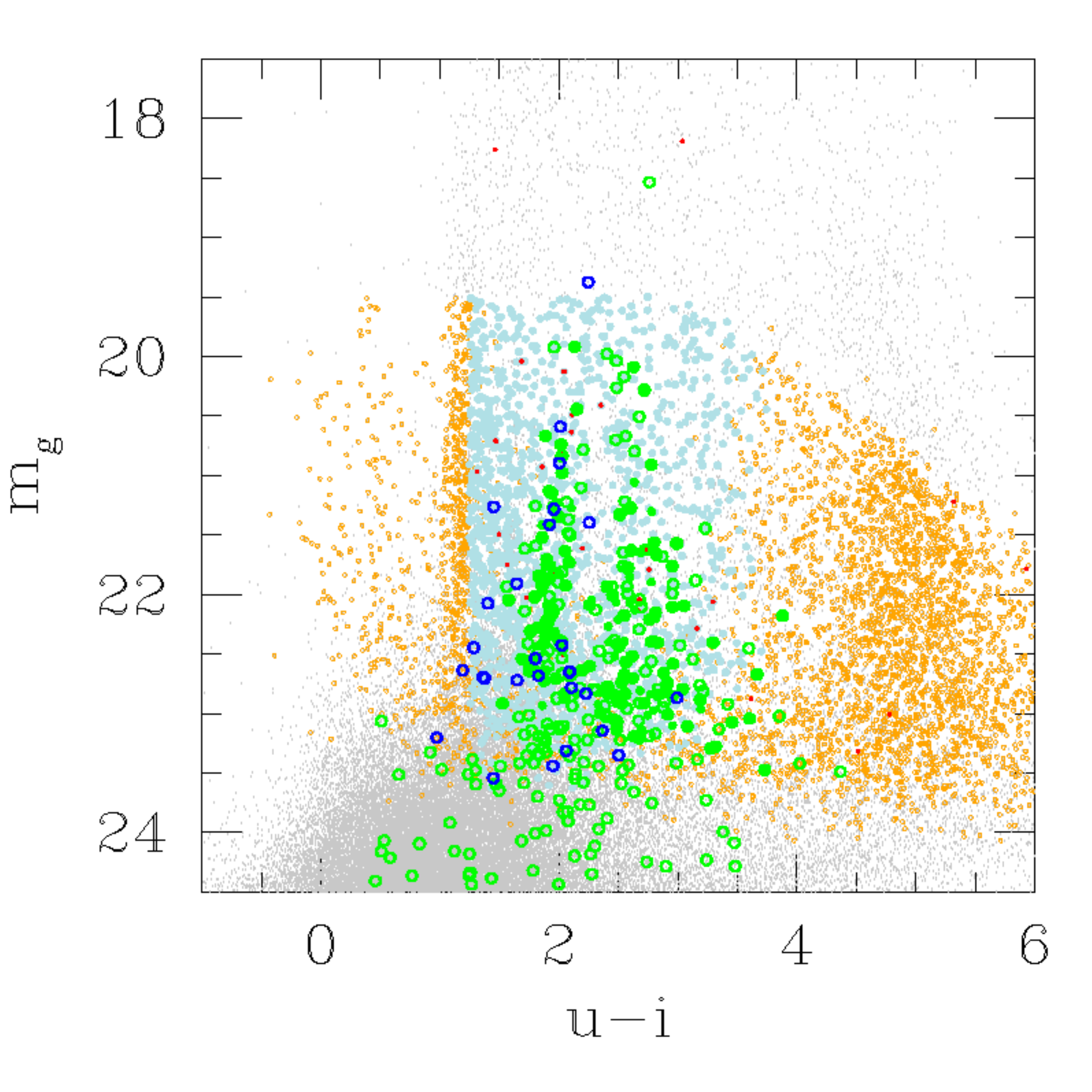}
   \includegraphics[width=8cm]{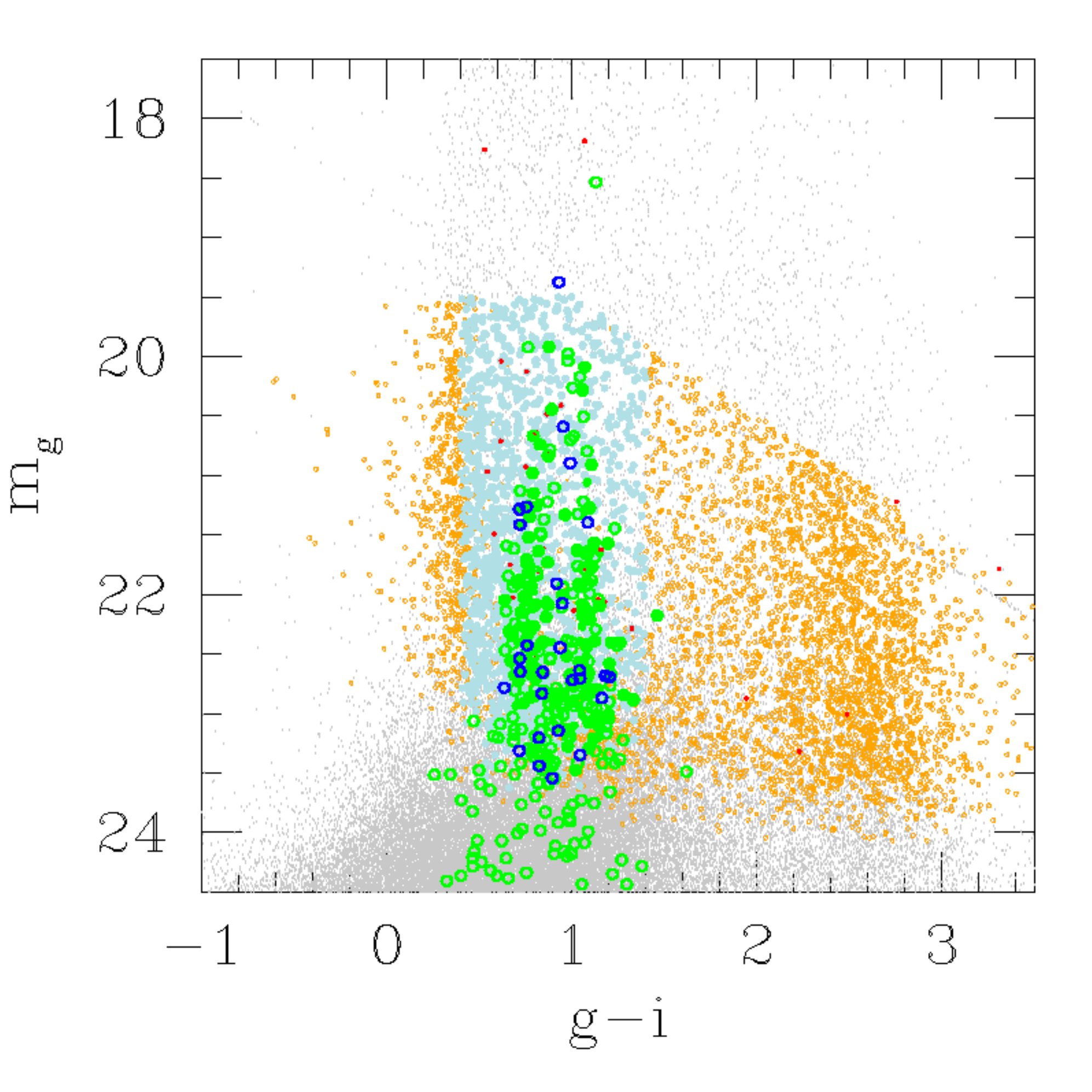}
   \caption{Color-magnitude diagram for the sources in the field of
     NGC\,3115. Symbols are the same as in Figure \ref{ugi}: gray dots
     refer to the full matched sample, orange empty circles to sources
     selected using the range of morphometric and photometric
     properties of confirmed GCs, pale-blue filled circles are the
     sources in the final catalog. Confirmed sources, taken from the
     literature, are shown with green circles (GCs), red dots (stars),
     blue empty circles (UCDs).}
   \label{colmag3115}
   \end{figure*}

\begin{figure*} 
   \centering
   \includegraphics[width=8cm]{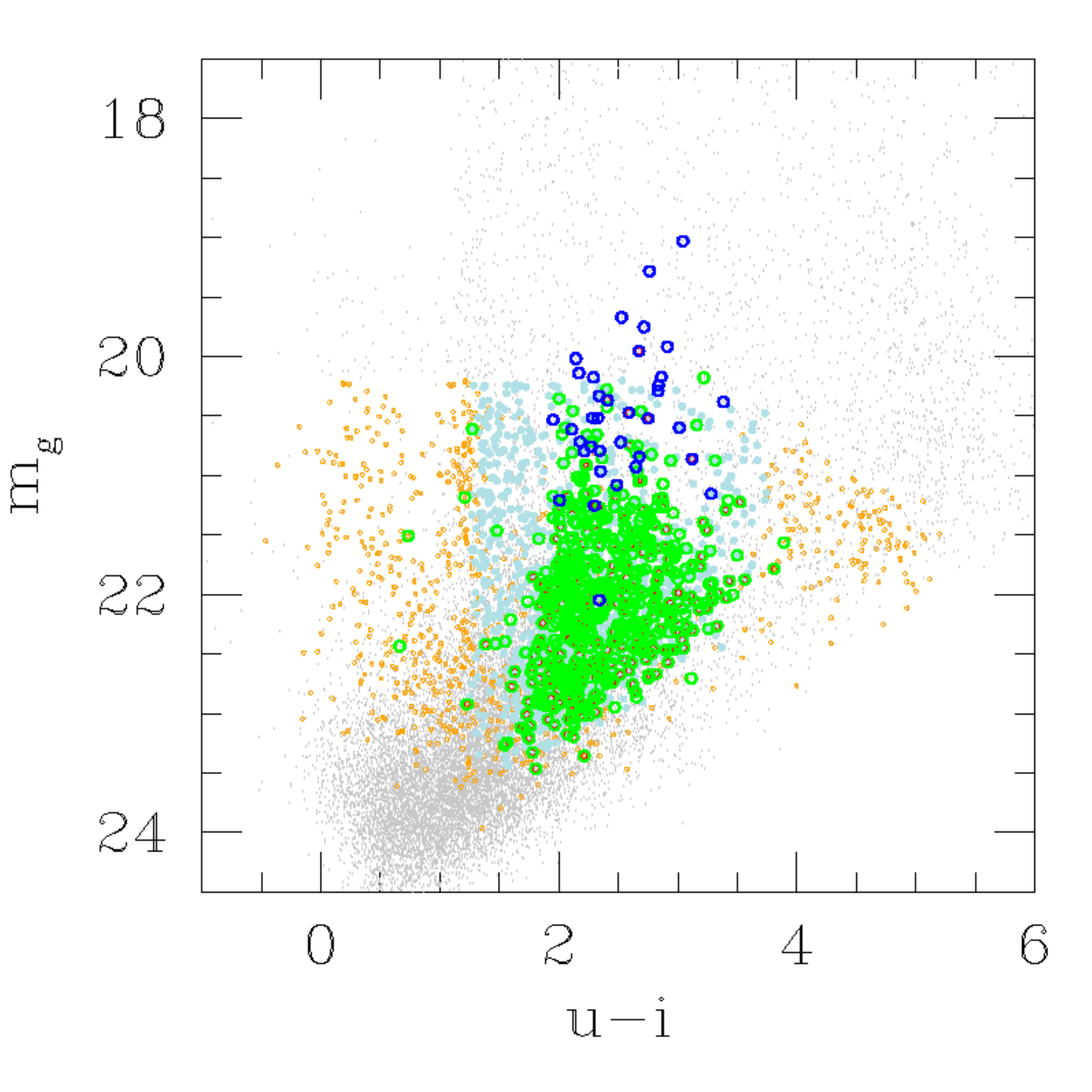}
   \includegraphics[width=8cm]{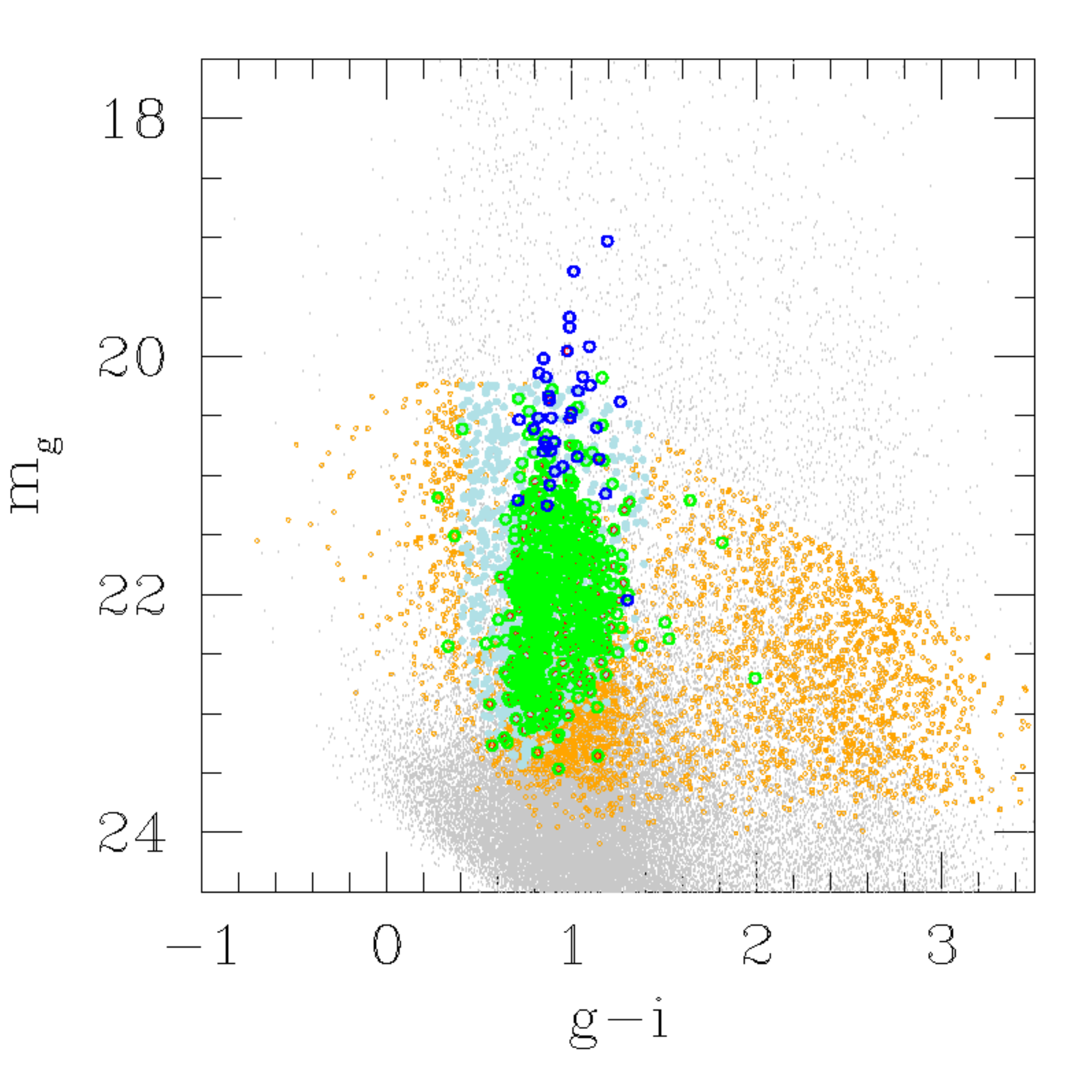}
   \caption{As in Figure \ref{colmag3115}, but for NGC\,1399.}
   \label{colmag1399}
   \end{figure*}

Using the final sample of matched $ugi$ sources, obtained from the
photometric, morphometric, and color selections described in the last
Section \ref{sec_sel}, we analyzed the two-dimensional spatial
distribution, the color distribution, and the azimuthally averaged
radial density profiles of GC candidates for the two targets.  The
color magnitude diagrams for the full matched sample of sources and
the selected GC candidates, together with the samples of confirmed
GCs, UCDs, and contaminating stars, are shown in Fig.
\ref{colmag3115} for NGC\,3115  and Fig. \ref{colmag1399} for NGC\,1399.

\subsection{Two-dimensional spatial distribution of GC candidates}

  \begin{sidewaysfigure*}
  \centering
  \includegraphics[width=0.9\hsize]{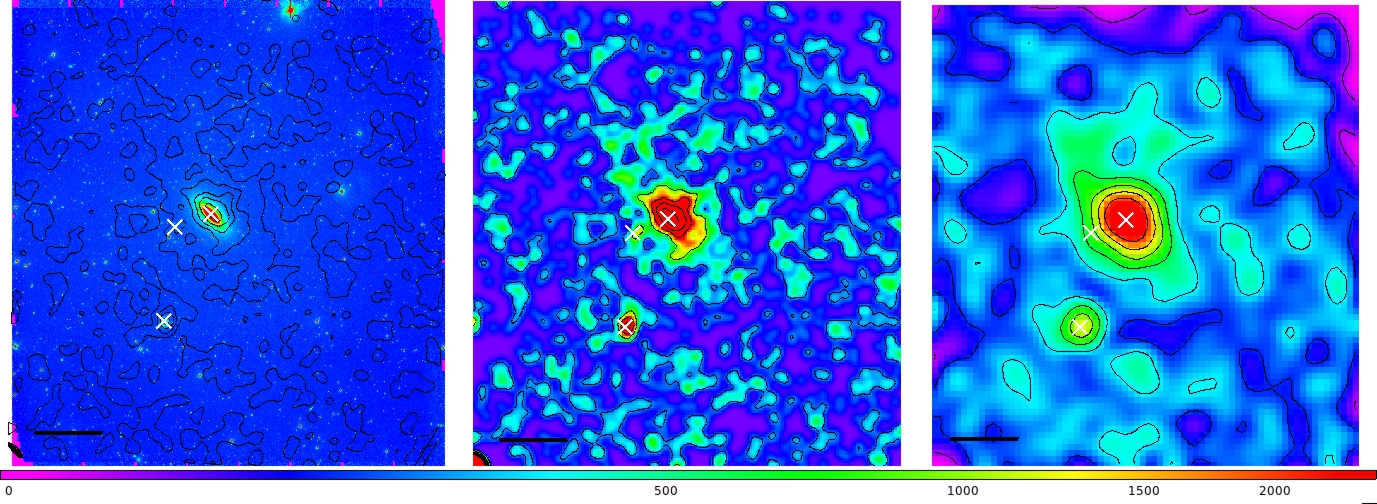}
  \vskip 1 cm
  \includegraphics[width=0.9\hsize]{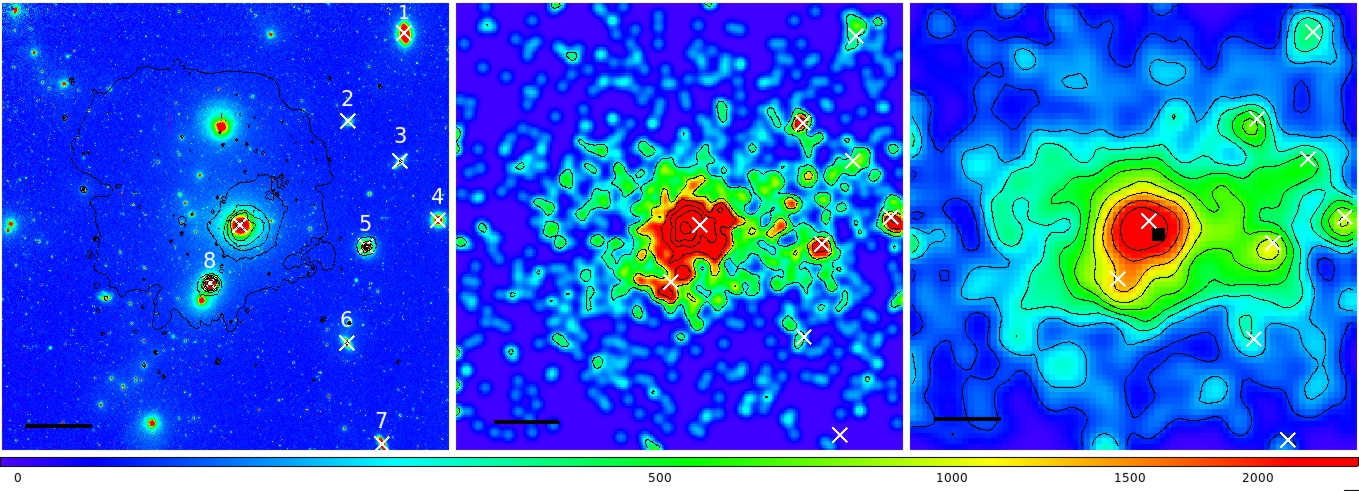}
   \caption{Upper panels. From the left: $g$ band image of NGC\,3115
     overlaid with contour levels derived from the GC density maps
     (from the $21\arcsec \times 21\arcsec$ grid map, see text). East
     is left, north is up. The white crosses indicate the position of
     NGC\,3115, NGC\,3115-DW01 ($\sim17\arcmin$ southeast of
     NGC\,3115), and KK\,084 ($\sim5.5\arcmin$ east). Middle panel:
     Density maps of GC candidates obtained from counting GCs in cells
     $21\arcsec$ each side are shown. Violet corresponds to zero level and red to
     highest counts, as indicated by the color bar. Density
     contours are also shown with black solid lines. Right panel: as
     middle panel, except that the grid spacing for GCs counting is
     $52.5\arcsec$ in this case. The black bar to the lower left of the
     panels corresponds to $10\arcmin$ length. Lower panels: as upper
     panels, but for NGC\,1399. In the left panel the X-ray contour
     map is overlaid on the $g$-band image, instead of the GC
     isodensity contours as for NGC\,3115. The black square in the
     lower right panel shows the centroid of the density distribution
     within isodensity radius $\sim25\arcmin$. The numbers in the
       left panel indicate, respectively, \#1$\equiv$NGC\,1380,
       \#2$\equiv$NGC\,1382, \#3$\equiv$NGC\,1381,
       \#4$\equiv$NGC\,1379, \#5$\equiv$NGC\,1387,
       \#6$\equiv$NGC\,1389, \#7$\equiv$NGC\,1386, and
       \#8$\equiv$NGC\,1404.}
   \label{gcmaps}
  \end{sidewaysfigure*}

Figure \ref{gcmaps} shows the surface density of GC candidates around
NGC\,3115 (upper, middle, and right panels) and NGC\,1399 (lower, middle, and right panels), together with the $g$-band image, shown in
the left panels. The GC maps shown in middle and right panels
are derived by dividing the surface area in equally spaced $RA$ and
$Dec$ bins, and then counting the number of GC candidates per unit
area. We adopted a binning size of $21\arcsec \times 21\arcsec$
($100\times100$ pixels) for the middle panels and $52.5\arcsec
\times 52.5\arcsec$ ($250\times250$ pixels) for the left panels of the
figure\footnote{At the adopted distances, $1\arcsec$ corresponds to
  $\sim45$ pc for NGC\,3115 and to $\sim97$ pc for NGC\,1399.}. The
resulting bidimensional maps were then smoothed with Gaussian kernels
adopting a standard deviation $\sigma=2$ in units of grid spacing.

For NGC\,3115, the first panel also shows the isodensity contour
levels obtained from the $21\arcsec\times 21\arcsec$ smoothed GC
density map. The surface density maps for this galaxy show several
noteworthy features. The presence of nonzero background (zero counts
correspond to violet color in the panels, highest GC count rates to
red), at even very large galactocentric radii, testifies the level of
contamination affecting the sample mainly caused by foreground Milky
Way stars. Even over the large area considered here, such stellar
contamination is either constant or changes very smoothly with respect
to the local density of GCs. Therefore, the GC system of NGC\,3115 and
of its dwarf companion, which is NGC\,3115-DW01 southeast of
NGC\,3115, clearly emerges over the contaminating sources. An
overdensity of GCs also appears in the region of the dwarf spheroidal
KK\,084, which is a dwarf galaxy with high specific frequency of GCs
\citep[$S_N\sim10$;][white cross $\sim5.5\arcmin$ east of
  NGC\,3115]{puzia08}. The GC overdensity is better seen in the upper
middle panel of Figure \ref{gcmaps}.

Two relevant points for the light and GC maps of NGC\,3115 seen in
Figure \ref{gcmaps} are the similarity in terms of elongation and
inclination between the GC density profile and light profile of
field stars, and the larger spatial extent of GCs compared to galaxy
light. While the latter is a long-known observational property of
GCSs, the common GCs and galaxy light geometry is more surprising, as
the system of halo GCs, at large galactocentric distances, is usually
pictured as spherical.  Nevertheless, several studies have found
azimuthal distributions of GCs mirroring the galaxy
\citep[][]{kp96,park13,wang13,hargis14,kartha16}. At the level of
surface brightness depth of our data
\citep[$\mu_g\sim29.5~mag~arcsec^{-2}$;][]{spavone17}, we do not
observe any obvious bridge of diffuse light between NGC\,3115 and
NGC\,3115-DW01, nor do we find any obvious enhancement of GC
counts between the two galaxies, even though the density
contours of GCs around the dwarf companion appear elongated in the
direction of NGC\,3115.

The surface density plots in the lower panels of the figure confirm
the results from previous studies on NGC\,1399, revealing the presence
of substructures in the spatial distribution of the GC population,
such as the overdensity bridging the GCS in NGC\,1399 with NGC\,1387,
or NGC\,1381, or the southeastern arc \citep[feature G
  in][]{dabrusco16}\footnote{In contrast to our previous analysis,
  presented in \citet{dabrusco16}, where we focused on large-scale
  ($\sim$degrees) substructures in the spatial distribution of bright
  GCs selected using PCA analysis, here we $i)$ apply different
  selection criteria for GCs, $ii)$ are more interested in the bulk
  appearance of the GCS on a smaller area than in
  \citeauthor{dabrusco16}, and $iii)$ revisit the analysis with
  specific focus on the comparison between the GCSs in NGC\,1399 and
  NGC\,3115.}.

As observed in the lower middle and left panels of Figure
\ref{gcmaps}, in addition to local density maxima associated with the
bright galaxies in the field (mainly around NGC\,1379, NGC\,1380,
NGC\,1381, NGC\,1382, and NGC\,1387, besides NGC\,1399), the spatial
extent of the GC overdensity covers a large portion of inspected
area, providing supporting evidence in favor of its intergalactic
origin. The morphology of the overdensity is asymmetric with an
elongated east-west shape. For the regions west of NGC\,1399,
\citet{iodice17} observed the presence of an overdensity in the
galaxy light profile as well, and suggested the presence of a stellar
stream between the galaxy and the close bright neighbor
NGC\,1387. We confirm the presence of the substructures, some already
presented and discussed in \citet{bassino06} and
\citet{dabrusco16}. To model the GC density profile, we ran ELLIPSE on
the density maps and found that the centroid of the distribution
within the isodensity radius $\sim25\arcmin$ is $\sim3\arcmin$
southwest offset with respect to the photometric center of NGC\,1399
(black square in the lower right panel in Figure \ref{gcmaps}), with
position angle $PA\sim-85\deg$ and ellipticity $\epsilon\sim0.4$. At
smaller galactocentric radii, within $\sim5\arcmin$, the offset is
$\lsim1\arcmin$ in the opposite direction, east of NGC\,1399 centroid,
consistent with the results from \citet{kim1399} with $PA\sim-70\deg$
and $\epsilon\sim0.15$. A similar GC distribution offset has also been
recently found in Coma by \citet{cho16}.  We also observe that
  centroid of NGC\,1399 GC distribution does not coincide with the
  maximum value of the GC distribution (close to NGC\,1399 center),
  mostly because the asymmetric east to west distribution of GCs is
  overdense on the west side of the cluster, where most of the bright
  companions of NGC\,1399 are located.
\newpage

Analyzing ROSAT X-ray data, \citet{paolillo02} highlighted the
presence of three distinct components in the core of Fornax: a central
cooling flow region, the galactic halo, and the cluster-wide X-ray
halo. In the first lower panel of Figure \ref{gcmaps} we overlaid the
$g$-band VST image of NGC\,1399 with the X-ray contours, derived from
\citet{scharf05}, showing the soft-band (0.3–1.5 keV) mosaic obtained
with Chandra/ACIS. On the scale of $\gsim30\arcmin$, the
southwest to northeast elongation of X-ray contours first observed from
ROSAT is also confirmed by the data from Chandra and provides
supporting evidence for a northeastern displacement of the hot gas in
the cluster with respect to the optical properties of galaxies
\citep{dabrusco16,iodice16}. Such geometry brought up the idea that
the Fornax core, and the subcluster hosting NGC\,1316, may lie along a
filamentary structure that is flowing in toward a common barycenter
\citep{drinkwater01,scharf05}.  The lack of immediate correspondence
between the X-ray southwest to northeast elongation and the east to west
elongation of the intracluster GCs is not surprising; given the
collisional/noncollisional nature of gas/galaxies, they react
differently to the possible ongoing interaction with the
\object{Fornax A} subgroup.

On smaller angular scales ($\lsim10\arcmin$) X-ray and GCs contour
maps for NGC\,1399 appear more similar, as both components appear
  approximately elongated in the southeast to northwest direction
  (lower left and middle panels in Figure \ref{gcmaps}),
  nearly orthogonal to that on a larger scale, connecting
  NGC\,1404 to NGC\,1399 and beyond. The large-to-small scale
  asymmetry of the gas distribution could be due to the motion of
  NGC\,1399 in the diffuse cluster halo, while for GCs the differences
  between large- and small-scale distributions might reflect the
  difference between intracluster GCs and the complex of interactions
  of the GCS in NGC\,1399 with the GCS in NGC\,1404 and other close
  galaxies \citep{bekki03}.

  \begin{sidewaysfigure*}
  \centering
  \includegraphics[width=0.9\hsize]{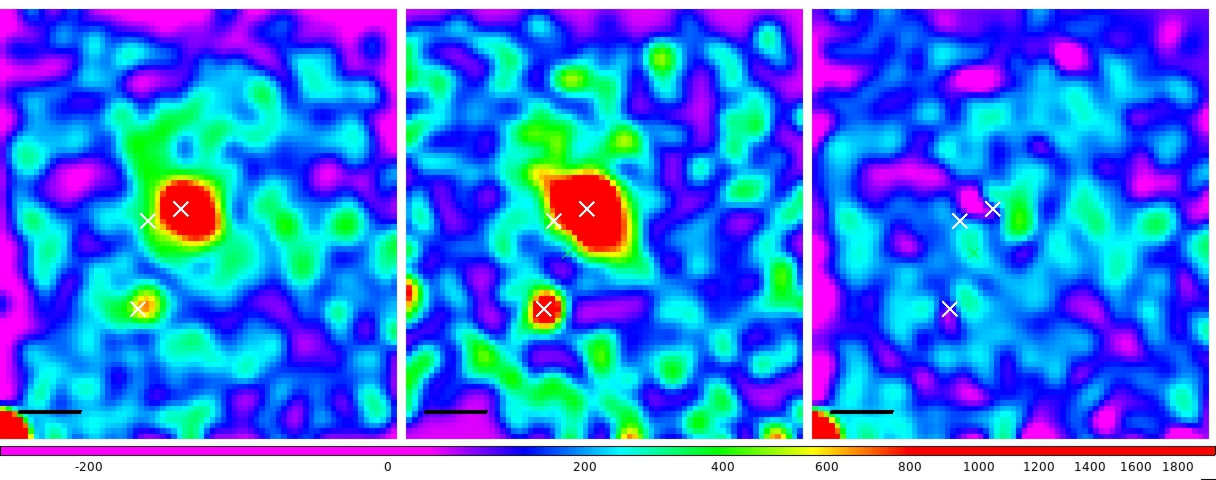}
  \vskip 1 cm
  \includegraphics[width=0.9\hsize]{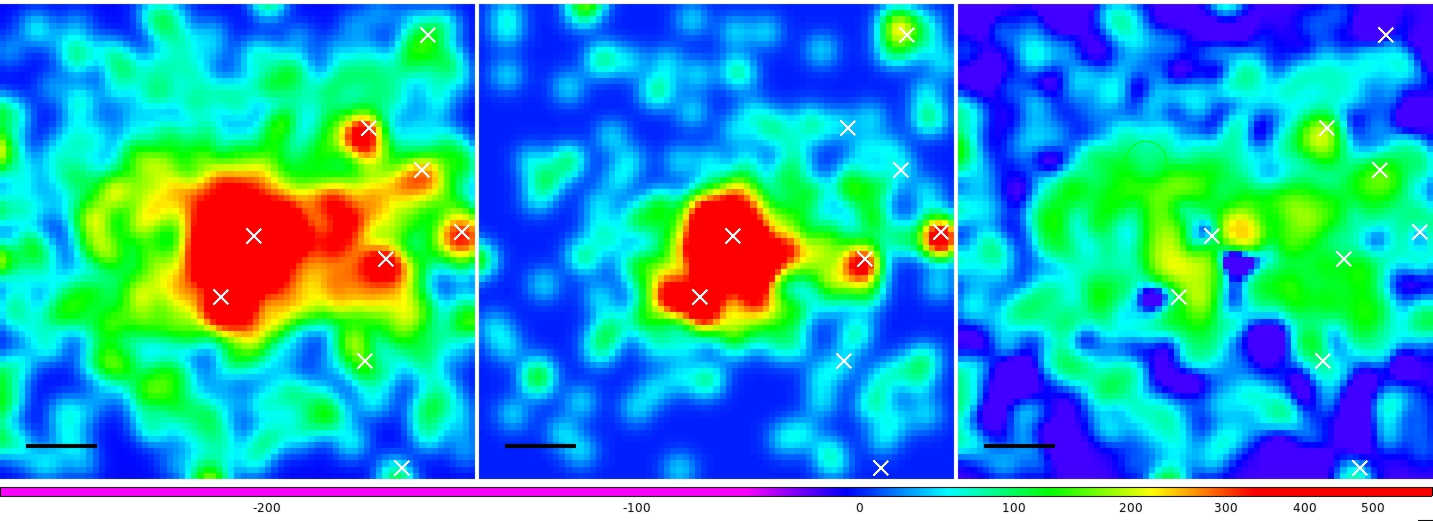}
  \caption{Upper left and middle panels: GC density maps in NGC\,3115
    as in right panel of Figure \ref{gcmaps}, except blue (left
    panel) and red (middle) GC candidates taken separately. Upper
    right panel: Blue to red residual image (see text) is shown. Violet
    corresponds to zero level, red to highest counts. Lower panels: as
    in upper panels, but for GC candidates in the field of
    NGC\,1399. }
   \label{blue_red}
  \end{sidewaysfigure*}

\subsection{Spatial distribution of blue and red GC candidates}

For both fields we also separately analyzed the surface density maps
of blue and red GCs, adopting the dividing color at \ui=2.3 for
NGC\,3115 GCs and \ui=2.5 for NGC\,1399 based on the average dip
between the Gaussian best matching the distribution of the two
subpopulations over the entire field of view (see next section
\ref{sec_ccr}).  The maps are shown in Figure \ref{blue_red} (upper
panels for NGC\,3115, lower panels for NGC\,1399).

The shape of blue and red GC maps for NGC\,3115 do not differ much
from each other, except for the larger angular extent of the blue
subcomponent, which is a well-known feature in individual galaxies
\citep[][]{geisler96b,cote01,cantiello07c}. The surface distribution
for red and blue GCs appears approximately concentric to galaxy light
and aligned with the major axis of the galaxy. This is illustrated in
the upper panels of Figure \ref{blue_red}, showing the density maps of
blue (left panel) and red (middle panel) GC candidates, adopting the
$52.5\arcsec$ binning grid.

The blue and red GC maps for NGC\,1399 are notably different form
each other. We find that the red component is mostly concentrated
around the bright galaxies closest to NGC\,1399 (white crosses in the
figure), and the blue GC component appears more extended, still
showing the densest peaks close to the positions of bright galaxies.

As further evidence for the intergalactic nature of the blue GC
component, we note that west of NGC\,1399 there are several bright
galaxies within a radius of $\sim30\arcmin$, thus the enhancement of
blue GCs might come from the overlap of the outer envelopes of blue
GCs host by the single galaxies in the region. However, for the GC
enhancement toward east, the overabundance of blue GCs cannot be
originated by the overlap of galaxy-hosted GCs because of the lack of
bright galaxies in the area.

Previous studies discovered a large population of GCs in galaxy
clusters, which appears not to be associated with individual galaxies,
numerically dominated by blue GCs, in relative fraction 4:1
\citep{peng10}. To further verify whether the blue GC overdensity in
NGC\,1399 (left panel in Figure \ref{blue_red}) is due to the overlap
between the intrinsically more extended blue GC distribution of
individual galaxies with respect to the red GC distribution, or if is truly an
extra intergalactic component, we made an attempt to generate a blue-to-red GC residual map. First, we rescaled the red GC density map
to match the blue GC density of NGC\,1399 at peak and at
galactocentric radius $r_{gal}=7\arcmin$. Then, by subtracting the
rescaled red GC map to the blue map, we obtained the residual map
shown in lower right panel of Figure \ref{blue_red}. With this
qualitative approach, we expected any overdensity of blue or red GCs to
appear in the residual image. To verify such expectation, the same
procedure was applied to NGC\,3115 (upper right panel of the figure) because for this galaxy the GCS is basically featureless. Indeed, the
residuals do not show the elongated structure associated with the GC
overdensities present in the first two panels of the figure. The GC
overdensity close to NGC\,3115-DW01 is reduced too, although the
red-to-blue scaling is properly derived only for the GCs in
NGC\,3115. Within the limits of the approach adopted, the GC residual
density map for NGC\,1399 clearly shows an overdensity stretching
along the east-west side of the cluster. Compared with the X-ray maps
discussed above, on cluster-wide scale blue GCs appear to cover the
region of the diffuse X-ray emission from hot gas, extending on the
east side of the Fornax galaxy cluster, where no bright galaxy with a
rich GCS exist. Following the results on the similarity between the
X-ray and blue GC profiles in NGC\,1399 \citep{hilker99c,forte05},
\citet{forbes12} recognized a general coherence of blue GC density
distributions with the X-ray profiles of bright early-type
galaxies. Blue GCs stretch on the east-west direction
with a relatively symmetric distribution around the cluster core,
while the hot gas from X-rays shows a northeast to southwest asymmetric
drift with respect to NGC\,1399. We speculate that the emerging
pattern is that the two Fornax subclusters are falling toward each
other with the galaxies and halo GCSs moving ahead of the gas
component because of their already mentioned noncollisional nature.

\subsection{Color bimodality and color-color relations}
\label{sec_ccr}

  \begin{figure*} 
  \centering
  \includegraphics[angle=0,width=9cm]{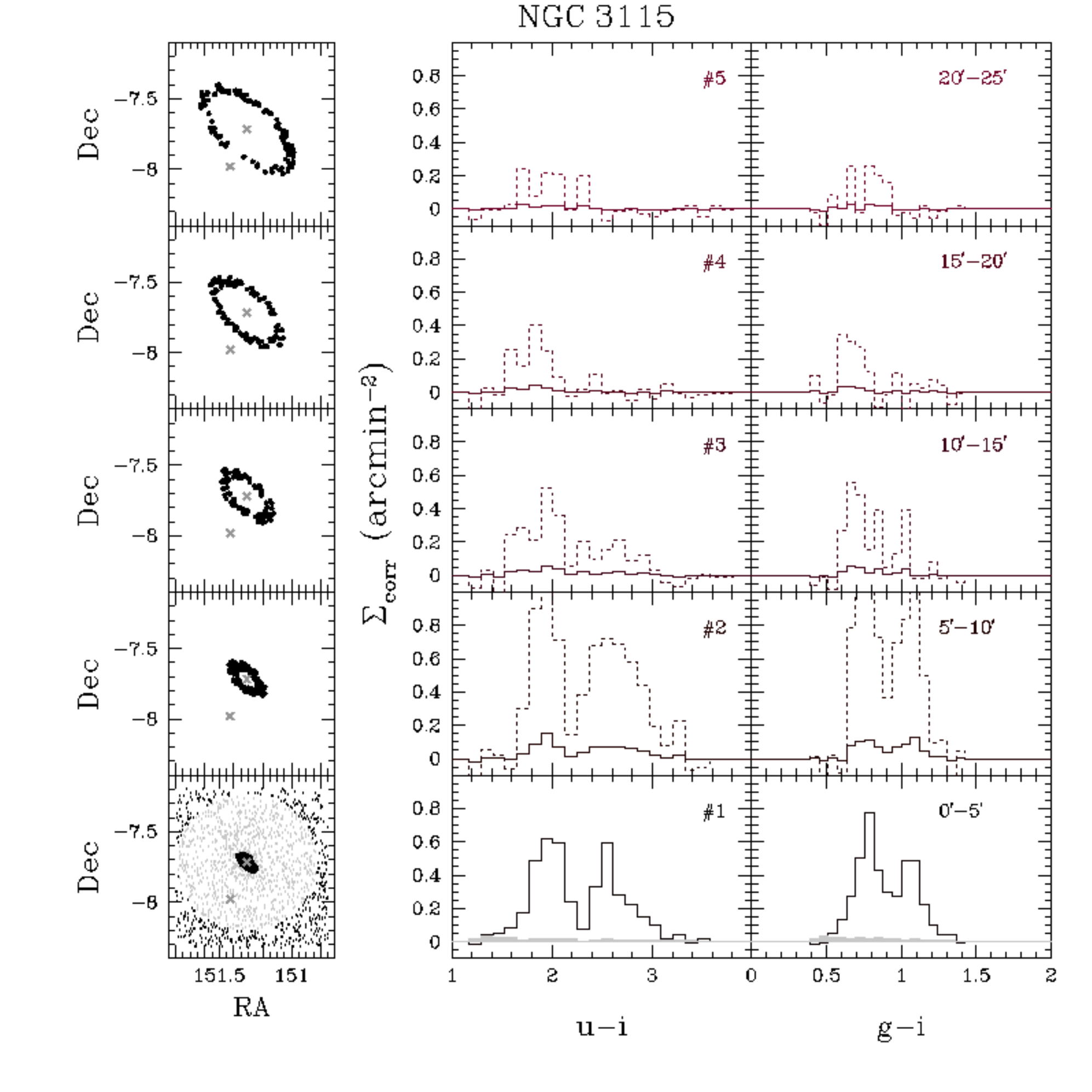}
  \includegraphics[angle=0,width=9cm]{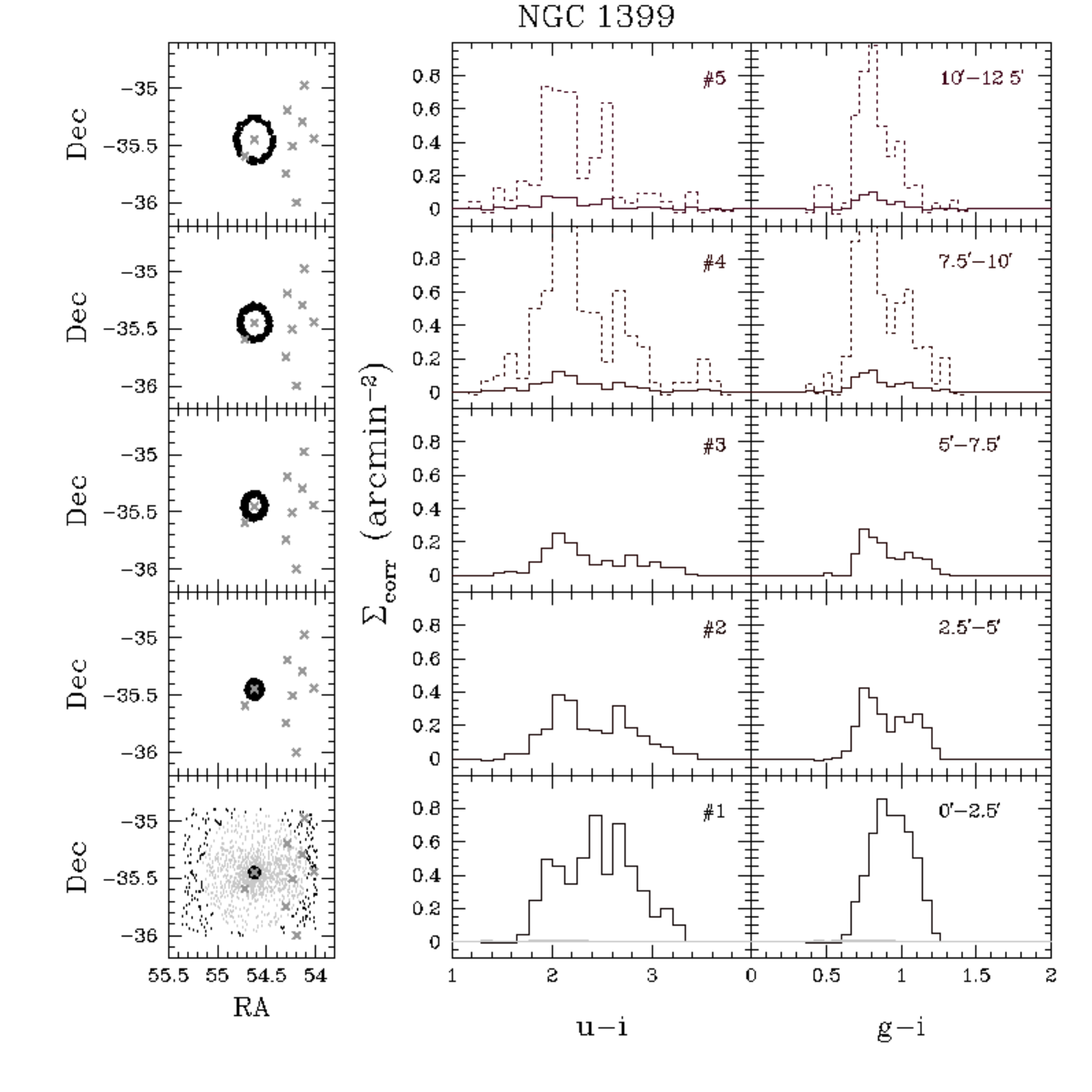}
 \caption{First column of panels from the left: position of GC
   candidates selected for color analysis. Each panel shows a
   different annular region, adopting the geometry of galaxy
   isophotes. In the lower panel we plot the full sample of selected
   GC candidates, adopting gray/black color for object within/outside
   the adopted background radius $r_{bg}=29\arcmin$. The gray crosses
   also indicate the centroid of NGC\,3115 and its companion dwarf
   NGC\,3115-DW01.  Second column of panels: \ui surface density
   distribution of GCs, corrected for background, at different
   galactocentric radii (as shown in left panel, and labeled in right
   panels). Dotted lines show the density histograms with maximum
   density $\Sigma_{corr}<0.2$ multiplied by a factor of 10. The gray
   histogram in the lowermost plot refers to background sources. Third
   column of panels: as second, but for \gi color. Fourth to last
   column: as first three columns of panels, but for the GC candidates
   in the field around NGC\,1399. The background radius adopted is
   $r_{bg}=32\arcmin$.}
   \label{posdhisto}
   \end{figure*}

As mentioned above, NGC\,1399 is one of the first galaxies suggested
to show evidence of color bimodality in its GCS and is among the most
thoroughly studied targets on this specific research topic
\citep{ostrov93,ostrov98,kp97b,kp99,dirsch03,bassino06}. For
  this galaxy, recent observational studies have shown the lack of
coherence of the color distribution with different color indices
\citep{blake12a,kim1399}, favoring the role of nonlinear
color-metallicity relations in shaping the observed color
distributions, against the classical scenario of a bimodal metallicity
distribution generating the observed CB.

NGC\,3115, in contrast, is the first external galaxy with CB for which all
observational evidence converges unambiguously toward the existence of
a bimodal metallicity distribution of the GCS
\citep{brodie12,cantiello14}.

In Figure \ref{posdhisto} we plot the \ui and \gi color distributions
for NGC\,3115 and NGC\,1399, obtained by inspecting various annular
regions with the same geometry of the galaxy isophotes [taken from
  \citet{arnold11,spavone17} for NGC\,3115; from \citet{iodice16}
  for NGC\,1399].

The first column of the panels shows the position of GC candidates in
NGC\,3115 starting from $r(\arcmin)=0$ (lowermost panel) out to
$25\arcmin$ (uppermost panel), with $5\arcmin$ steps.  The GC
candidates in each annulus are plotted as filled black circles. The
lowermost panel also shows the position of the full sample of GC
candidates at galactocentric distance larger/smaller than
$r_{bg}=29\arcmin$ with black/gray dots. The second and third columns
of the panels in the figure show surface density histograms
versus \ui and \gi color, corrected for background density using
the candidates at $r\geq r_{bg}$: $\Sigma_{corr}=[\Sigma
  (r_{out})-\Sigma (r_{in})]-\Sigma (r>r_{bg})$, where
$\Sigma(r)=N_{GC~cand.}(r)/Area(r)$ is given in units of counts per
square arcminute. For each color the background density
distribution is shown with gray shaded histogram in the lowermost
panels. The density distributions are corrected for the masked area of
the GCs close to the companion galaxy NGC\,3115-DW01.

Each row of the panels refers to a different annulus, as shown in the
spatial distribution panel (left) and also labeled in the right
panels. For sake of clarity, the density histograms with maximum
density $\Sigma_{corr}<0.2$ are multiplied by a factor of 10 and are
shown with dotted lines\footnote{The residuals for some histogram bins
    are $<0$, as is easily seen in the dotted histograms. Such negative
    counts are consistent with zero within estimated
    uncertainties.}.

The diagram for the innermost annulus of NGC\,3115 shows the
unambiguously bimodal color distribution, especially for the \ui color
that reveals a marked dip at \ui$\sim2.3$ mag. At increasing radii,
the distribution shows the behavior already discussed in
\citet{cantiello15}: the blue peak moves to bluer colors and the
density of red GC decreases with respect to blue GCs. In the last
annulus shown, $20\leq r(\arcmin)\leq 25$, there is only a residual of
candidates with \ui$\sim2.0$ mag and the corrected density is
$\Sigma_{corr}\sim0.022$ GCs per square arcmin, which can be compared
with the statistical error $\delta \Sigma_{corr}\sim0.011$
GCs/sq. arcmin; this error is estimated assuming Poisson statistic for
GC counts and assuming $5\%$ uncertainty on the estimates of the area
for both the annular region and the background.

The rightmost three columns of Figure \ref{posdhisto} show the same
analysis for NGC\,1399. Also in this case we rejected all sources
close to the bright galaxies in the field (indicated with gray crosses in
the left panels of the figure). For the correction of background
contamination, we adopted $r_{bg}=32\arcmin$, although the
distributions do not change substantially for larger background
radii. Given the nearly twice larger distance of NGC\,1399 compared to
NGC\,3115, to preserve the linear size of the area inspected for the
comparison, the angular size adopted for the annuli is half that
used for NGC\,3115.

The first obvious feature that emerges for the \ui and \gi color
distributions of NGC\,1399 (last two columns in Figure
\ref{posdhisto}) is the less marked gap between the distributions of
the blue and red GC peaks.

To quantify the differences between the two GCSs in terms of the color
distributions, we used the Gaussian mixture modeling code
\citep[GMM;][]{muratov10}\footnote{GMM uses the likelihood-ratio test
  to compare the goodness of fit for double Gaussian versus a single
  Gaussian.  For the best-fit double model, it estimates the means and
  widths of the two components, their separation DD in terms of
  combined widths, and the kurtosis of the overall distribution. In
  addition, the GMM analysis provides the positions, relative widths,
  and fraction of objects associated with each peak. It also provides
  uncertainties based on bootstrap resampling.  \citet[][see their
    section 4.2]{blake12a} provide a discussion on the issues inherent
  to bimodality tests.}.  The analysis is limited to the two innermost
annuli, where the relative fraction of background contamination is
much more negligible than at larger galactocentric radii.

\begin{table*}
\caption{\label{tab_gmm} GMM and Dip test results.}
\centering
\begin{tabular}{ccccccccccc}
\hline\hline
  Target & Color & Ann. & $N_{GC}$  & Blue &  Red                & $f_{red}$ &   kurt & DD            & p($\chi^2$) & p(Dip) \\
  (1)  & (2)   & (3)  & (4)       & (5)  &  (6)                & (7)   &  (8)   & (9)           &  (10)         &  (11) \\ 
\hline\hline
NGC\,3115 & \gi & $1$ & 166 & 0.765$\pm$0.015 & 1.064$\pm$0.022 & 0.417 & -0.723 & 3.02$\pm$0.43 & 0.001  & 0.78 \\       
NGC\,3115 & \gi & $2$ & 116 & 0.743$\pm$0.025 & 1.079$\pm$0.021 & 0.415 & -0.747 & 3.01$\pm$0.45 & 0.010  & 0.95 \\       
\hline                                                                                                                           
NGC\,1399 & \gi & $1$ & 103 & 0.834$\pm$0.056 & 1.036$\pm$0.059 & 0.492 & -0.818 & 2.38$\pm$0.34 & 0.275  & 0.02 \\    
NGC\,1399 & \gi & $2$ & 149 & 0.777$\pm$0.012 & 1.059$\pm$0.020 & 0.465 & -1.149 & 3.48$\pm$0.35 & 0.001  & 0.81 \\    
\hline\hline                                                                                                                   
NGC\,3115 & \ui & $1$ & 166 & 1.909$\pm$0.029 & 2.619$\pm$0.110 & 0.472 & -0.439 & 2.81$\pm$0.75 & 0.001  & 1.00 \\    
NGC\,3115 & \ui & $2$ & 116 & 1.883$\pm$0.050 & 2.698$\pm$0.072 & 0.463 & -0.812 & 2.94$\pm$0.36 & 0.014  & 0.83 \\       
\hline                                                                                                                           
NGC\,1399 & \ui & $1$ & 103 & 1.992$\pm$0.112 & 2.570$\pm$0.142 & 0.797 & -0.772 & 2.34$\pm$0.34 & 0.077  & 0.50 \\       
NGC\,1399 & \ui & $2$ & 149 & 2.079$\pm$0.038 & 2.738$\pm$0.065 & 0.475 & -0.772 & 2.82$\pm$0.40 & 0.001  & 0.81 \\       
\end{tabular}
\tablefoot{Columns list: (1-3) Target, color and annulus analyzed; (4) number of GC candidates in the annulus; (5) mean and uncertainty of the first, blue, peak; (6) mean and uncertainty of the second, red, peak; (7) fraction of the GC candidates GMM associated to the second peak; (8-9) kurtosis, Col. 8, and separation, Col. 9, between the peaks in units of the fitted $\sigma$ parameters. Values of DD larger than $\sim$2, and negative  kurtosis are necessary but not sufficient conditions for
  bimodality; (10) GMM value indicating the significance of the preference for a double Gaussian distribution over a single Gaussian (lower p($\chi^2$) are more significant;  p($\chi^2$) close to unity means unimodal distribution is preferred over bimodal); (11) significance level with which a unimodal distribution can be rejected based on the Dip test (high p(Dip) are more significant). }
\end{table*}

Table \ref{tab_gmm} presents the results of the GMM analysis, which is
accompanied by the results of the dip test\footnote{The dip statistic
  is a further test of unimodality, proposed by \citet{hartigan85},
  based on the cumulative distribution of the input sample, and its
  maximum distance with a best-fitting unimodal distribution. The test
  searches for a flat step in the cumulative distribution of the input
  function, which corresponds to a dip in the histogram
  representation.}. For seven of eight inspected GC samples, negative
kurtosis, large peak separations (in units of the fitted $\sigma$s for
the two Gaussian), and low p($\chi^2$) values from the GMM tests
confirm the preference for a bimodal Gaussian model over a unimodal
distribution for both galaxies. As also revealed by visual inspection,
the first annulus of NGC\,1399 shows that color distributions are
different in \gi and \ui. For the \gi color, the distribution visually
appears to be unimodal and, despite negative kurtosis and the relative
separation of the fitted peak positions DD$>$2, $p(\chi^2)\sim0.27$
indicates a marginal preference for the unimodal over the bimodal
distribution. For the \ui, the GMM tests on the selected GCs sample
show that the distribution is best represented by a multimodal
function with three peaks at $\sim2.0$, $\sim2.4$, $\sim2.8$, hosting
a fraction of $\sim31\%$,$\sim27\%$,$\sim42\%$ of the GC
population. For a further check, we adopted the catalog of GC
candidates from \citet{kim13}, selecting only sources within
$\sim2.5\arcmin$ from NGC\,1399 center; we found that for such sample
a distribution with three Gaussian with peaks at
$U{-}I\sim1.6,~1.9,~2.5$ mag, with 26\%, 27\%, and 47\% of the GC
population, also provides a good representation of the observed color
distribution, although it is not preferred over the bimodal
distribution. The complex structure of color distributions for GCs in
the innermost regions of NGC\,1399 ($r<2.5\arcmin$) was also discussed
by \citet{blake12a}, in a study based on HST ACS and WFC3 data. The
authors recognized differing bimodalities in different colors,
including the preference for unimodal color distribution in selected
GCs subsamples and for given color selections. The two colors
inspected here, \ui and \gi, have different relative sensitivity to
metallicity changes (more pronounced for \ui), and the lack of
correspondence between different color distributions of the same GC
population was one of the motivations that led to the projection
scenario aforementioned \citep[e.g.,][]{cantiello07d}.

Furthermore, by comparing of the dip statistics for the same annuli
and colors of both targets, the test highlights the presence of a more
obvious step in the cumulative distribution function of NGC\,3115
compared to NGC\,1399 in all inspected cases, indicating a more
pronounced dip between the two color distribution peaks for the first
GCS. The dip statistics, reported in last column of
  Table \ref{tab_gmm}, does not find a dip in the \gi color for the
  GCs in first annulus of NGC\,1399.

\begin{figure*} 
  \centering
  \includegraphics[width=9cm]{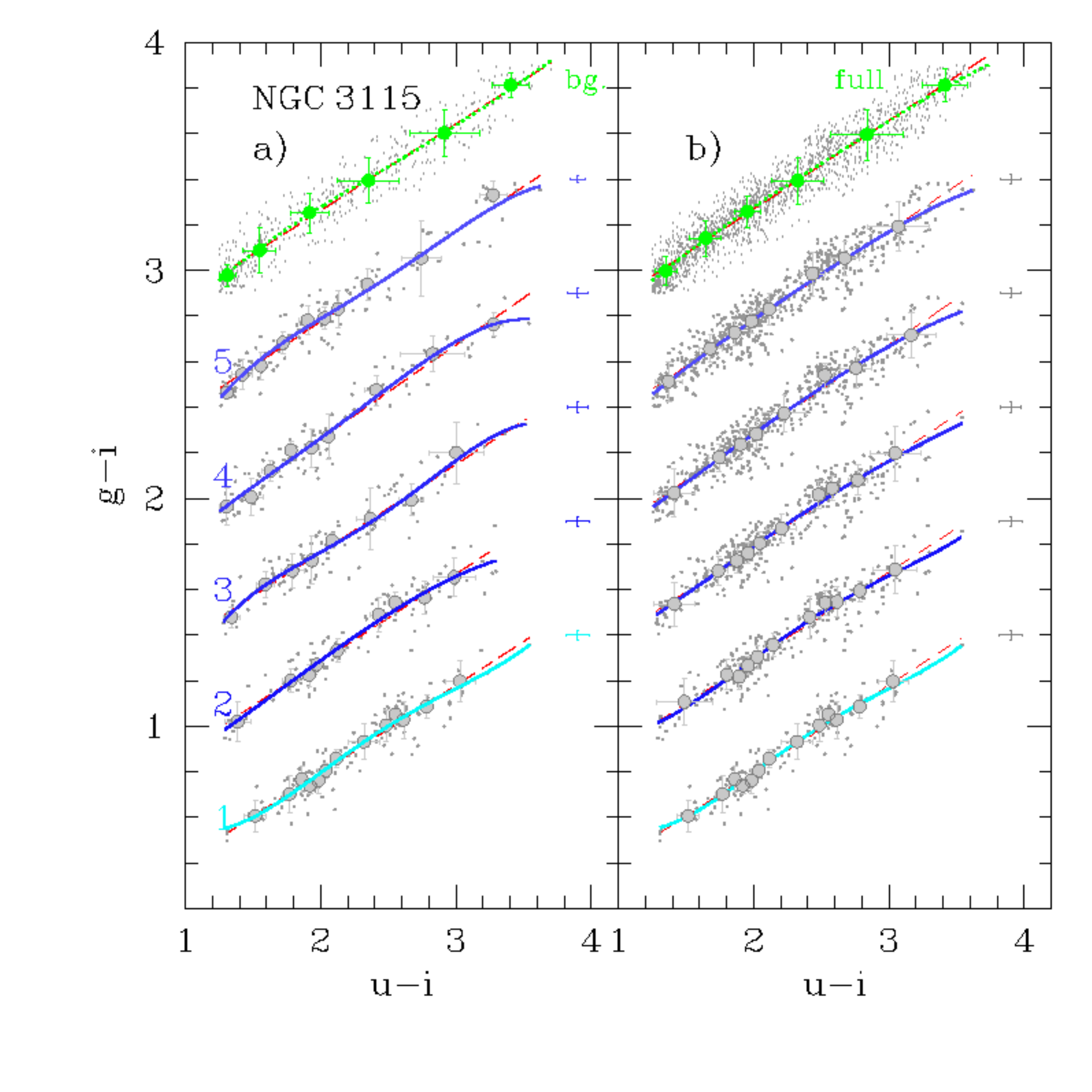}
  \includegraphics[width=9cm]{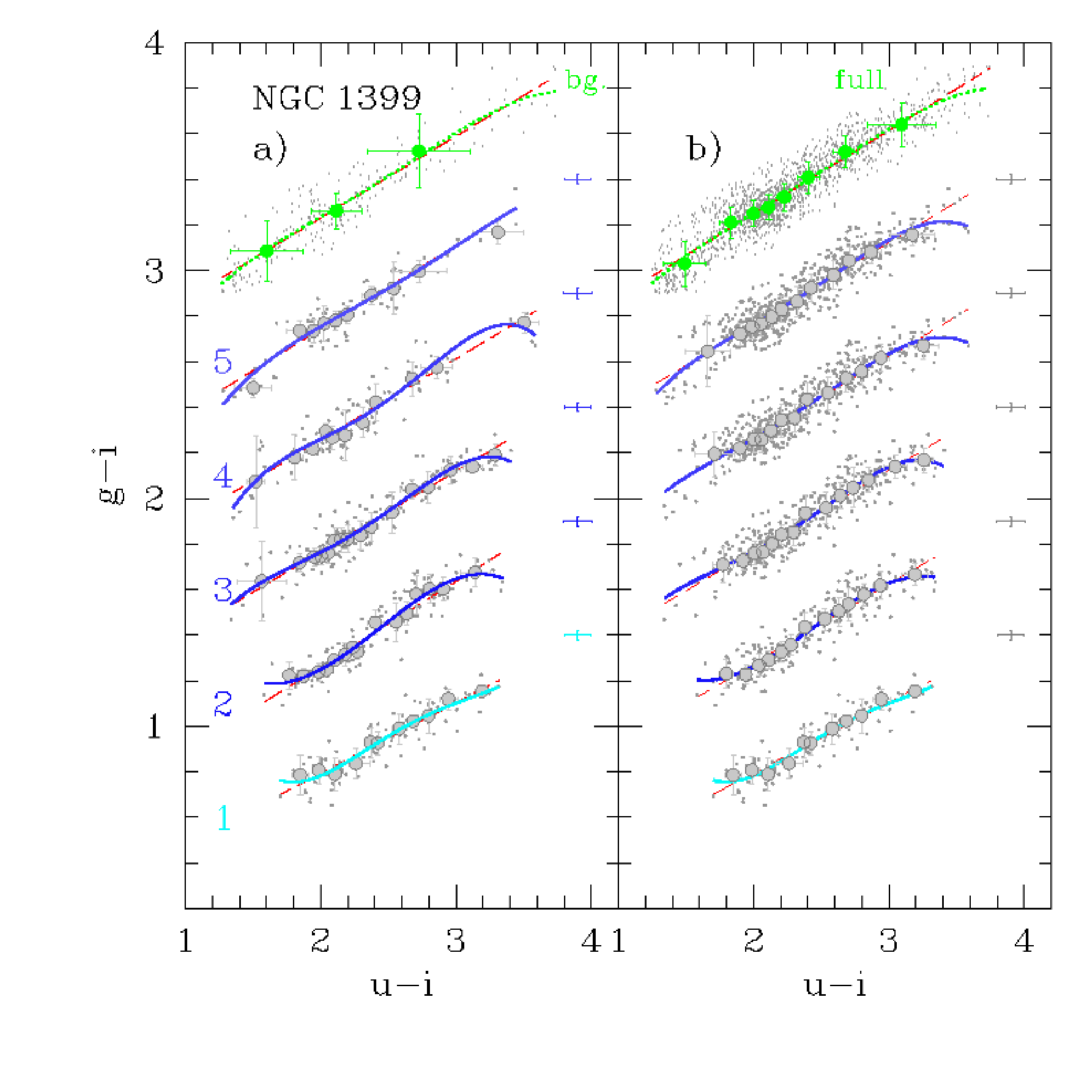}
  \caption{Left. Panel $a)$: Color-color relations fits for the GC
    candidates in the five annular regions analyzed in NGC\,3115.  The
    CCR for the innermost annulus is the lowermost. Single GC
    candidates are shown with dots; filled gray circles mark the
    running median, with $rms$ error bars derived using the maximum
    absolute deviation. The \gi of other annuli are shifted by
    multiples of +0.5 mag at increasing radii. Annuli are also labeled
    in the figure.  Linear fits are shown with red dashed lines,
    polynomial fits with solid curves of different shades of blue. The
    uppermost fits show the CCRs of background sources (red dashed and
    green-dotted line for linear and polynomial fits; {\it bg} label).
    The median color error for the sources used to derive the CCR fits
    is reported on the right of each fit. Panel $b)$: as in panel $a$,
    except that the cumulative sample of candidates is taken within
    each radius (see text). The uppermost curves are obtained from the
    full sample of candidates. Right: as in left panels, but for
    NGC\,1399.}
   \label{fithisto}
   \end{figure*}

Figure \ref{fithisto} shows the color-color relation (CCR hereafter)
for the selected GC candidates obtained in each one of the five annuli
analyzed above.

For each annulus we derived a linear fit to the CCR (shown with dashed
red lines in Fig. \ref{fithisto}), and a fourth-degree polynomial fit
(solid curves in various shades of blue). Only the color interval
spanned by real sources is plotted for each radius. In the figure we
plot two cases: In the first case, linear and polynomial fits are
derived from only the GC candidates in the annulus (left panels $(a)$
in the figure) and, in the second case, all GC candidates inside the
given outer radius are used for the fits (right panels $(b)$). For
sake of clarity, the CCR of the annuli beyond the first are shifted by
multiples of +0.5 mag in \gi color.

In each panel, we draw a sixth linear and polynomial fit obtained from
all candidates beyond the background radius (assuming circular shape,
left plots panel $a$; green dotted line for polynomial fit and red
dashed line for the linear fit) and from all GC candidates on the
frame (right plots, panel $b$).

The same analysis is carried out for NGC\,3115 and NGC\,1399 GCs with
the only difference that the radii adopted have half the angular size
for the latter, for the reasons described above.

\begin{figure*} 
  \centering
  \includegraphics[angle=0,width=9cm]{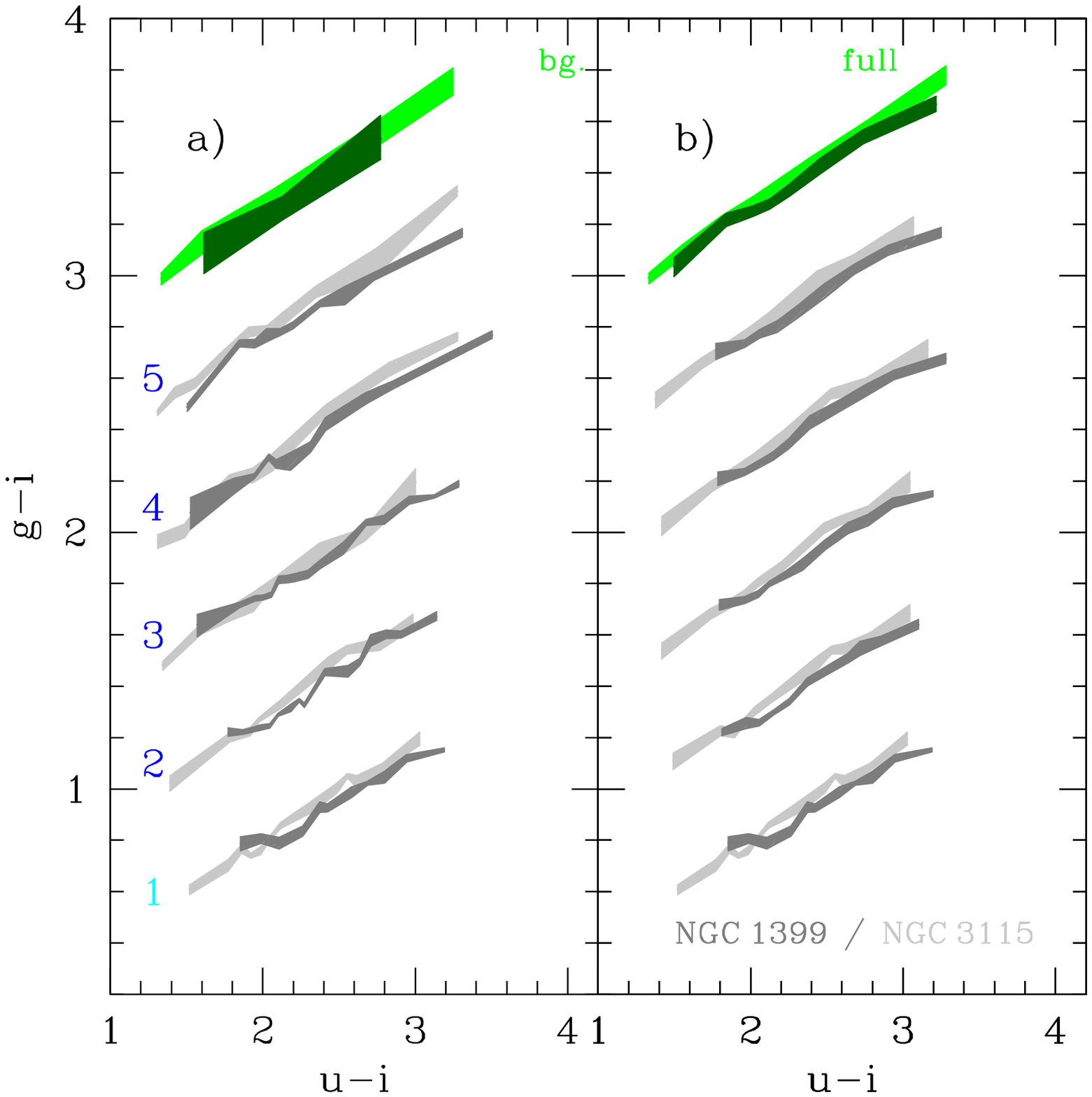}
  \includegraphics[angle=0,width=9cm]{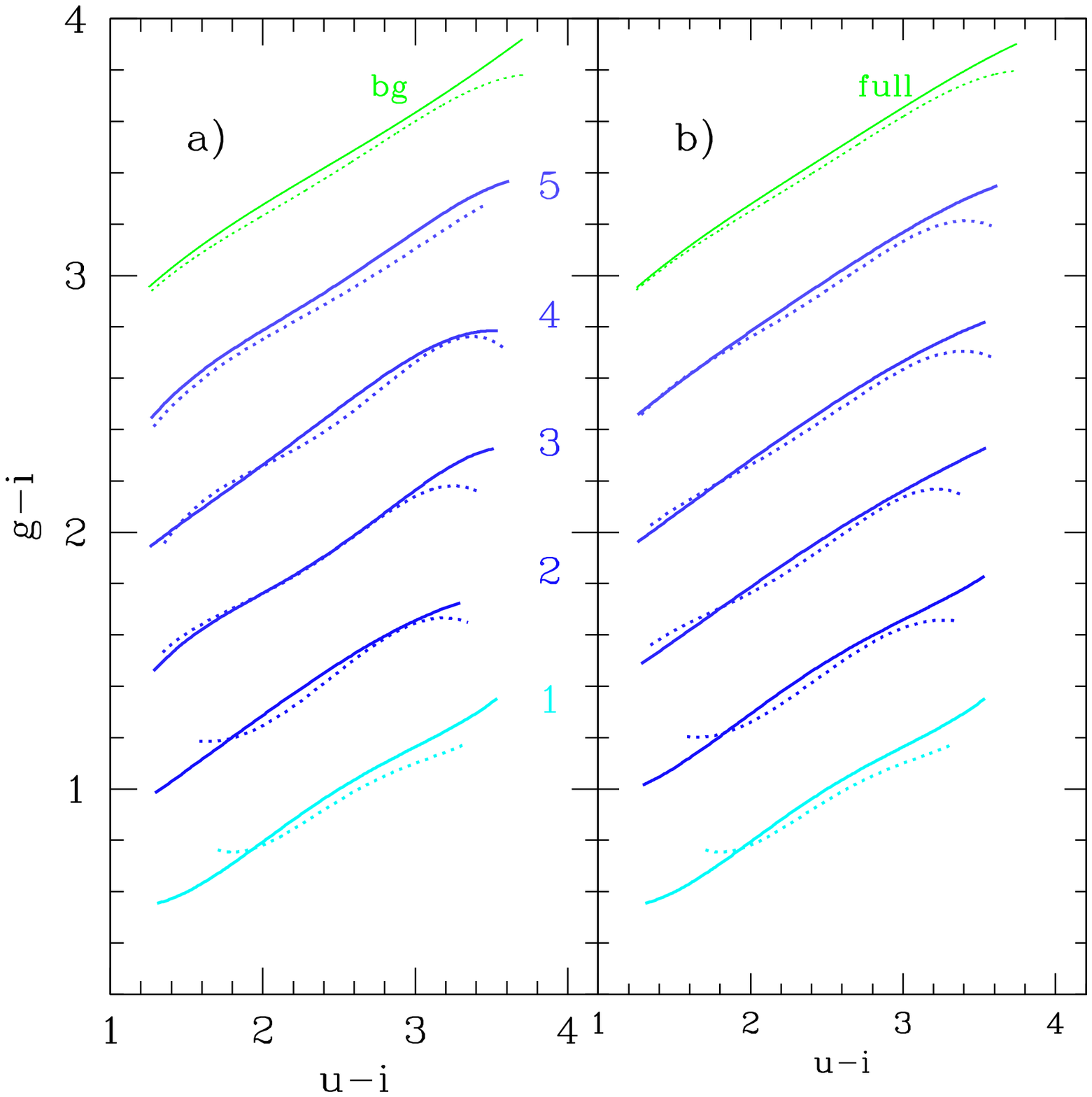}
  \caption{Left: as in Figure \ref{fithisto}, except that both
    galaxies are plotted together and only the areas identified by
    the running median are shown. The width of the shaded areas is
    given by the standard deviation of the mean. Darker colors refer
    to NGC\,1399; light gray and green to NGC\,3115. Right: as in
    Figure \ref{fithisto}, except that the polynomial fits for the
    CCRs of NGC\,3115 (solid lines) are superposed on NGC\,1399
    (dotted lines).}
   \label{combined}
   \end{figure*}

Some differences between the two targets appear more pronounced for
the innermost annuli, where the GC density is higher and hence the role
of residual contamination is smaller. The visual impression of
the difference between the two GCSs is even more evident in Figure
\ref{combined}. The first two panels in the figure (from left)
overplot the CCRs running mean area for NGC\,3115 (light gray areas)
and for NGC\,1399 (dark gray). The GC candidates in the labeled annulus
are shown in panel $(a)$, all candidates within the $i$-th annulus in
panel $(b)$. The right two panels in the figure plot the fourth-degree
fits for same data with solid blue lines for NGC\,3115 and dotted
blue lines for NGC\,1399.  The data for background regions are shown
with green areas/lines, according to the symbols in the
panel\footnote{The CCR for NGC\,3115 extend to bluer colors than
  NGC\,1399, mostly due to our choice of using annuli of similar
  linear sizes for the two galaxies. As shown in Figure
  \ref{posdhisto}, the outermost annuli used for NGC\,1399 do not
  extend to the farthest intergalactic regions that are richer in blue
  GCs. Furthermore, because of the much larger mass of NGC\,1399 with
  respect to NGC\,3115, a larger portion of red GCs is expected at
  similar physical sizes \citep{peng06}.}

Consistent with visual impression, the Spearman correlation
coefficient $r_{xy}$ -- a nonparametric measure of statistical
dependence between two variables \citep{spearman04}, reported in
Table \ref{tab_rxy} for the \ui-\gi correlation, is in all cases
larger for NGC\,3115 than for NGC\,1399 at a given annulus.\footnote{
  Because of the presence of a strong blue tilt in the GC
  population of NGC\,1399 \citep{harris06a,mieske10}, we tested the
  robustness of the differences between the relative shape of CCRs
  in the two galaxies versus changes of the bright magnitude cut. In
  spite of the reduced number of GCs selected, placing the bright
  magnitude limit $m_{bright}$ at $2\sigma$ or $1\sigma$ brighter than
  the turnover magnitude, rather than $4\sigma$, did not change the
  result, i.e., a higher degree of nonlinearity for NGC\,1399 CCRs is
  found in both tests.}

%
\begin{table}
\caption{\label{tab_rxy} Spearman correlation coefficients for the \ui-\gi relation.}
\centering
\begin{tabular}{lcc}
  \hline\hline
Ann. &  NGC\,3115 & NGC\,1399 \\
\multicolumn{3}{c}{$r_{xy}$ in single annuli}\\
\hline
1   &0.940 & 0.887  \\
2   &0.943 & 0.928  \\
3   &0.943 & 0.931  \\
4   &0.959 & 0.921  \\
5   &0.959 & 0.930  \\
bg  &0.958 & 0.927  \\
\hline
\multicolumn{3}{c}{$r_{xy}$ incremental sample}\\ 
\hline
1  & 0.940 & 0.887 \\ 
2  & 0.941 & 0.912 \\ 
3  & 0.942 & 0.921 \\
4  & 0.947 & 0.922 \\
5  & 0.951 & 0.925 \\
All&  0.960 & 0.931\\
\hline
\end{tabular}
\end{table}

Overall, our results show that the CCRs of the two galaxies are
  very similar in general but have some differences. In spite of the
identical ranges for color selections, the running mean in Figures
\ref{fithisto}-\ref{combined} shows the GC candidates population in
NGC\,1399 is shifted toward red colors and has a different shape with
respect to GCs in NGC\,3115, which underlines different median
properties of the two GCSs.

Assuming, as reasonable, nearly uniform old ages for both GCSs, the
color differences might mark slightly larger content of metals in
NGC\,1399 GCSs, which is not unexpected given the much larger
luminosity, and hence mass, of the Fornax cluster giant galaxy. Our
analysis also supports various degrees of nonlinearity in the CCRs
of the two targets with the CCRs in NGC\,1399 appearing more
nonlinear. Such difference is reduced for the CCRs at larger
galactocentric distances, as expected, owing to the increased
relative contamination from MW stars.

Since old GC colors are mainly driven by metallicity, at least
one of the color-metallicity relations involved must be nonlinear to
some degree. Hence, making it doubtful any analysis of the metallicity
distribution drawn from the simplistic assumption of bimodal
metallicity distributions evidenced by bimodal color distributions,
i.e., by assuming linear color-metallicity relations. Nevertheless,
given the dependence of the shape of the CCRs on the presence of few
very red or very blue GCs, the results shown here are not robust
enough to be generalized. We will further study this specific issue on
other galaxies of the VEGAS sample with different masses and in
different environments.

\newpage

\subsection{Radial density distributions and (local) specific frequency}
\label{radensn}

  \begin{figure*} 
  \centering
  \includegraphics[bb=20 420 400 720,width=8cm]{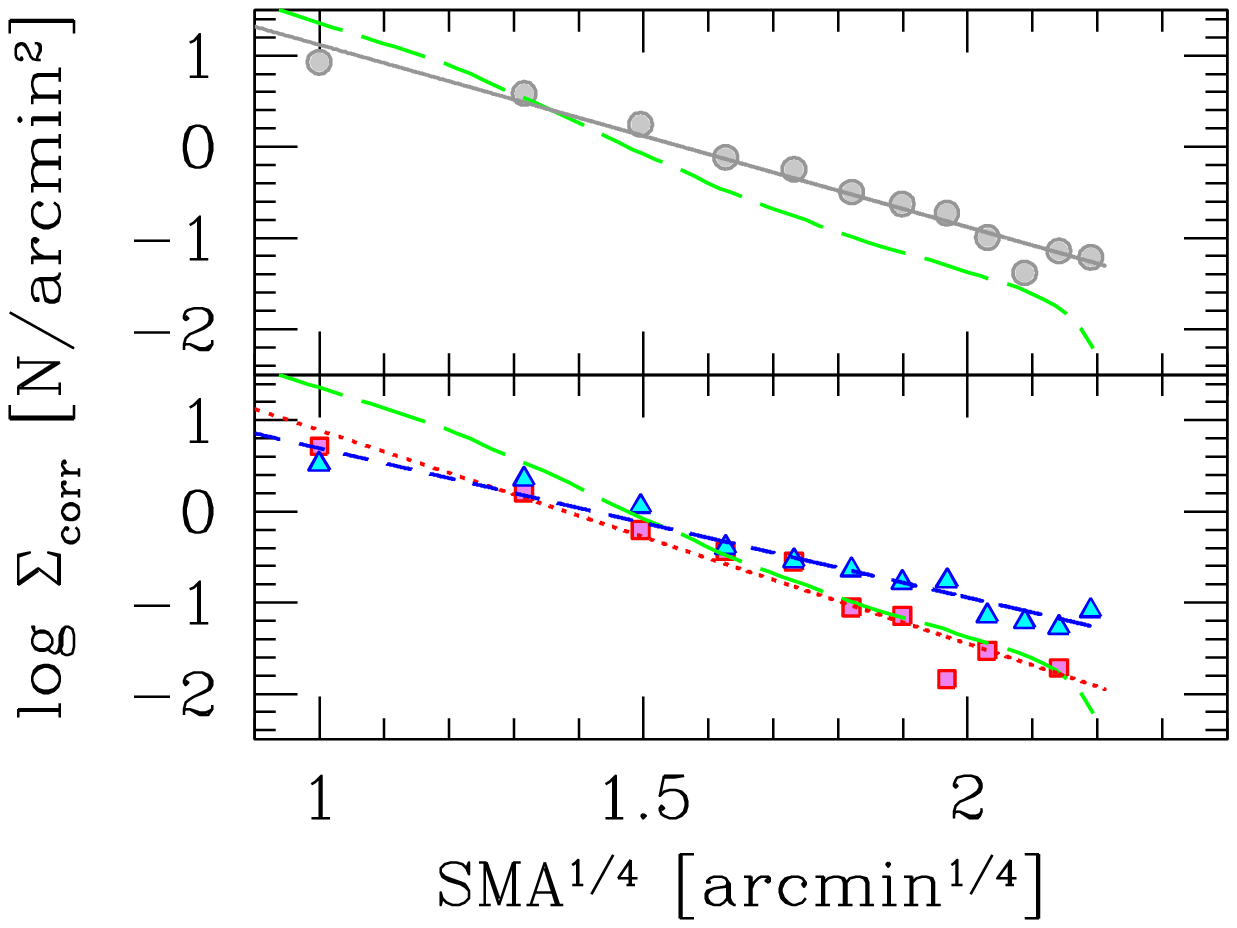}
  \includegraphics[bb=20 420 400 720,width=8cm]{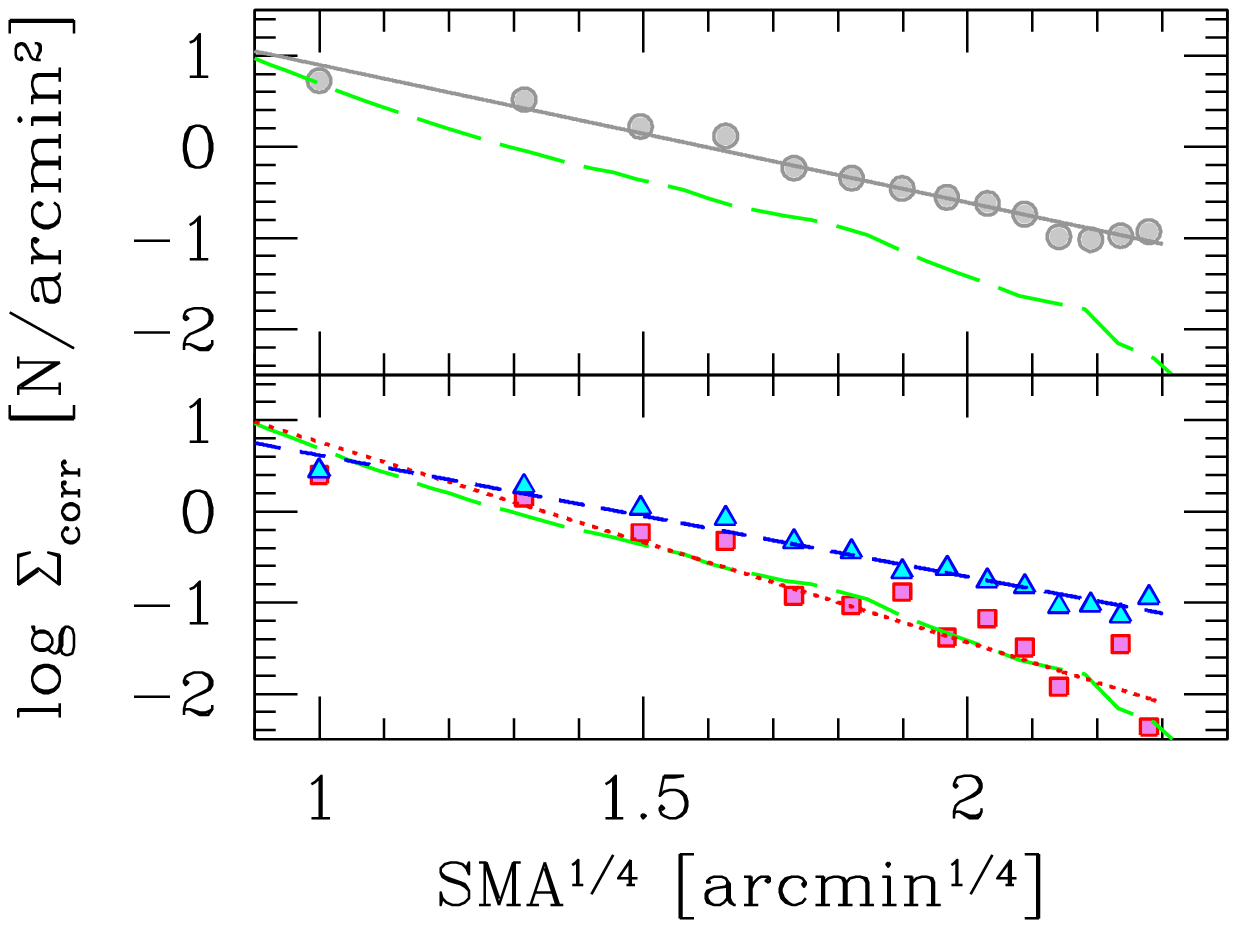}
  \vskip 2.5cm
  \caption{Upper left panel: surface density of GC candidates in
    NGC\,3115, corrected for background contamination, vs. galaxy
    semimajor axis length, SMA. The binned data are shown with gray
    circles, the $r^{1/4}$ law is fit with a solid line. The green
    long-dashed line denotes the $g$-band light profile from
    \citet{spavone17}, arbitrarily scaled vertically to match the red
    GC density profile. Lower left panel: as upper, but separately
    for blue GCs (blue triangles and dashed line) and red GCs (red
    squares and dotted line). Right panels: as left, but for the GCs
    in NGC\,1399.
    }
   \label{capaccio_v3}
   \end{figure*}

The radial density profiles of GCs in both our targets out to large
galactocentric distances have been thoroughly analyzed in other
studies [for NGC\,3115 \citet{cantiello15}, and references therein;
  for NGC\,1399 \citet{dirsch03,bassino06,schuberth10,kim1399}].

Now we compare the two targets
highlighting how, in spite of the differences emerged for
two-dimensional spatial distribution and for the color distributions
discussed above, some degree of homogeneity in the radial density
profiles of the two GC systems appears, especially for the red GC
component.

The background-corrected azimuthally averaged radial density of GC
candidates in each field was fitted to a \citet{devaucouleurs48}
$r^{1/4}$ law plus a constant background. We also fitted the $r^{1/4}$
profile to the blue and red GC subpopulations separately\footnote{We
  fit the surface density profiles with a Reynolds-Hubble law
  and a Sersic law. The improvement of the matching for the
  best-fitting curve to data is marginal, in particular using the
  Reynold-Hubble profile the inner radii of the distributions get a
  better match to data. However, the distributions add further
  parameters, i.e., degrees of freedom, to the fits, while the general
  trends observed do not change notably.}.

The results are shown in Figure \ref{capaccio_v3}. For both GCSs, our
analysis confirms the results of previous similar studies on these and
other targets: first, GCs have a shallower light profile than the
  galaxy field star light and second, the density profile of red GCs is
more concentrated than that of blue GCs and also closely follows
the underlying galaxy light profile.

For NGC\,3115, we determined the semimajor axis where the density of
GCs drops below 1 over an isodensity annulus one arcminute wide,
$SMA_1$, in addition to the radial profiles in Figure
\ref{capaccio_v3} and given the observed larger radial extent of the
GC population with respect to the field stars light distribution, to
obtain a rough estimate of the radius in which the GC density is
negligible Using an extrapolation to the total GC radial density
profile shown in Figure \ref{capaccio_v3} (right panels) and adopting
an approximate incompleteness factor of $\sim$2 (see below), we
estimated $SMA_1\sim61\arcmin$. Such a crude estimate relies on
various assumptions: $i)$ linear extrapolation to the GC density
profile (from Figure \ref{capaccio_v3}); $ii)$ similarity between the
geometry of galaxy light and GC distribution, which is only motivated
by the comparable profiles within $\sim25\arcmin$; and $iii)$
negligible contamination of the GCs beyond the assumed background
radius, $r_{bg}=29\arcmin$. In comparison, the surface brightness
profile drops to $\mu_g\sim30$ mag $arcsec^{-2}$ at $SMA\sim25\arcmin$
\citep{spavone17}.

\begin{figure} 
  \centering
  \includegraphics[bb=40 290 250 500,width=4.2cm]{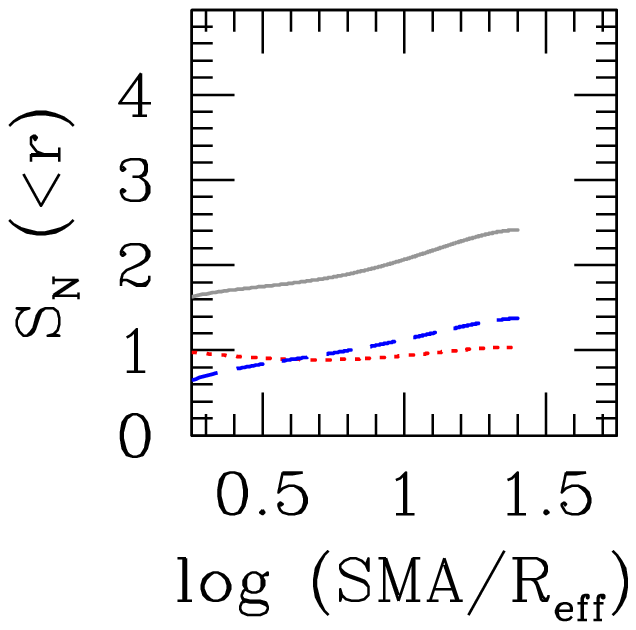}
  \hskip -0.5cm
  \includegraphics[bb=40 290 250 500,width=4.2cm]{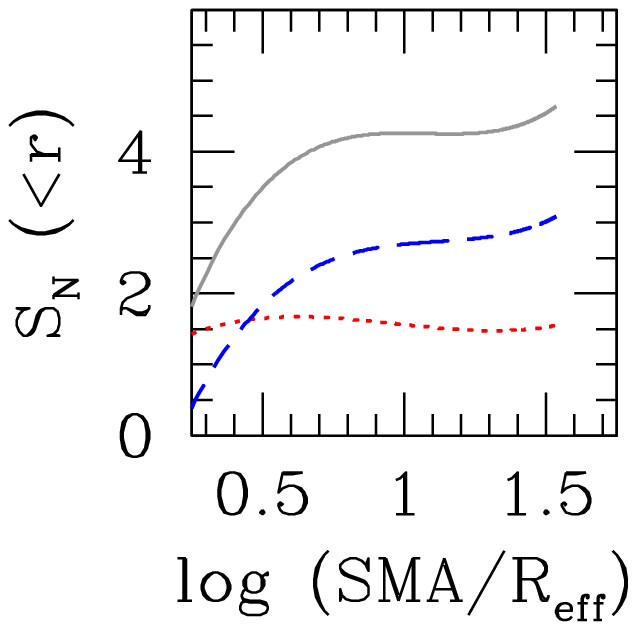}
  \vskip 3cm
  \caption{Left panels: local value of the specific frequency of GCs,
    $S_N(<r)$, for NGC\,3115. The solid gray line shows the full GC
    population, the blue dashed and red dotted lines show the local
    $S_N^{blue}(<r)$ and $S_N^{red}(<r)$ for blue and red GCs
    separately. Right panel: as left, but for NGC\,1399.}
   \label{capaccio_int}
   \end{figure}

A final relevant comparison between the two GCSs is the specific
frequency, $S_N$. By definition, $S_N$ is a global quantity
characterizing the properties of the galaxy and its host
GCs. Nevertheless, we analyzed the behavior of $S_N$ versus radius,
$S_N(<r)$, obtained from the total magnitude within the given radius
and total number of GCs enclosed within the same area. Because of
the incompleteness, in particular for NGC\,1399, to obtain absolute
$S_N$ approximately comparable with existing literature values, we
corrected the total number of enclosed GCs by a scaling factor
obtained from the ratio of the GC counts within a common detection
area between our study and the studies from deep imaging of the same
targets \citep[for NGC\,3115][for NGC\,1399]{jennings14,jordan15}. We
found that the scaling factor for incompleteness is $\sim2$ ($\sim11$)
for NGC\,3115 (NGC\,1399) with an estimated number of $\sim40$
($\sim150$) unobserved GCs in the central galaxy regions. In Figure
\ref{capaccio_int}, we show the results of the analysis of the local
$S_N(<r)$, which are derived using the total number of GCs and total
magnitude within a given radius in place of $N_{tot}$ and
$M_V^{tot}$. We corrected the determinations of $S_N(<r)$ for the
rough incompleteness factors given above, although we are interested
in the relative comparison of the two GC systems.

As for the asymptotic limit of $S_N$, we find the specific frequency
for NGC\,1399 is a factor of $\sim2$ higher than NGC\,3115, mainly
because of extra blue GCs.

The $S_N(<r)$ for the total GC population (gray lines in left panel
for NGC\,3115, right panel for NGC\,1399) reveals very different
radial trends: relatively smooth for NGC\,3115, and rapidly increasing
then smooth at SMA$\geq10\arcmin$ for NGC\,1399, with possible signs
of a further increase at larger radii.

It is instructive to inspect the local specific frequency for the blue
and red GCs, $S_N^{blue}(<r)$ and $S_N^{red}(<r)$, separately; these
are shown with dashed blue and dotted red lines in Figure
\ref{capaccio_int}. The estimated local frequency for the red GCs is
nearly constant over the range of inspected radii for both
galaxies. Furthermore, the similarity of the observed $S_N^{red}(<r)$
between the two galaxies implies that, contrary to expectation,
NGC\,3115 has a relatively high proportion of red GCs than
NGC\,1399. The blue GC component, hence, is the dominant, although is
not the only cause of the difference between the $S_N(<r)$ of the
total GC population. The fraction of blue GCs relative to field stars
in NGC\,3115 grows smoothly and steadily with galaxy radius, while it
shows a quasi-inflection point in NGC\,1399, where the value is
basically constant at around $\sim10\arcmin$ and then shows signs of a
further increase at larger radii.

  \section{Discussion}
\label{sec_discussion}

  \subsection{Differences and similarities}
It is an interesting question to ask to what degree the level of
galaxy clustering and the inherent hierarchical assembly affects the
observed GCSs properties \citep[see,
  e.g.,][]{blake97,cote98,blake99gc,peng08,hudson14}. We started to
answer this question by presenting the data for two very dissimilar
targets with the purpose of further extending the study as the sample
of galaxies observed within VEGAS and the area covered by FDS
increase.

If compared to the GC population in NGC\,1399, the GCs in NGC\,3115
formed and evolved in a very isolated environment. The present status
of the GC system, then, is less contaminated by the evolutionary
effects active on a central cluster galaxy, and there have been much
fewer changes to the original properties of the system, which could be
considered more pure than for the GCS in NGC\,1399.  Our work
indicates that the GC systems of NGC\,3115 and NGC\,1399 are markedly
different in some aspects, while sharing similar behavior on some
others.

Based on the surface density maps and radial profiles of field
stars and GCs, we find that both stellar components in NGC\,3115 are
well thermalized with each other.  The observed similarity between the
red GC density profile and galaxy stellar light profile out to
$\sim25\arcmin$, the resemblance of the geometry of the distribution
of red and blue GCs with the galaxy isophotes, the combined
observational evidence that blue GCs follow the X-ray profile of the
galaxy \citep{forte05,forbes12} and that X-rays trace the dark matter
in the galaxy point toward a one-to-one correlation between galaxy
light and GCS, suggesting a very closely related formation history of
the systems.  Furthermore, the analysis of resolved stars in the
galaxy based on HST data of three fields between 7 and 21 effective
radii, has shown that the densities of halo stars and GCs also appear to
decrease in a similar fashion with increasing radius, providing
evidence of the common formation of blue GCs and halo stars
\citep{peacock15}.

For NGC\,1399, we confirm and corroborate the previous observations for
markedly different distribution of blue and red GCs. Red clusters are
closely associated with the spheroid of the central giant galaxy and
other bright galaxies angularly close to NGC\,1399. Blue GCs are more
diffuse and elongated along the east-west direction of the cluster. On
the eastern side the angular position in the sky of such a diffuse
blue GC component is not obviously associated with any bright
galaxy; the intergalactic nature of the blue GC cloud does not tell
much about its origin, whether the GCs come from primordial mini
dark-matter halos, or are stripped from local satellites or from any
other process. Nevertheless, a later accretion of blue GCs from local
dwarfs would explain the observed differences of azimuthal
distributions with respect to the red GCs. West of NGC\,1399, some
evidence from the bridges of GCs toward NGC\,1387 and NGC\,1381
suggest that those GCs are stripped from the halos of their host
galaxies.

The GC color distributions and color-color relations of the two
galaxies also present differences worth highlighting. Both galaxies
share the property of a markedly bimodal color distribution in \ui and
\gi, a common property of bright galaxies. While for NGC\,3115 there is
universal consensus for a bimodal metallicity distribution generating
the observed color bimodality, the case of NGC\,1399 is less
obvious. By comparing the CCRs for the sources in approximately
equivalent spatial intervals of the two galaxies, we observe
non-negligible differences. The CCRs for NGC\,1399 annuli appear
shifted toward redder \ui for a given \gi and these CCRs are slightly, but
  noticeably, more nonlinear than in NGC\,3115. Assuming all GCs are
approximately equally old, it is easy to associate the redder colors
with on average higher metal content. The slightly different shapes of
the CCRs might be related either to the presence of age structures,
such as age gaps, or to other physical parameters affecting
the horizontal branch morphology, for example \citep[e.g., helium
  content;][]{dantona05,gratton10}. The efficiency of
galaxy and GCSs cannibalism in the core of Fornax for NGC\,1399 may have led to a
stratification of GC subcomponents, producing as a result a
collection of GCs with a broad range of metallicity and no distinctive
age structure; instead, there may be a gap between the
ages of metal-rich and metal-poor populations which appear
more distinctly separated for NGC\,3115.

Together with previous results from \citet{blake12a}, the differences
between the two sets of CCRs might be pointing out a possible role of
nonlinear color-metallicity relations in shaping the observed color
distribution in NGC\,1399. The complex nature of the GCS around this
bright galaxy hampers the simple equivalence of
color-bimodality$\equiv$metallicity-bimodality, hence raising key
questions: how frequently is the color bimodality we observe due to
real metallicity bimodality? Or, viceversa, how often is an artifact
generated by some projection effect on a parent distribution that
would otherwise have any other shape (multimodal, smooth, and broad,
asymmetric, etc.)?

The MW GCS is another system we know has a well-known bimodal \feh
distribution. If placed at the distance of 10/20 Mpc, i.e., at the
distance of NGC\,3115/Fornax cluster, the MW with its bright companion
M\,31 at $\sim1.15\degr/2.3\degr$ - and a number of companions
brighter than $M_V\sim-15$ mag - appears in an environment with
intermediate density that is not as isolated as NGC\,3115 nor as
companion rich as NGC\,1399. Hence, it would be tempting to look for
purely bimodal \feh distributions in GCS within galaxies in small
groups of galaxies, such as the MW, or in isolated galaxies, such as
NGC\,3115. However, the case of the GCS in M\,31 is not as obvious as
the MW, with a multimodal rather than bimodal \feh distribution
\citep{galleti09,caldwell11}.  NGC\,5128 and NGC\,4494 are similar
cases of galaxies in small groups or isolated from large groups with
well-studied GCSs and bimodal color distributions, where the
spectroscopically derived GCS metallicity shows complex distributions
with perhaps three peaks
\citep{beasley08,woodley10,foster11,usher12}. The complex GCS for
these three large galaxies gives evidence that these galaxies had more
active formation histories than the MW. Nevertheless, given the
locations in sparsely populated regions, their formation histories
were not as turbulent as those of cluster ellipticals, such as
NGC\,1399 \citep{puzia14,webb16}

A further difference we observed between NGC\,3115 and NGC\,1399 is
related to the local specific frequency $S_N(<r)$, i.e., the $S_N$
value obtained from the magnitude and GC number integrated within a
given radius, a relative comparison largely insensitive to any GC
magnitude incompleteness. We observe a steady increase of
$S_N(<r)^{blue}$ for the blue GCs in NGC\,3115, which is not
unexpected given the way $S_N$ is defined, because the blue GC density
profile is more extended than the galaxy light. In the case of
NGC\,1399 we first observe a steep increase of $S_N(<r)^{blue}$
followed by a relatively flat region and then a possible further rise
of the local specific frequency at large galactocentric
radii. Although our analysis of $S_N(<r)$ is not quantitatively as
refined as the surface brightness analysis presented in
\citet{iodice16}, the flat region we identify in the local specific
frequency diagram matches with the break radius found by
\citeauthor{iodice16} at $r\sim10\arcmin$, which marks the transition
between the galaxy and its stellar halo fading into the intracluster
light.

The behavior of the local specific frequency of red GCs,
$S_N^{red}(<r)$, appears very similar between the GC systems of the
two galaxies. The step-wise behavior of red GCs and galaxy field
stars is commonly assumed to be evidence of their common formation
history \citep{forbes12,kartha16}. Furthermore, our results on $i)$
the nearly flat $S_N^{red}(<r)$ out to $\sim25$ effective radii; $ii)$
the similar in behavior of red GCs and the galaxy with radius,
and within the limits of the approximate estimate of the magnitude
incompleteness, and; $iii)$ the close absolute values of $S_N^{red}(<r)$
despite the large differences existing between the two galaxies
compared here (we find $S_N^{red}(<r)$ ranging between $\sim1$ and
$\lsim1.7$ for both galaxies), imply a deeper similarity not
only of the red GCs and galaxy field stars within the galaxy, but also
from one galaxy to the other, even in extremely different
environments. We will further inspect and discuss the appearance of
such similarity with the future VEGAS and FDS datasets.

A final noteworthy similarity of the two galaxies is the crossing
  radius of the $S_N^{blue}(<r)$ and $S_N^{red}(<r)$  curves,
  which appear very similar in the adopted normalized distance units.

\subsection{Cosmological context}

The literature on GCSs formation and evolution, in connection with the
history of galaxy formation and evolution, is very rich.  As already
mentioned in the introductory section, in a cosmological context the
scenarios proposed for GC formation can be broadly classified into two
families. In one case the blue and red GCs have an age gap of $\lsim$2
Gyr and in the second they are approximately coeval.

In the age-gap case, there seems to be consensus toward a scenario
in which the earliest stellar populations in a protogalaxy, halo field
stars, and metal-poor GCs form early with GCs growing from the rare
density peaks of the primordial dark matter field, which is a process
interrupted by cosmic reionization \citep{diemand05,moore06}. In this
context, blue GCs would have nearly constant abundances with small
deviations caused by different local reionization epochs, an
expectation that is compatible with the small color scatter of blue
GCs \citep{peng06,brodie06,harris06a}.
The metal-rich GC population subsequently formed after a dormant period in
the dissipational processes that built up the bulk starlight of the
massive seed galaxies with mean \feh of the system following a
mass-metallicity relation.

Other authors, still within the blue and red GCs age-gap scenario, adopted
slightly different schemes more or less explicitly rejecting the role
of reionization in stopping the formation of metal poor clusters
\citep[see][and references therein]{forbes15}.
In general, any scenario proposing different mechanisms and/or epochs
of formation for the two GC subpopulations should explain the strong
similarities of the two subpopulations.

In the family of models in which the red and blue GCs are coeval,
\citet{cote98,cote00,cote02} proposed GCSs formation from
dissipationless hierarchical growth. The metal-poor GC component is
accreted to the massive seed galaxy through mergers or tidal
stripping. Also in this scenario the blue GCs are predicted to have
common properties with the galaxy halo. The metal-rich clusters
represent the intrinsic GC population of massive galaxy seeds.

To date, observational evidence does not allow us to firmly establish the
existence of an age gap between the blue and red GCs, not even in the
MW GC system \citep{strader05,marin09,vandenberg13}. Furthermore, as
discussed in previous sections, the last decade has seen increasing
interest toward the projected bimodality, i.e., the possibility that a
given fraction of the observed color bimodality of extragalactic GCSs
is not necessarily a one-to-one match of the \feh distribution.
Hence, currently there is no consensus about the existence of age gaps
between GC subpopulations and on the ubiquitousness of metallicity
bimodality.

Summarizing from above, two approximate scenarios can be drawn.

In the first case, blue metal-poor GCs form at the very early stages
in galaxy evolution from primordial density fluctuations. The
formation process is spatially very uniform and is halted by a spatially
extended event taking place on a relatively short timescale, such as
cosmic reionization. Red metal-rich GCs form later in a dissipational
or dissipative process that builds up the red GCs together with the
bulk of the host galaxy starlight.

In the second case, all GCs are coeval, form from primordial
fluctuations of the dark matter density field, and mean abundances of
the forming star clusters are directly affected by the depth of the
potential well of the protogalaxy. Then, the dissipationless
accretion of blue GCs from dwarf galaxies completes the shaping of the
present day GCSs.

In both scenarios, there are two requirements. First, the blue GCs
share common history with the galaxy halo, avoid feedback processes,
and thereby explain the relatively uniform properties of GCs and their
strong ties to the galaxy dark matter halo properties, such as the
essentially constant mass ratio $\eta=M_{GCS}/M_{halo}\sim4\times
10^{-5}$ \citep{hudson14} \citep[see
  also][]{peng08,spitler09,georgiev10}. Second, red GCs share common
history with the bulk starlight of the massive seed galaxy.

The results we presented in this paper, on the comparison between the
GCSs of NGC\,3115 and NGC1399, are compatible with both such
broadview scenarios.

In either the single age or age-gap case, the GCs in the isolated
NGC\,3115 had evolved after the last GC star forming episode without
much interaction with other massive companion galaxies. Even supposing
the latest GC formation ages to $\sim10$ Gyr
\citep{vandenberg13,forbes15}, the blue and red GCs, dark matter halo,
and galaxy stellar systems had a long common stage of evolution, which
was sufficient to virialize all the dark, stellar, and GC matter
components.  The presence of a clear \feh bimodality, hence of a
well-defined, metal-rich GC component with mean properties clearly
separated from the metal-poor component, could be the result of the
presence of a unique deep potential well, in which the red GCs formed
and evolved in a common epoch with the blue metal-poor GCs residing in
the halo; or this presence could result from multiple nearly
same-sized lower mass wells, each with similar red GCs, summing up in
the present-day GC populations.

Also the red GCs in NGC\,1399 share common properties with the galaxy
field stars. As for blue GCs, they appear to fade into the cluster
halo density distribution, a property common to other galaxy clusters
as well \citep{peng10,durrell14}. The GC metallicity bimodality
appears less obvious than in NGC\,3115, as would be expected in a
picture of hierarchical growth in a dense environment such as
Fornax. The accretion of red GCs associated with galaxy progenitors
with different masses, hence with a different position of the red GC
metallicity peak, generates the superposition of metallicity
distributions; after the addition of the more uniform blue GC
component, this superposition is hard to reconcile with a simple
bimodal \feh pattern.

A plausible scenario that seems to emerge is that the asymmetric blue
GC distribution visible in Figure \ref{gcmaps} is connected to the
large-scale cluster potential well \citep[the cluster halo, one of the
  three X-ray components detected by][]{paolillo02} preceding the bulk
of the cluster X-ray emitting gas, which is slowed by ram pressure
stripping due to the interaction with the halo of the NGC\,1316
subgroup. Red GCs instead follow the stellar mass density profiles of
the individual galaxies more closely and, in particular, the galactic
halo component from \citet{paolillo02} in the NGC\,1399 case.

\section{Summary and conclusions} 
\label{sec_summa}

In this study we compared the properties of the GCSs in NGC\,3115 and
NGC\,1399 as derived from the analysis of one square degree $u-$,
$g-$, and $i-$band images of the fields centered on the galaxies,
taken with the VST telescope as part of the VEGAS and FDS surveys.

The two galaxies analyzed are very differently from each other.
NGC\,3115 is one of the closest lenticular galaxies, at $\sim 10$ Mpc; this galaxy
is very isolated with only one close companion galaxy brighter than
$M_V\sim-15$ mag within a projected area of 100 $deg^2$.

NGC\,1399, located near the dynamical center of the Fornax cluster, is
the second brightest early-type galaxy of the cluster located in a densely
populated region hosting 43 galaxies brighter than $M_V\sim -15$ mag
over a projected area of 100 $deg^2$.

For the two galaxies we analyzed the surface density maps, color
distributions, and radial density profiles of GC candidates,
selecting the list of candidates using as reference the
morpho-photometric properties and colors of confirmed GCs available in
the literature. Our main conclusions are the following:

\begin{enumerate}

\item The field stellar light and GC density maps of NGC\,3115
  closely follow each other in terms of position of the center, elongation, and inclination.

\item For NGC\,1399, GC density maps confirm the known presence of
  substructures.  The GC overdensity covers a large portion of
  inspected area, providing supporting evidence in favor of its
  intergalactic nature. Globally, the morphology of the overdensity is
  asymmetric with an elongated east-west shape, and rich with
  substructures already discussed by, for example,
  \citet{dabrusco16}. On large scales, $\sim30\arcmin$, we do not find
  obvious correspondence between the GC surface density and the
  cluster-halo X-ray component first discussed by \citet{paolillo02},
  which has a southwest to northeast elongation. In particular, we
  observe a spatial offset in the NGC\,1399 GC centroid with respect
  to galaxy field stellar light that does not match the X-ray density
  contours. On smaller angular scales ($\lsim10\arcmin$) X-ray and GC
  contours appear more similar.

\item For both galaxies the GC system has larger spatial extent
  compared to galaxy light. For NGC\,3115, the semimajor axis where
  the density of GCs drops below 1 over an isodensity annulus one
  arcminute wide is $SMA_1\sim61\arcmin$, for comparison the surface
  brightness profile of field stars drops to $\mu_g\sim30$ mag
  arcsec$^{-2}$ at $SMA\sim25\arcmin$.

\item The shape of blue and red GC maps for NGC\,3115 do not differ
  much from each other.

\item The blue and red GC surface density maps for NGC\,1399 are
  notably different from each other: red GCs are mostly concentrated
  around bright galaxies and blue GCs occupy the entire observed region
  and generate the observed east-west overdensity.


\item Both galaxies show bimodal color distributions in \ui and \gi\ and are more prominent in \ui, especially for NGC\,3115.

\item The color-color relations of the two galaxies do not overlap.
  The GC candidate population in NGC\,1399 is shifted toward red
   and has CCRs that are are slightly different with respect to GCs in
  NGC\,3115, which is quantitatively more nonlinear.

\item Our results support existing results on GC metallicity
  bimodality as the main cause of the bimodal color distribution in
  NGC\,3115.  The case of NGC\,1399 is less obvious, as would be
  expected in a dense environment like Fornax.


\item The azimuthally averaged radial density profile of GC candidates,
for both galaxies, reaches larger galactocentric radii than the field
  stellar light distribution. Moreover, the density profile of red GCs
  is more concentrated than that of blue GCs and follows the
  underling galaxy light profile.

\item The local specific frequency for the total GC populations is
  notably different for the two galaxies. For NGC\,1399 $S_N$ is a
  factor of $\sim2$ higher than that for NGC\,3115, mainly because of
  extra blue GCs. By inspecting the local specific frequency for
    red GCs, the radial trend and  absolute values of
  $S_N^{red}(<r)$ appear very similar from one galaxy to
    another. Such similarity implies a deeper similarity not only of
  the red GCs and galaxy field stars within the galaxy, but also from
  one galaxy and the other, even in extremely different environments.

\item Whether or not red and blue GCs are coeval, our observations
  confirm, and further strengthen, the need for blue GCs to share a
  common history with the galaxy dark matter halo, and for red GCs to
  be more closely bound to the galactic stellar field component.


\item Overall, our analysis shows that field stars and GCs in
  NGC\,3115 are well thermalized with each other. In the isolated host
  environment the blue and red GCs, dark matter halo, and galaxy stellar
  systems had a long common stage of evolution that was sufficiently long to
  virialize all the components.

\item For NGC\,1399 the red GCs share common properties with the
  galaxy field stars, while blue GCs appear either associated with the
  galaxy halo or to fade into an intracluster GC component.  We
  speculate that the emerging pattern is that the two Fornax
  subclusters are falling toward each other, with the galaxies and
  the (halo) GCSs moving ahead of the gas component because of their
  noncollisional nature.

\end{enumerate}

Additional wide-field imaging studies of GC populations for a large
number of galaxies of different masses and in various environments,
together with new constraints on the relative ages of various GC
subpopulations (if any), along with revised formation models free from
unneeded observational constraints, will be important in assessing the
formation process of GCs and of all the galaxy components sharing
common evolutionary paths with GCs. As the VEGAS and FDS surveys
proceed, more multiband imaging data over large areas for early-type
galaxies will be available; this resource is extremely useful for this
purpose.

\begin{acknowledgements}

The data reduction for this work was carried out with the
computational infrastructures of the VST Center at Naples (VSTceN). We
gratefully acknowledge INAF for financial support to the VSTceN.  We
acknowledge the usage of the HyperLeda database \citep{makarov14},
\url{http://leda.univ-lyon1.fr}.  This research has made use of the
NASA Astrophysics Data System Bibliographic Services, the NASA
Extragalactic Database, the SIMBAD database, operated at CDS,
Strasbourg, and of Aladin sky atlas developed at CDS, Strasbourg
Observatory, France.  Based on observations collected at the European
Organisation for Astronomical Research in the Southern Hemisphere
under ESO programs 088.B-4012(A), 090.B-0414(B), 090.B-0414(C),
090.B-0414(D), 092.B-0623(C), 092.B-0623(D), 094.B-0496(A).

Michele Cantiello acknowledges John L. Tonry for providing the code
colmerge, used to match the photometric tables, and Lucia Panico for
useful discussions.
\end{acknowledgements}


\bibliographystyle{aa}
\bibliography{cantiello_jan17}



\begin{appendix}
\label{appendix}

\section{Sources in the fields}

The VST images used for the present study are extremely rich in
details. The complete catalog of selected sources in the three bands
used is available both via CDS interface, and through the VEGAS survey
web pages. Here we report several interesting objects, in particular
for the field of NGC\,3115.

\subsection{Ultra Compact Dwarfs (UCDs)}
\label{app_ucd}
The selection of UCDs was carried out based on the same approach
adopted for GCs, i.e., we rely on the observed properties of known UCD
to select a sample of candidates. As already discussed in
\citet{cantiello15}, the sample of UCDs present in \citet{jennings14}
reveals significant contamination from non-UCDs. Hence, we decided to
use as reference the sole sample of UCDs from the Fornax cluster
sample. As shown in the color-color diagrams in Figure \ref{ugi}, the
UCD candidates from NGC\,3115 are highly scattered around the GCs and
the galaxies sequences, while the UCDs candidates close to NGC\,1399
-- all spectroscopically confirmed -- line up with the GC sequence,
with negligible scatter around it.

The evidence shown in \citet{cantiello15} that some UCDs selected by
\citeauthor{jennings14} are clearly background objects, together with
the large color-scatter around the GC and SSP sequence, and the purely
photometric nature of the selection criterion adopted by the authors,
motivated our choice to reject the sample of UCDs from
\citeauthor{jennings14} for defining the selection criteria of
UCDs. We rather adopted the properties of UCDs in Fornax for selecting
UCD candidates around NGC\,3115, i.e., we assumed that the expected
locus of UCDs in (nearly) all the parameter spaces does not change
significantly for the two galaxies. In particular, with respect to the
GC selection parameters reported in Table \ref{tab_selparam}, we
adopted larger values for the morphological parameters, magnitude
limits $M_g\geq-10$ mag and $M_i\geq-11$ mag, while the same color
selection as for GCs was used for UCDs. No size cut has been applied
to UCDs.

  \begin{figure} 
   \centering
   \includegraphics[width=0.9\hsize]{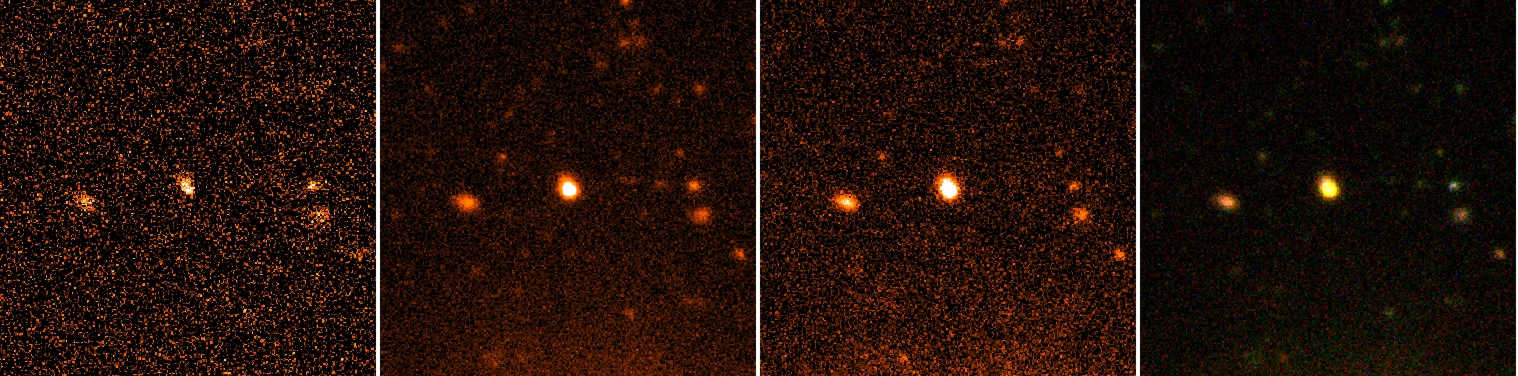}
   \caption{From left to right: $ugi$ and RGB thumbnail of the UCD
     candidate selected in the $\sim$1 sq. degree area centered on
     NGC\,1399. The obvious asymmetry of the source, especially
     seen in the $u$ band cutout, is shown. The boxes are $1\arcmin$ on
     each side.}
   \label{ucd1399}
   \end{figure}
 
  \begin{figure} 
   \centering
   \includegraphics[width=0.9\hsize]{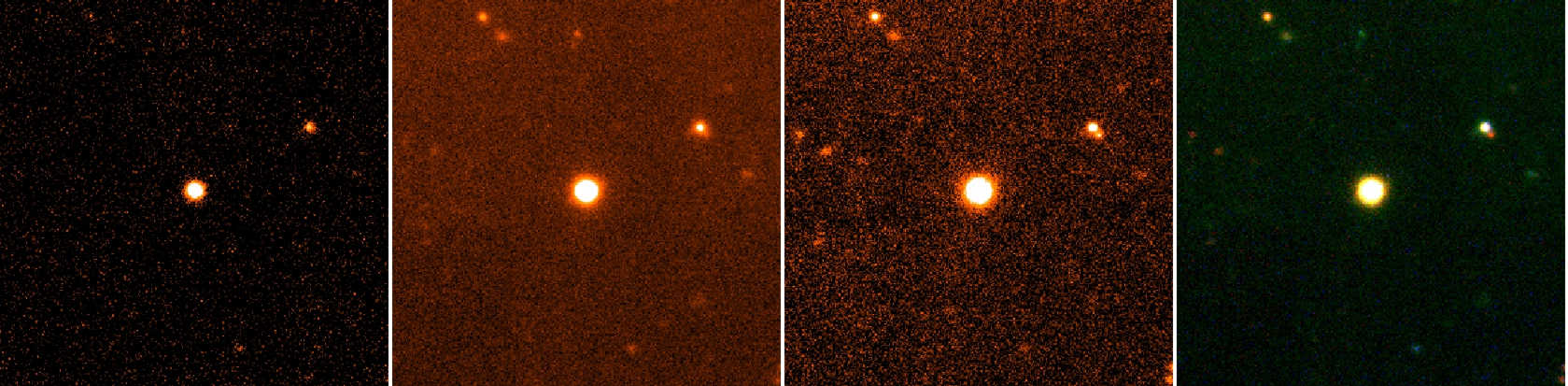}
   \caption{From left to right: $ugi$ and RGB thumbnail of one UCD candidate
     in the $\sim$1 sq. degree area centered on NGC\,3115. The boxes
     are $1\arcmin$ on each side.}
   \label{ucd3115}
   \end{figure}

For NGC\,1399, adopting such criteria calibrated on known UCDs and
after rejecting objects too close to bright contaminants (either stars
or other galaxies) or to the edges of the frame, we ended up with a
selection of 86 candidates. To further clean the sample, we analyzed
the azimuthally averaged radial profiles of the 86 UCD candidates (and
of their power spectra), comparing these candidates with the profiles
of known UCD candidates. The culled sample contained 13 objects. By
cross-correlating the sample of 13 objects with available catalogs
(NED, VizieR), we obtained that 4 of these objects were observed and
spectroscopically classified as foreground stars by
\citet{mieske02,mieske04}, and 8 more are UCDs from other
catalogs. Hence, the UCD selection adopted has a success rate of
$\sim67\%$. Only one of the candidates, at R.A.(J2000)=03h39m51.2s
Dec.(J2000)=-35d37m28.6s, does not match with any previously known
extragalactic source, and has measured proper motion compatible with
zero within uncertainties \citep{smart13}. The source overlaps with
the GALEX Medium Imaging Survey Catalog source GALEXMSC
J033951.22-353728.2, with $FUV{-}NUV_0\sim1.09,$ which is within the
range of observed UV colors for UCDs \citet{mieske08}, but is also
consistent with the locus of MW stars. Figure \ref{ucd1399} shows the
$ugi$ and RGB color thumbnail centered on the UCD candidate. Given the
significant asymmetry of the source, similar to the cases discussed in
\citet{voggel16} and \citet{wittmann16}, and its proper motion
consistent with zero, we are inclined to consider it an extragalactic
source. However, whether it is a UCD or not cannot be concluded from
the present dataset. Spectroscopic observations are needed to
definitely classify the source.

For NGC\,3115, a sample of 76 UCD candidates was preselected. After
inspecting the azimuthal averaged radial profiles (and the profiles of
the power spectra) and cross-correlating with available catalogs, the
final list of UCD candidates contains 24 objects listed in table
\ref{tab_ucd}.  One of the selected candidates is shown in Figure
\ref{ucd3115}.
\begin{table*}
  \tiny
\caption{\label{tab_ucd} Coordinates of UCD candidates in the field of NGC\,3115}
\centering
\begin{tabular}{lccccc}
  \hline\hline
ID & R.A.(J2000) &  Dec(J2000) & $m_u$ (mag) & $m_g$ (mag) & $m_i$ (mag) \\
\hline
1  &  151.35565  &   -8.24734 &   20.387 $\pm$ 0.008   &   19.529 $\pm$ 0.002   &   19.090 $\pm$ 0.003   \\
2  &  151.37987  &   -8.17344 &   21.257 $\pm$ 0.012   &   19.572 $\pm$ 0.002   &   18.654 $\pm$ 0.002   \\
3  &  151.39560  &   -8.15749 &   19.800 $\pm$ 0.005   &   18.417 $\pm$ 0.001   &   17.715 $\pm$ 0.001   \\
4  &  151.40596  &   -8.05649 &   19.609 $\pm$ 0.005   &   18.753 $\pm$ 0.002   &   18.229 $\pm$ 0.002   \\
5  &  151.35355  &   -7.98409 &   19.577 $\pm$ 0.005   &   18.397 $\pm$ 0.001   &   17.643 $\pm$ 0.001   \\
6  &  151.39183  &   -7.98296 &   19.990 $\pm$ 0.006   &   18.652 $\pm$ 0.002   &   17.832 $\pm$ 0.002   \\
7  &  151.58919  &   -7.96235 &   20.365 $\pm$ 0.008   &   19.537 $\pm$ 0.002   &   19.027 $\pm$ 0.003   \\
8  &  151.42009  &   -7.85778 &   20.337 $\pm$ 0.007   &   18.705 $\pm$ 0.001   &   17.739 $\pm$ 0.001   \\
9  &  151.50733  &   -7.78802 &   19.587 $\pm$ 0.005   &   18.323 $\pm$ 0.001   &   17.561 $\pm$ 0.001   \\
10 &  151.00620  &   -7.74170 &   20.475 $\pm$ 0.008   &   19.072 $\pm$ 0.002   &   18.220 $\pm$ 0.002   \\ 
11 &  151.56052  &   -7.72850 &   19.382 $\pm$ 0.004   &   18.141 $\pm$ 0.001   &   17.469 $\pm$ 0.001   \\ 
12 &  151.11798  &   -7.70166 &   18.682 $\pm$ 0.003   &   17.704 $\pm$ 0.001   &   17.000 $\pm$ 0.001   \\ 
13 &  151.55324  &   -7.61832 &   20.807 $\pm$ 0.009   &   19.428 $\pm$ 0.002   &   18.522 $\pm$ 0.002   \\ 
14 &  151.42659  &   -7.51542 &   19.863 $\pm$ 0.006   &   18.866 $\pm$ 0.002   &   18.355 $\pm$ 0.002   \\ 
15 &  151.16905  &   -7.47111 &   20.706 $\pm$ 0.010   &   19.726 $\pm$ 0.003   &   19.030 $\pm$ 0.003   \\ 
16 &  151.40532  &   -7.46064 &   20.193 $\pm$ 0.007   &   18.988 $\pm$ 0.002   &   18.292 $\pm$ 0.002   \\ 
17 &  151.25995  &   -7.45713 &   21.118 $\pm$ 0.015   &   19.587 $\pm$ 0.003   &   18.797 $\pm$ 0.002   \\ 
18 &  151.78131  &   -7.37360 &   20.158 $\pm$ 0.007   &   18.787 $\pm$ 0.002   &   17.919 $\pm$ 0.002   \\ 
19 &  151.77539  &   -7.26576 &   21.850 $\pm$ 0.020   &   19.572 $\pm$ 0.003   &   18.193 $\pm$ 0.002   \\ 
10 &  151.53886  &   -7.21417 &   20.742 $\pm$ 0.016   &   19.837 $\pm$ 0.005   &   19.233 $\pm$ 0.006   \\ 
21 &  151.60184  &   -7.83037 &   20.543 $\pm$ 0.008   &   18.677 $\pm$ 0.001   &   17.296 $\pm$ 0.001   \\ 
22 &  151.25984  &   -7.81241 &   19.896 $\pm$ 0.006   &   18.416 $\pm$ 0.001   &   17.491 $\pm$ 0.001   \\ 
23 &  151.12245  &   -7.76488 &   18.353 $\pm$ 0.002   &   17.404 $\pm$ 0.002   &   16.547 $\pm$ 0.001   \\
24 &  151.43657  &   -7.76087 &   20.565 $\pm$ 0.008   &   19.491 $\pm$ 0.002   &   18.930 $\pm$ 0.003   \\ 
\hline 
\end{tabular}
\end{table*}

\begin{table}
\caption{\label{tab_peculiar} Coordinates of peculiar objects in the
  field of NGC\,3115} \centering
\begin{tabular}{lcc}
  \hline\hline
ID & R.A. (J2000) &  Dec (J2000) \\
\hline
LSB\#1          & 151.55289  &   -7.5001689  \\ 	     
LSB\#2          & 151.61118  &   -7.5495657  \\            
LSB\#3          & 151.6404   &   -7.5095855  \\            
LSB\#4          & 151.3558   &   -7.7283793  \\            
LSB\#5          & 151.36609  &   -7.7305557  \\            
LSB\#6          & 151.29081  &   -7.5397223  \\            
LSB\#7          & 151.30012  &   -7.5492535  \\            
LSB\#8          & 151.47394  &   -7.8804395  \\            
Interacting\#1  & 151.41344  &   -7.4736593  \\            
Interacting\#2  & 150.78087  &   -8.1403071  \\            
Interacting\#3  & 150.94656  &   -7.4637473  \\            
Interacting\#4  & 151.68343  &   -7.7048583  \\            
Ring\#1         & 150.77724  &   -7.4205627  \\            
Ring\#2         & 151.30856  &   -8.0436752  \\            
Ring\#3         & 151.62259  &   -7.6092057  \\            
Ring\#4         & 151.60656  &   -7.5072465  \\	     
Lens          & 151.71811  &   -7.9269669  \\	           
\hline 
\end{tabular}
\end{table}

\subsection{Other peculiar objects in the frame of NGC\,3115}

\begin{figure*} 
  \centering
   \includegraphics[width=1.\hsize]{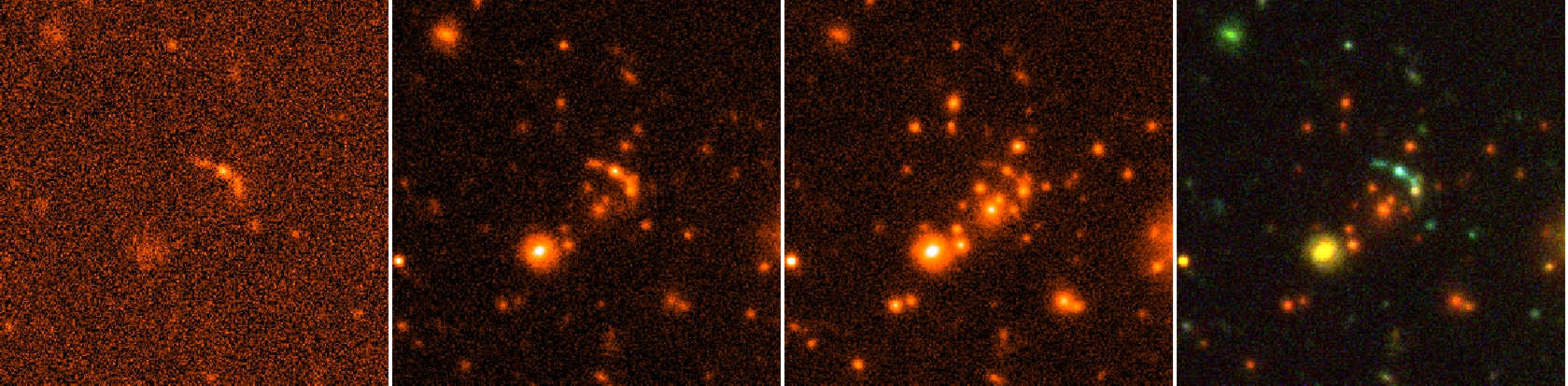}
   \caption{From left to right: $ugi$ and RGB thumbnail of a lens
     candidate. The cutout size is $1\arcmin\times 1\arcmin$.}
   \label{lens}
\end{figure*}

\begin{figure} 
   \centering
  \includegraphics[width=0.9\hsize]{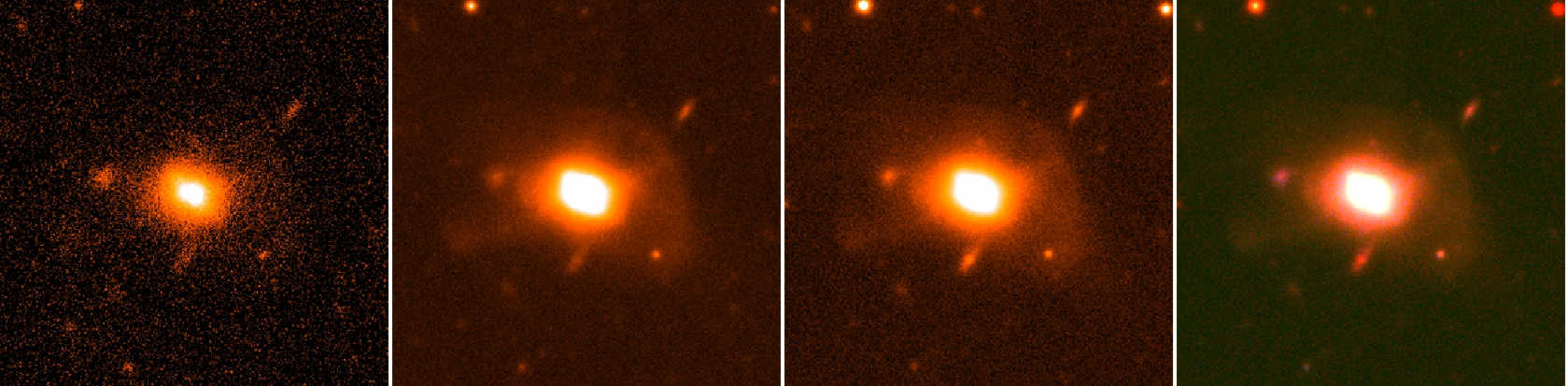}
  \includegraphics[width=0.9\hsize]{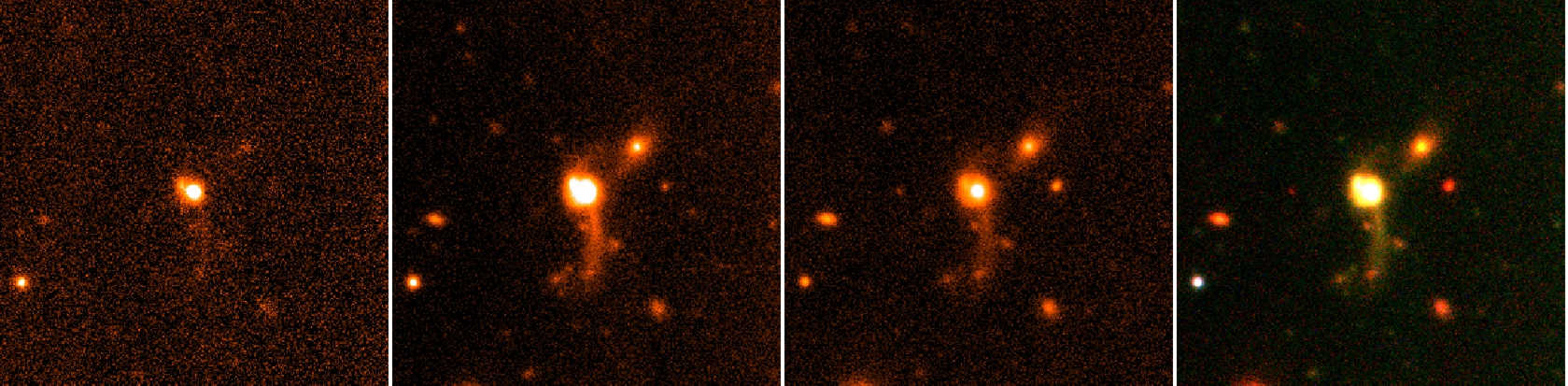}
  \includegraphics[width=0.9\hsize]{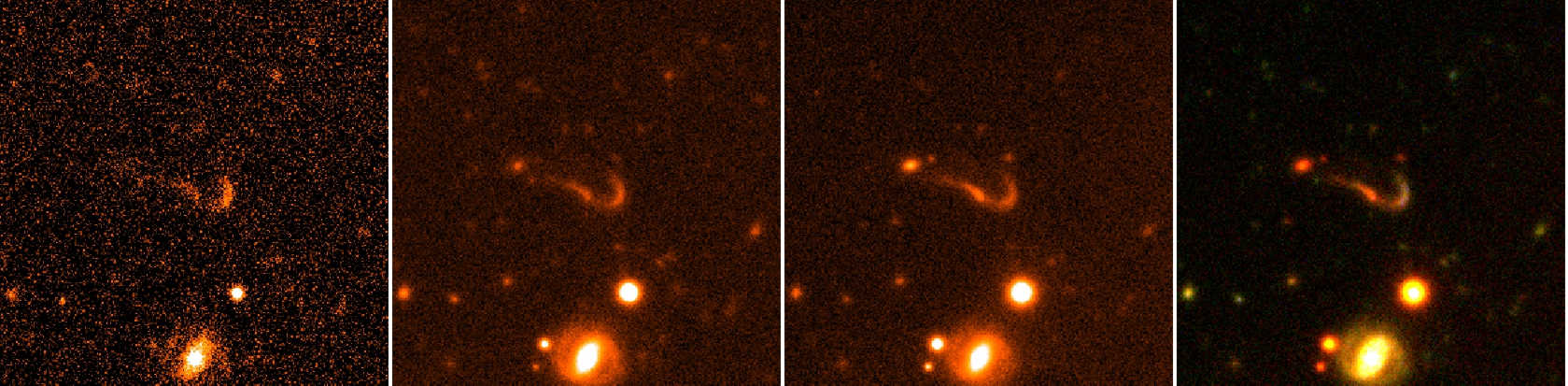}
  \includegraphics[width=0.9\hsize]{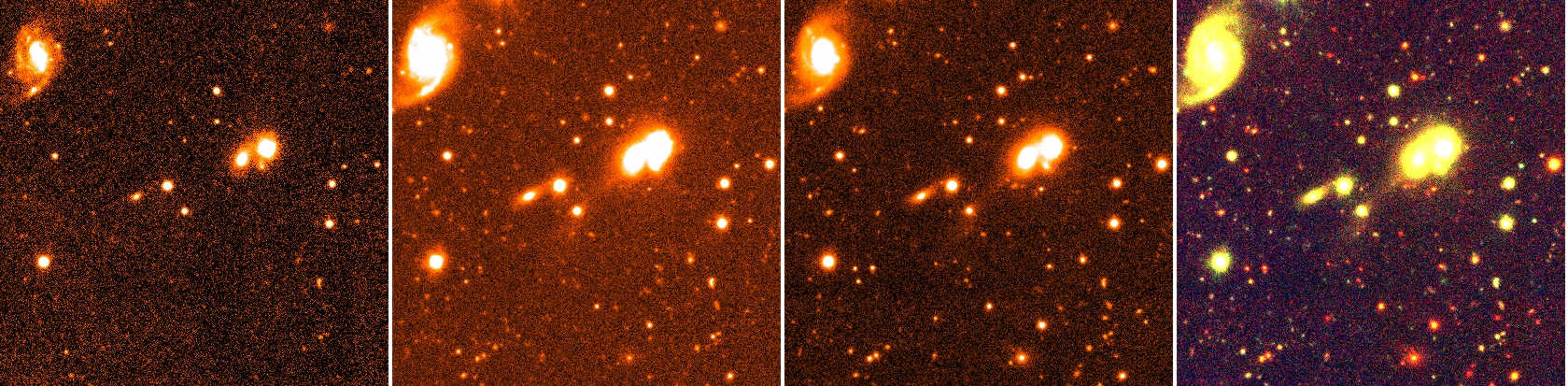}
  \caption{From left to right: $ugi$ and RGB thumbnail of interacting
    candidates from ID\#1 to \#4. All boxes are $1\arcmin\times
    1\arcmin$, except for the lowermost cutout showing an area of
    $3\arcmin\times 3\arcmin$.}
  \label{interacting}
\end{figure}

\begin{figure} 
   \centering
  \includegraphics[width=0.9\hsize]{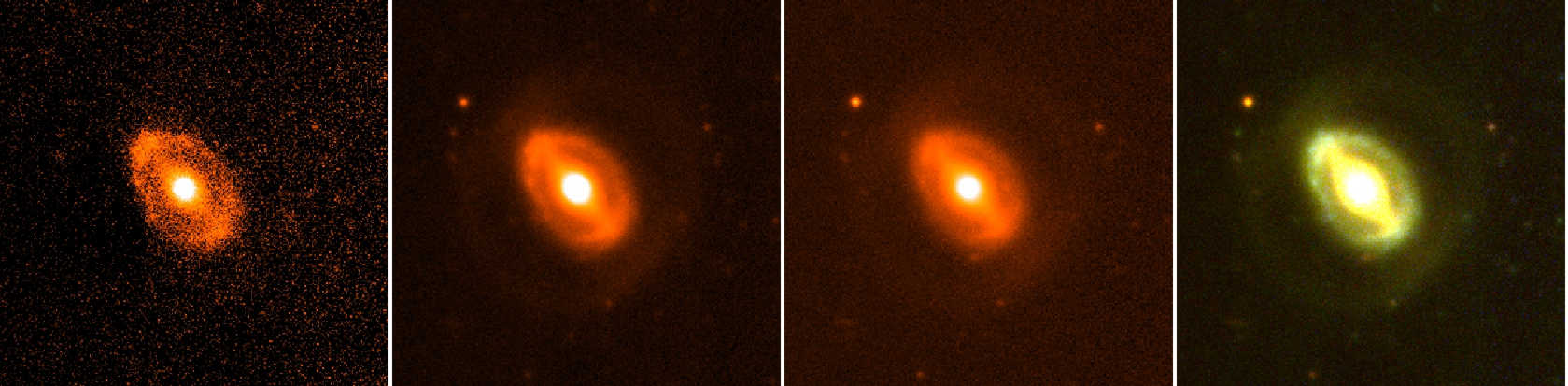}
  \includegraphics[width=0.9\hsize]{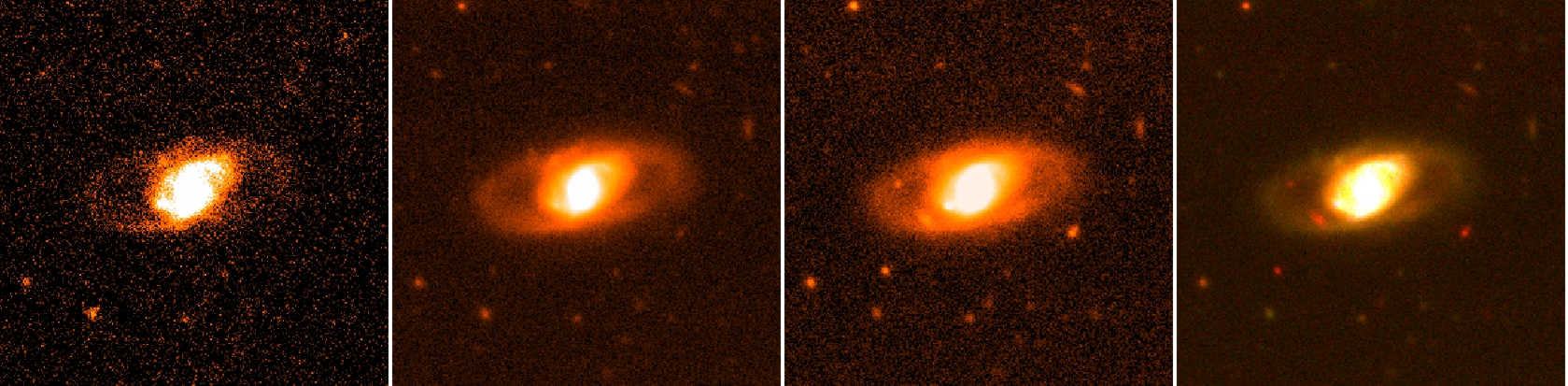}
  \includegraphics[width=0.9\hsize]{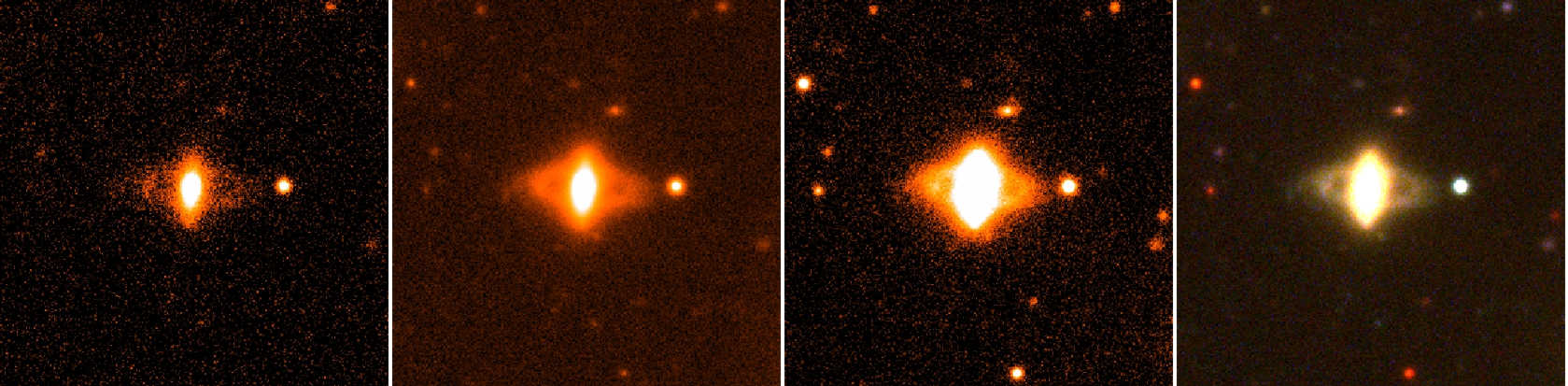}
  \caption{From left to right: $ugi$ and RGB thumbnail of ring
    galaxies candidates from ID\#1 to \#3. The ringed galaxy with
    ID\#4, close to LSB\#2-LSB\#3 candidates, appears in the second
    row of thumbnail in Figure \ref{lsb}. The cutout size is
    $1\arcmin\times 1\arcmin$. }
   \label{ring}
   \end{figure}

\begin{figure} 
   \centering
  \includegraphics[width=0.9\hsize]{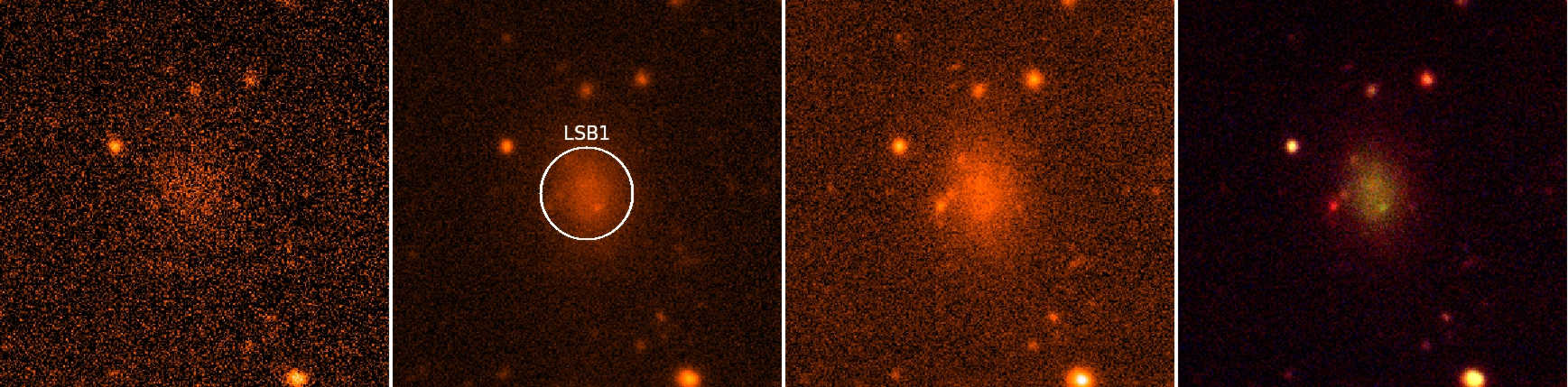}
  \includegraphics[width=0.9\hsize]{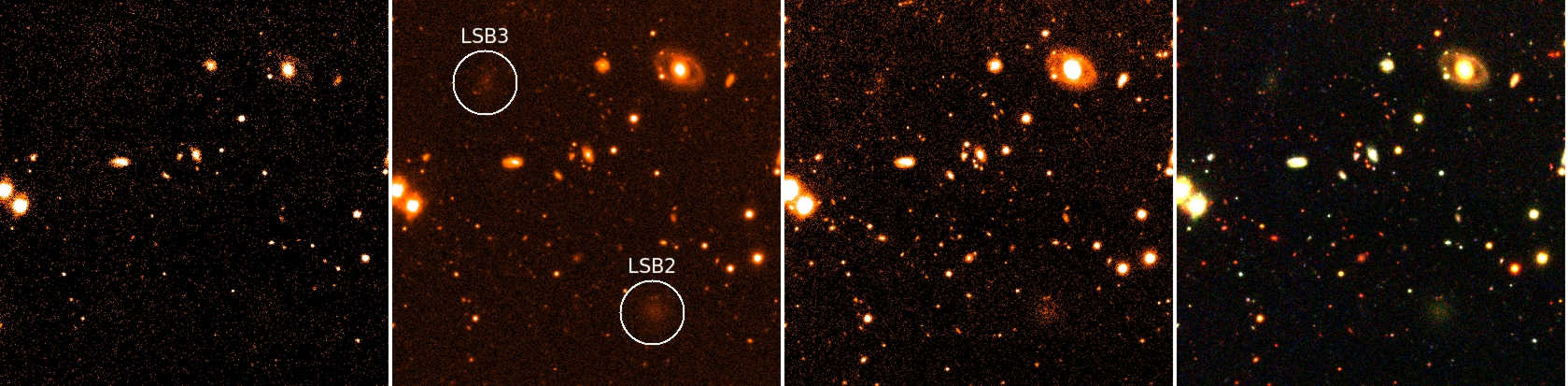}
  \includegraphics[width=0.9\hsize]{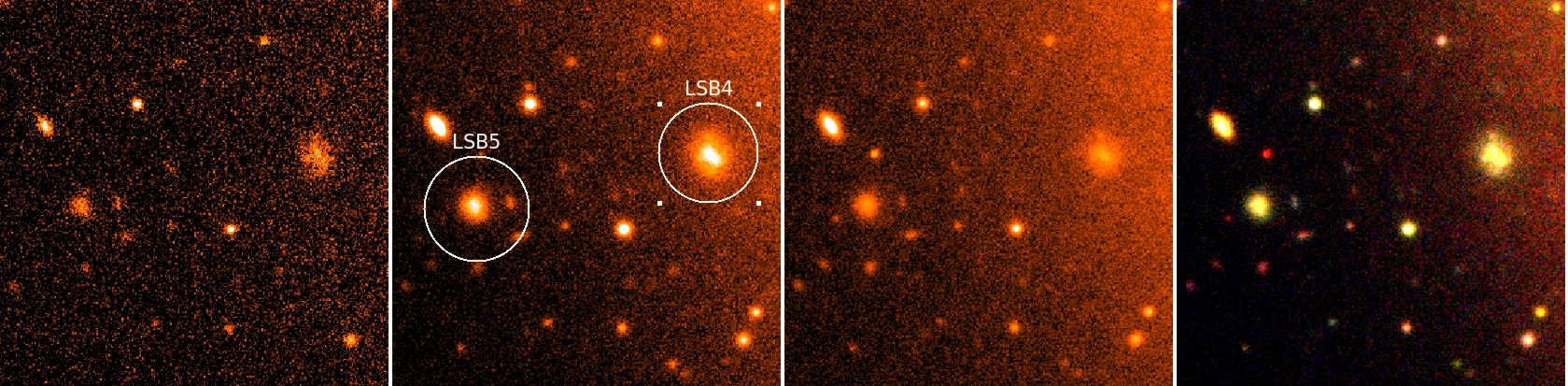}
  \includegraphics[width=0.9\hsize]{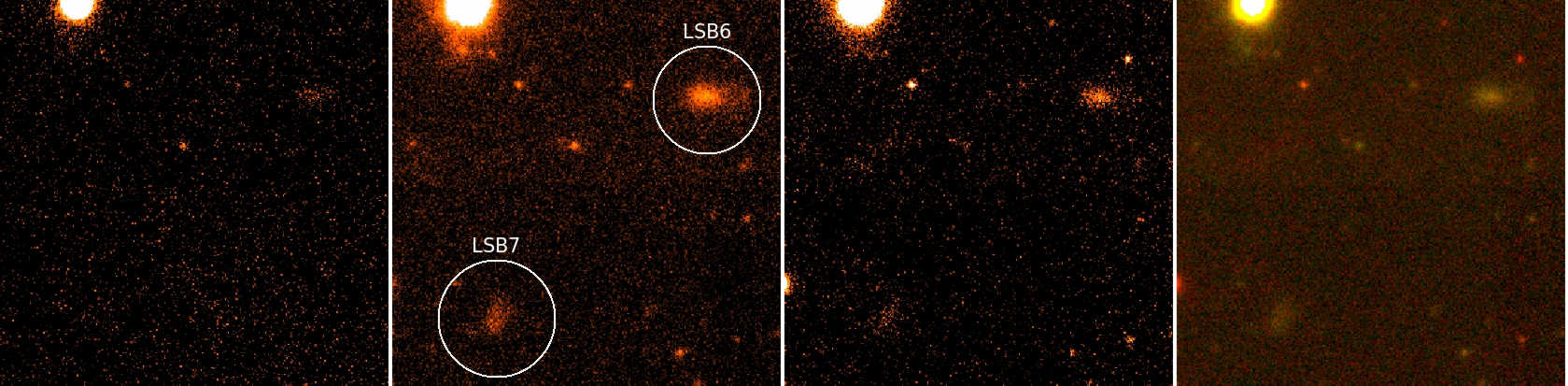}
  \includegraphics[width=0.9\hsize]{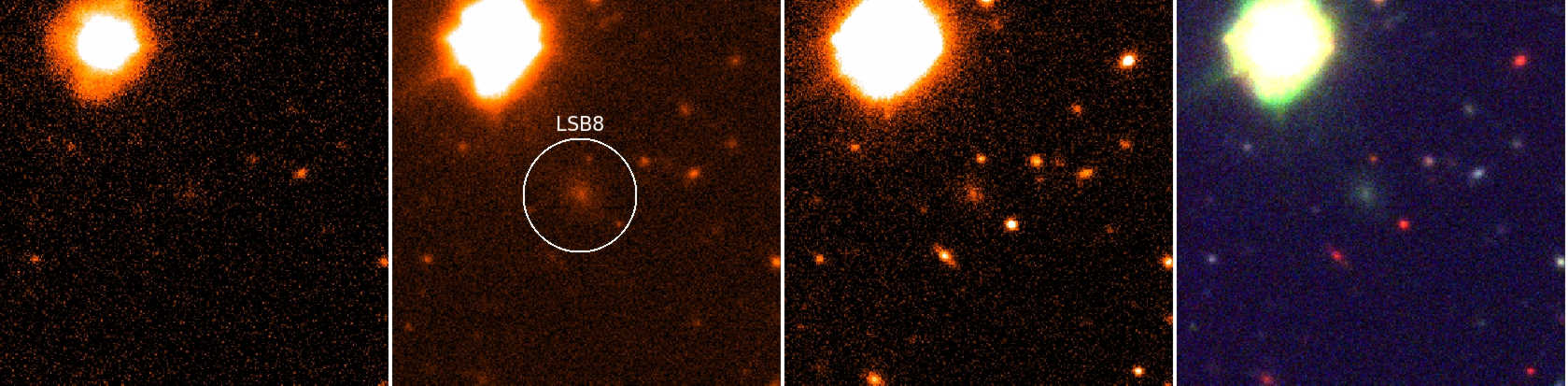}
  \caption{From left to right: $ugi$ and RGB thumbnail of LSB
    candidates. The cutout size is $1\arcmin\times 1\arcmin$, except
    for the second cutout showing an area of $4\arcmin\times
    4\arcmin$.}
   \label{lsb}
   \end{figure}

We also inspected the VST frame in search of low surface brightness
(LSB) objects and other peculiar sources. The coordinates of the
selected objects are given in table \ref{tab_peculiar}. The thumbnail
of the objects are reported in the Figures \ref{lens}-\ref{ring}. In
particular, we show what seems to be a cluster of galaxies with a
strong gravitational lens (Figure \ref{lens}), several interacting
galaxies (Figure \ref{interacting}), some galaxies with rings and bars
(Figure \ref{ring}) and, finally, the selected LSB candidates (Figure
\ref{lsb}).

\newpage

\end{appendix}

\listofobjects

\end{document}